\documentstyle[11pt,a4wide,cite,epsfig]{article}

\def\no{\noindent}
\def\sy{\scriptscriptstyle}
\def\dps{\displaystyle}

\def\square{\kern1pt\vbox{\hrule height 1.2pt
\hbox{\vrule width 1.2pt\hskip 3pt
\vbox{\vskip 6pt}\hskip 3pt
\vrule width 0.6pt}\hrule height 0.6pt}\kern1pt}
\newcommand{\be}{\begin{equation}}
\newcommand{\ee}{\end{equation}}
\newcommand{\bear}{\begin{eqnarray}}
\newcommand{\ear}{\end{eqnarray}}

\def\O{{\cal O}}
\def\P{{\cal P}}
\def\D12{{\tau_1 - \tau_2}}
\def\D13{{\tau_1 - \tau_3}}
\def\D23{{\tau_2 - \tau_3}}
\def\half{{1\over 2}}

\def\fourth{{1\over 4}}

\def\g12{{\dot {G_B}_{12}}}
\def\g13{{\dot {G_B}_{13}}}
\def\g23{{\dot {G_B}_{23}}}

\def\d{\mbox{d}}
\def\e{\mbox{e}}
\def\g{\mbox{g}}

\def\ka{\kappa}
\def\la{\lambda}
\def\ro{\rho}
\def\si{\sigma}

\begin{document}

\renewcommand{\thefootnote}{\fnsymbol{footnote}}

\pagestyle{empty}

\setlength{\parindent}{0pt}

\hskip 10.5cm {\sl HD-THEP-97/24}
\vskip.0pt
\hskip 10.5cm {\sl HUB-EP-97/42}
\vskip .7cm

\begin{center}

\vskip2cm
{\Large The Higher Derivative Expansion of the Effective Action\\}
{\Large by the String Inspired Method. Part II.\\}

\vskip.7cm
{\large D. Fliegner$^{\rm \,\,a}$
\footnote{e-mail address D.Fliegner@ThPhys.Uni-Heidelberg.De},
        P. Haberl$^{\rm \,\,b}$
\footnote{e-mail address haberl@physik.rwth-aachen.de}, 
        M. G. Schmidt$^{\rm \,\,a}$
\footnote{e-mail address M.G.Schmidt@ThPhys.Uni-Heidelberg.De}, 
        C. Schubert$^{\rm \,\,c}$ 
\footnote{e-mail address schubert@qft2.physik.hu-berlin.de}}

\vskip.7cm 

{\it $^{\rm a}$Institut f\"ur Theoretische Physik,
Universit\"at Heidelberg\\
Philosophenweg 16,
D-69120 Heidelberg, Germany}

\vskip.2cm

{\it $^{\rm b}$Institut f\"ur Theoretische Physik,
RWTH Aachen,\\
Sommerfeldstr. 26-28,
D-52056 Aachen, Germany}

\vskip.2cm

{\it $^{\rm c}$Institut f\"ur Physik,
Humboldt Universit\"at zu Berlin\\
Invalidenstr. 110, D-10115 Berlin, Germany}

\vskip1.4cm

{\large\bf Abstract}

\end{center}

\begin{quotation}

\noindent
We apply the string-inspired worldline formalism to 
the calculation of the higher derivative expansion
of one-loop effective actions in non-Abelian gauge 
theory. For this purpose, we have completely 
computerized the method, using the symbolic 
manipulation programs FORM, PERL and M.
Explicit results are given to sixth order in the 
inverse mass expansion, reduced to a minimal basis 
of invariants specifically adapted to the method. 
Detailed comparisons are made with other gauge-invariant 
algorithms for calculating the same expansion.  
This includes an explicit check of all coefficients
up to fifth order.
\end{quotation}

\clearpage

\renewcommand{\thefootnote}{\protect\arabic{footnote}}
\setcounter{footnote}{1}

\pagestyle{plain}

\setcounter{page}{1}

\noindent

{\bf 1. Introduction}
 
\vskip.5cm

It is a well-known fact that one-loop amplitudes in 
quantum field theory can often be represented by 
worldline path integrals over the space of closed loops.
For instance, the one-loop effective action induced by
a spinor loop in an Abelian or non-Abelian background gauge
field may be written in the following way  
\cite{feynman,fradkin,brink,bordi,barducci,gitman,polybook}:

\begin{equation}
\Gamma\lbrack A\rbrack = -2 {\dps\int_0^{\infty}}
{dT\over T} \e^{-m^2T} {\rm tr} {\dps\int} 
{\cal D} x {\cal D} \psi \; \P
\exp\Bigl [- \int_0^T d\tau \Bigl 
(\fourth{\dot x}^2 + {1\over 2}\psi\dot\psi
+ ig A_{\mu}\dot x^{\mu} 
- ig \psi^{\mu}F_{\mu\nu}\psi^{\nu}
\Bigr )\Bigr ]\,.
\label{spinorpi}
\end{equation}

\no
Here the $x^{\mu}(\tau )$ are the periodic 
functions from the circle with fixed circumference  
$T$ (known as the Schwinger proper-time) into 
$d$-dimensional Euclidean spacetime, and the 
$\psi^{\mu}(\tau )$ their antiperiodic Grassmann valued 
(supersymmetric) partners. 
$\P$ denotes path ordering in the non-Abelian case.

This type of path integral representation has already 
proven useful for various types of calculations in 
quantum field theory  \cite{cks,hjs}. In particular, 
for the calculation of anomalies and index densities 
\cite{alv-g,alvwit,fw,bast,boer} 
superparticle path integrals have shown to be a remarkably 
powerful alternative to heat kernel methods  \cite{abp}. 

Recently, renewed interest in these integral 
representations has been triggered by the work of 
Bern and Kosower. In 1992 these authors established a 
new set of rules for one-loop calculations by 
representing amplitudes in ordinary quantum field
theory as the infinite string tension limits of
certain (super) string amplitudes  \cite{bk}. Those
rules are equivalent to standard Feynman rules \cite{bd},
but lead to a significant reduction in the number of
terms to be computed both in one-loop gauge theory 
 \cite{5glu} and quantum gravity  \cite{dannor}
calculations.

Strassler later showed  \cite{str1} that, 
for many cases of interest, the same integral 
representations can be obtained by evaluating 
worldline path integrals of the type eq.~(\ref{spinorpi})
in analogy to the Polyakov path integral,
i.e.~using one-dimensional perturbation theory.

This reformulation turned out to be well-suited 
to the calculation of one-loop effective actions
in general ( \cite{str2,ss,chd,adlsch,rss,sch}; 
see also  \cite{mckeon}), 
and highly efficient for the calculation of 
their inverse mass expansions  
\cite{fss,fhss1,fhss2,fhss3,flinter}. 

The inverse mass expansion (or, more generally, 
the higher derivative expansion) is a standard 
tool for the approximative calculation of one-loop
effective actions, and considerable work has gone 
into the determination of its coefficients.
It is applied in fields as different as chiral 
perturbation theory  \cite{ball}, high temperature 
physics and the theory of phase transitions.
In the latter context, it has been applied 
both to bubble nucleation during the electroweak 
phase transition  \cite{klsch} and to baryon number 
violation by sphaleron processes  \cite{khsch}. 

Little seems to be known, however, about the high order 
behaviour of this expansion  \cite{dodgrad,zhitnitsky}, 
and the question of its convergence. A recent all-order
calculation of the higher derivative expansion for
a specific example in three-dimensional quantum 
electrodynamics indicates that, at least in the 
case considered, this expansion is an asymptotic rather 
than a convergent series  \cite{dunne}.

Due to the rapidly growing capacity of computers to 
handle large numbers of terms in symbolic calculations,
in recent times there has been growing interest
in the explicit form of higher order coefficients of 
the inverse mass expansion 
 \cite{carson,vdv,abc,avra,bles,carsal,bargus}. 

In the first paper of this series \cite{fss}, we used the
`string-inspired method' to calculate this expansion 
to order $\O(T^7)$ in the proper-time parameter for the 
simplest case of an external scalar potential.

In the present paper, we consider the more general problem 
of constructing the inverse mass expansion for the case 
of both a background gauge field and a scalar potential.
Moreover, we have completed the computerization of the 
method, using the algebraic manipulation programs FORM 
\cite{form}, PERL \cite{PERL} and M \cite{M}.
This allows us to push the calculation to $\O(T^{12})$ in 
the pure scalar case, and to $\O(T^6)$ in the general case 
(the result for the scalar case has been presented to order 
$\O(T^{8})$ in  \cite{fhss1}).
With conventional methods, the inverse mass expansion has 
been obtained to order $\O(T^5)$ in the general 
case  \cite{vdv}, and only recently to order $\O(T^7)$ in 
the scalar case  \cite{lanyov}.

At those high orders it is, of course, essential to 
represent the result in the most compact form possible.
In the scalar potential case, our method turns out to have 
the very useful property of leading to {\sl minimal} bases 
of operators automatically, once cyclically equivalent terms 
have been identified. In the general case the result for the 
effective Lagrangian has to be further reduced using Bianchi 
identities and (possibly) transposition symmetry.
Usually those operations would have to be combined with 
partial integrations in spacetime, however this turns out 
not to be necessary in the present scheme.

In Chapter 2, we will explain our method of calculating the 
inverse mass expansion  \cite{ss}, which is a pure $x$--space 
version of the one proposed by Strassler  \cite{str1,str2},
made manifestly gauge invariant by using Fock--Schwinger gauge.

The results of this calculation will be presented in Chapter 3, 
reduced to a minimal basis of invariants specifically adapted 
to the algorithm. We explicitly present the effective Lagrangian 
for the pure gauge theory case, calculated to order $\O(T^6)$ in 
the inverse mass expansion.

Chapter 4 contains a technical comparison with previous 
calculations of higher derivative expansions. A considerable 
number of different algorithms for this type of calculation can 
already be found in the literature  \cite{avra,kikkawa,mwz,
fraser,tH,onofri,fow,zuk1,zuk2,nepo,chan,bog}, and we cannot 
possibly discuss all of them. We will therefore pay attention 
mainly to those methods which have already proven suitable for 
higher order calculations in gauge theory. Those are:

\begin{itemize}

\item The method developed by Onofri  \cite{onofri}, Fujiwara 
et al. \cite{fow} and Zuk  \cite{zuk1,zuk2} (Section 4.1).

This approach is also the one most closely related to our work, 
a fact which becomes particularly conspicuous in the path integral
formulation of  \cite{fow}. We will therefore spend some effort on 
a detailed comparison with that method.

\item A modified version of the method proposed by Nepomechie
 \cite{nepo} (Section 4.2).

This technique is not manifestly gauge invariant as it stands and 
thus less convenient for the present purpose. Still we will present 
a modified version of it, which turns out sufficiently efficient for
a explicit check of our results to $\O(T^5)$ completely and of $\O(T^6)$
partially.

\item The method invented by 't Hooft \cite{tH} and elaborated 
by van de Ven  \cite{vdv} (Section 4.3).  

\end{itemize}

Our conclusions will be offered in Chapter 5, where we will also 
discuss further possible generalizations.

Appendix A deals with the rather technical problem of constructing 
minimal bases of invariants for a background consisting of a scalar
field and/or a non-Abelian gauge field. In Appendix B we discuss 
the impact of different choices for the Green function used in the
evaluation of the path integral.

\newpage

\vskip.8cm
{\bf 2. The inverse mass expansion from the worldline path integral}
\vskip.5cm

First let us set up some terminology. We refer by the 
`higher derivative expansion' of an effective action to an 
expansion both in the number of external fields and the number 
of derivatives acting on the fields. This expansion exists in 
several versions, which differ by the grouping of terms. The 
one we will consider here is the `inverse mass expansion', 
which is usually obtained by writing the one-loop determinant 
in the Schwinger proper time representation

\begin{equation}
\Gamma [A,V] = -  {\rm log}({\rm det}\,M) =
- {\rm Tr}({\rm log}\,M) = {\dps\int_0^{\infty}}{dT\over T}
{\rm Tr}\; \e^{-TM}
\label{ime}
\end{equation}

\no
and expanding in powers of the proper-time parameter $T$. 
This groups together terms of equal mass dimension. 
Up to partial integrations in space-time, it coincides 
with the (diagonal part of the) `heat-kernel expansion' 
for the second order differential operator in question. 
In particular, every coefficient in this expansion is 
separately gauge invariant. 

Alternatively, one may calculate the same series up to a 
fixed number of derivatives, but with an arbitrary number 
of fields or potentials  \cite{carsal,zuk1,chan}.
Yet another option is to keep the number of external 
fields fixed, and sum up the derivatives to all orders. 
This leads to the notion of Barvinsky-Vilkovisky form 
factors \cite{ss,bv}.

Note that in general the proper-time integral eq.~(\ref{ime}) 
need not be convergent. It has to be regularized in some way,
e.g.~using a simple cut-off, $\zeta$-function or dimensional 
regularization. We will therefore work in $d$ dimensions from 
the beginning. Throughout this paper we will not perform the 
final $T$-integration, because we are interested in the explicit 
form of the coefficients of the inverse mass expansion only.  

In the present paper we consider the case of massive scalars 
in the loop and a background consisting of both a gauge field 
and a (possibly matrix valued) scalar potential. 
This corresponds to the following choice of the fluctuation 
operator $M$:

\begin{equation}
M = - D^2 + m^2 + V(x),
\label{defM}
\end{equation}

\no with $D_{\mu} = \partial_{\mu} + ig A_{\mu}$. 

The case of a matrix valued scalar potential is by far more
general than the pure scalar case. In particular it allows to
treat particles with spin in the loop and can therefore be
used to calculate fluctuation determinants, e.g.\ around 
the electroweak sphaleron \cite{carmcl,khsch}.
 
In the case of a fluctuation operator (\ref{defM}) the one-loop 
effective action can be expressed in terms of the worldline path 
integral (see e.g. \cite{polybook})

\begin{equation}
\Gamma\lbrack A,V\rbrack =  
{\dps\int_0^{\infty}} {dT\over T}
\e^{-m^2T} {\rm tr} \;
{\dps\int} {\cal D} x \; \P
\exp\Bigl [- \int_0^T d\tau
\Bigl (\fourth{\dot x}^2 
+ ig A_{\mu}\dot x^{\mu} + V(x)
\Bigr )\Bigr ].
\label{scalarpi}
\end{equation}

\no
The path integral will be path-ordered, except if both $A$ 
and $V$ are Abelian. The method applied generalizes to the 
spinor loop case without difficulty \cite{ss,rss}. 

To obtain an effective Lagrangian from this path 
integral we split it into an ordinary integral 
over the center of mass, and a path integral
over the relative coordinate:

\begin{eqnarray}
{\dps\int}{\cal D}x &=&
{\dps\int}d^d x_0{\dps\int}{\cal D} y\label{pathsplit}\\
x^{\mu}(\tau) &=& x^{\mu}_0  +
y^{\mu} (\tau )\\
\int_0^T d\tau\,   y^{\mu} (\tau ) &=& 0\, .
\label{splitcond}
\end{eqnarray}

\no
This leads to an expression for the effective action
in terms of an effective Lagrangian $\cal L$,

\begin{equation}
\Gamma [A,V] = \int d^d x_0 \,{\cal L}(x_0)\quad,
\label{defL}
\end{equation}

\noindent
where
${\cal L} (x_0)$ is represented as
an integral over the space of all loops with a
fixed common center of mass $x_0$.

To obtain the higher derivative expansion, we 
Taylor--expand both $A$ and $V$ around $x_0$,
and use $\dot x^{\mu} = \dot y^{\mu}$ to rewrite

\begin{eqnarray}
V(x) &=& e^{y\partial}V(x_0)\nonumber\\
\dot x^{\mu}A_{\mu}(x) &=&
{\dot y^{\mu}}e^{y \partial}A_{\mu}(x_0)\quad .
\label{taylor}
\end{eqnarray}

\no
The path-ordered interaction exponential is then 
expanded with the result

\begin{eqnarray}
\Gamma[A,V] &=& \int_0^\infty \!{dT\over T}
{\rm e}^{-m^2 T} {\rm tr}\int \! d^d x_0 \sum_{n=0}^\infty
\frac{(-1)^n}{n} T \int \! {\cal D}y \;{\rm exp} \Bigl[ -
\int_0^T \! d\tau {\dot{y}^2\over 4} \Bigr]\\
&& 
\int_0^{\tau_1 = T} \hspace{-0.4cm}d\tau_2 \ldots 
\int_0^{\tau_{n-1}} \hspace{-0.4cm}d\tau_n
\prod_{j=1}^n \Bigl[ {\rm e}^{ y(\tau_j)\partial_{(j)}} 
V^{(j)}(x_0)\!+\! ig \dot{y}^{\mu_j}(\tau_j) 
{\rm e}^{y(\tau_j) \partial_{(j)}} A_{\mu_j}^{(j)}(x_0) \Bigr]
\nonumber .
\label{expand}
\end{eqnarray}

\no
Here we have labeled the background fields, and the first
$\tau$--integration has been eliminated by using the freedom
of choosing the $0$ somewhere on the loop. This is also the
origin of the factor of $T\over n$.
Our normalization is such that for the free path integral

\begin{equation}
{\dps\int} {\cal D} y\,
 {\rm exp}\Bigl [- \int_0^T \d\tau
{{\dot y}^2\over 4}\Bigr ]
 =  {\lbrack 4\pi T\rbrack}^{-{d\over 2}} \,.
\label{freepi}
\end{equation}

Individual terms in this expansion are now generated by
Wick contractions in the one-dimensional worldline 
theory at fixed $T$, using the Green function for the second 
derivative on the circle,

\begin{equation}
G(\tau_1,\tau_2) =
 \mid \tau_1 - \tau_2\mid -
{{(\tau_1 - \tau_2)}^2\over T}\quad .
\label{defG}
\end{equation}

\no
Elementary fields are thus contracted by

\begin{equation}
\langle y^{\mu}(\tau_1)y^{\nu}(\tau_2)\rangle
    = - g^{\mu\nu}G(\tau_1,\tau_2) \,,
\label{contract}
\end{equation}

\no
and exponentials of fields
using formulas familiar from string theory,

\begin{eqnarray}
\langle e^{y({\sy\tau_1})\partial_{(1)}}
e^{y({\sy\tau_2})\partial_{(2)}}\rangle &=&
e^{{\scriptstyle -G}({\sy\tau_1},{\sy\tau_2})
\partial_{(1)}\partial_{(2)}} \nonumber\\
\langle \dot y^{\mu}({\tau_1})
e^{y({\sy\tau_1})\partial_{(1)}}
e^{y({\sy\tau_2})\partial_{(2)}}\rangle &=&
-\dot G({\tau_1},{\tau_2})\partial_{(2)}^{\mu}
e^{{\scriptstyle -G}({\sy\tau_1},
{\sy\tau_2})\partial_{(1)}\partial_{(2)}}
\nonumber\\
\label{expcon}
\end{eqnarray}

\no
etc. (a dot always denotes a derivative with respect 
to the first variable). We will often abbreviate
$G_{ij}:=G(\tau_i,\tau_j)$ etc.

To make this procedure manifestly gauge-invariant, 
we now take the background gauge field to be in 
Fock-Schwinger gauge with respect to $x_0$, imposing 
the gauge condition

\begin{equation}
y^\mu A_{\mu}(x_0+y(\tau))\equiv 0 \,.
\label{deffs}
\end{equation}

\no
In this gauge,

\begin{equation}
A_{\mu}(x_0+y) = y^{\rho}
\int_0^1 \d\eta\eta F_{\rho\mu}(x_0+\eta y) \,,
\label{AtoF}
\end{equation}

\no
and $F_{\rho\mu}$ and V can be {\sl covariantly}
Taylor-expanded as (see e.g. \cite{shif})

\begin{eqnarray}
F_{\rho\mu}(x_0 + \eta y)&=& 
e^{\eta y D}F_{\rho\mu}(x_0)\nonumber\\
V(x_0 + y) &=&
e^{y D}V(x_0) \,.
\label{expFV}
\end{eqnarray}

\no
This leads also to a 
covariant Taylor expansion for A:

\begin{equation}
A_\mu(x_0+y)=
\int^1_0 \d\eta\eta\, y^\rho 
e^{\eta y\cdot D} F_{\rho\mu}(x_0) =
\half y^\rho F_{\rho\mu}+{1\over 3}y^{\nu}y^{\rho} 
D_\nu F_{\rho\mu} + ...
\label{Aexpand}
\end{equation}
 
\no
Using these formulas, we obtain the following manifestly
covariant version of eq.~(\ref{expand}):

\begin{eqnarray}
\Gamma[F,V] &=& \int_0^\infty \!{dT\over T}
{\rm e}^{-m^2 T} {\rm tr}\int \! d^d x_0 \sum_{n=0}^\infty
\frac{(-1)^n}{n} T \int \! {\cal D}y \;{\rm exp} \Bigl[ 
- \int_0^T \! d\tau {\dot{y}^2\over 4} \Bigr]
\int_0^{\tau_1 = T} \hspace{-0.7cm}d\tau_2 \ldots 
\!\int_0^{\tau_{n-1}} \hspace{-0.7cm}d\tau_n
\nonumber\\
&& \!\!\!\!\!\!\prod_{j=1}^n \!\Bigl[\;{\rm e}^{ y(\tau_j)D_{(j)}} 
V^{(j)}(x_0)\!+\! ig \dot{y}^{\mu_j}(\tau_j) y^{\rho_j}(\tau_j) 
\int_0^1 d\eta_j \eta_j {\rm e}^{\eta_j y(\tau_j) D_{(j)}}
F_{\rho_j\mu_j}^{(j)}(x_0)  \Bigr].
\label{master}
\end{eqnarray}

\no
From this master formula, the inverse mass expansion of the one-loop
effective action to some fixed order $N$,

\begin{equation}
\Gamma[F,V] = \int_0^\infty \!{dT\over T} \; 
\frac{{\rm e}^{-m^2 T}}{[4\pi T]^{d/2}} \; 
{\rm tr} \; \int \! d^d x_0 \; \sum_{n=1}^N \; 
\frac{(-T)^n}{n!} \; O_n[F,V] \,,
\end{equation}

is obtained in three steps:

\begin{enumerate}

\item Wick contractions: Truncate the master formula to 
$n=N$, and the covariant Taylor expansion eq.~(\ref{Aexpand})
accordingly. Wick contract the integrand, which is now a 
polynomial.

\item Integrations:
Perform the $\tau$-integrations. The integrand is a 
polynomial in the worldline Green function $G$ and 
its first two derivatives,

\begin{eqnarray}
\dot G(\tau_1,\tau_2) &=& {\rm sign}(\tau_1 - \tau_2) 
- 2 {{(\tau_1 - \tau_2)}\over T}\nonumber\\
\ddot G(\tau_1,\tau_2) 
&=& 2 {\delta}(\tau_1 - \tau_2)
- {2\over T} \,.
\label{GdGdd}
\end{eqnarray}

\no
It is useful to first rescale all $\tau$-integrals to the 
unit circle, $\tau_i=Tu_i$, using the scaling properties 
$G(Tu)=TG(u),\dot G(Tu)=\dot G(u),\ddot G(Tu)={1\over T}\ddot G(u)$.
The $\delta$-function in $\ddot G(u_i,u_j)$ only contributes
if $u_i$ and $u_j$ are neighbouring points on the loop, which
also includes the case $\ddot G(1,u_n)$. In the non-Abelian case the 
coefficient $2$ in front of the $\delta$-function has to be omitted, 
since only half of the $\delta$-function contributes to the ordered 
sector under consideration.

\item Reduction to a minimal basis:
The result of this procedure is the effective Lagrangian 
at the required order, albeit in redundant form. To be 
maximally useful for numerical applications, it still needs 
to be reduced to a minimal set of invariants, using all 
available symmetries. Those are

\begin{itemize}
\item
Cyclic invariance under the trace.
\item
Bianchi identities.
\item
Antisymmetry of the field strength tensor.
\end{itemize}

\end{enumerate}

\noindent
Usually those symmetry operations would have to be preceded 
by judiciously chosen partial integrations performed on the 
effective action. 
It is a remarkable property of the present calculational 
scheme that the reduction of our result for the effective 
action to a minimal basis of invariants can be achieved 
without any partial integrations. In particular, for the 
pure scalar case the reduction process amounts to nothing 
more than the identification of cyclically equivalent 
terms. We will come back to this point in Chapter 4.1.
In the general case, the reduction to a minimal basis 
of invariants is much more involved. The method adopted here 
follows a proposal by M\"uller  \cite{muellerbasis} and is
explained in Appendix A.

\vskip.8cm
{\bf 3. Computerization and explicit results}
\vskip.5cm

For a computation of higher coefficients in the inverse mass
expansion starting from the master formula (\ref{master}) one 
clearly has to computerize the three steps described in the
Chapter 2. 

The first step (expanding the interaction exponentials, 
truncating them to a given order and performing all possible
Wick contractions) can be done very conveniently with FORM 
\cite{form} for both the pure scalar and the general case.
In the pure scalar case the contraction of exponentials 
eq. (\ref{expcon})
is used, for it yields the more compact intermediate expressions.

The second step (integrations) is also done with FORM in the pure
scalar case, where the integrations are purely polynomial. 
The general (gauged) case involves $\delta$-functions stemming from 
the contractions yielding a second derivative of the Green function. 
Once the corresponding integrations are done, the remaining integrand 
is again purely polynomial.

In the pure scalar case the second step almost completes the
computation. The remaining cyclic redundancy is fixed using a PERL 
\cite{PERL} program. In the general case the reduction algorithm
described in Appendix A is used. It consists of a set of nontrivial 
substitution rules and therefore requires a symbolic manipulation
program, which contains rule based programming and flexible pattern 
matching. For this purpose we chose a new system for symbolic manipulation
called M \cite{M}, which turned out to be much faster than comparably 
flexible systems. We also used M in performing the integrations in
the general case.

The coefficients were calculated to order $\O(T^{12})$ in the
pure scalar case (they can be found at \cite{flinter}) and to order 
$\O(T^6)$ in the general case. After the reduction into the minimal
basis the results to order $\O(T^5)$ read (absorbing the coupling 
constant $\mbox{g}$ into the fields, $F_{\kappa\lambda\mu\nu} \equiv 
D_\kappa D_\lambda F_{\mu\nu}$ etc.):

\bear
 O_1 &=& V \nonumber
\ear

\bear
 O_2 &=& V^2 + \;\frac{1}{6}\; F_{\kappa\lambda}F_{\lambda\kappa} 
\nonumber
\ear

\bear
 O_3 &=& V^3 
 + \;\frac{1}{2}\;V_{\kappa}V_{\kappa} 
 + \;\frac{1}{2}\;VF_{\kappa\lambda}F_{\lambda\kappa} 
 \nonumber \\ 
  &&  
 + \;\frac{1}{20}\;F_{\kappa\lambda\mu}F_{\kappa\mu\lambda} 
 - \;\frac{2}{15}\;\mbox{i}\;F_{\kappa\lambda}F_{\lambda\mu}F_{\mu\kappa} 
 \nonumber
\ear

\bear
 O_4 &=& V^4 
 + 2\;VV_{\kappa}V_{\kappa} 
 + \;\frac{1}{5}\;V_{\kappa\lambda}V_{\lambda\kappa} 
 + \;\frac{2}{5}\;VF_{\kappa\lambda}VF_{\lambda\kappa} 
 + \;\frac{3}{5}\;V^2F_{\kappa\lambda}F_{\lambda\kappa} 
 \nonumber \\ 
  &&  
 - \;\frac{4}{5}\;\mbox{i}\;F_{\kappa\lambda}V_{\lambda}V_{\kappa} 
 + \;\frac{1}{3}\;F_{\kappa\lambda}F_{\mu\lambda\kappa}V_{\mu} 
 + \;\frac{1}{3}\;F_{\kappa\lambda}V_{\mu}F_{\mu\lambda\kappa} 
 + \;\frac{1}{5}\;VF_{\kappa\lambda\mu}F_{\kappa\mu\lambda} 
 \nonumber \\ 
  &&  
 -\;\frac{2}{15}\;F_{\kappa\lambda}F_{\lambda\mu}V_{\mu\kappa} 
 -\;\frac{8}{15}\;\mbox{i}\;VF_{\kappa\lambda}F_{\lambda\mu}F_{\mu\kappa} 
 -\;\frac{1}{21}\;F_{\kappa\lambda}F_{\lambda\mu}F_{\mu\nu}F_{\nu\kappa} 
 \nonumber \\ 
  &&  
 + \;\frac{1}{70}\;F_{\kappa\lambda\mu\nu}F_{\lambda\kappa\nu\mu} 
 + \;\frac{2}{35}\;F_{\kappa\lambda}F_{\lambda\kappa}F_{\mu\nu}F_{\nu\mu} 
 + \;\frac{4}{35}\;F_{\kappa\lambda}F_{\lambda\mu}F_{\kappa\nu}F_{\nu\mu} 
 \nonumber \\ 
  &&  
 -\;\frac{6}{35}\;\mbox{i}\;F_{\kappa\lambda}F_{\mu\lambda\nu}F_{\mu\nu\kappa} 
 -\;\frac{8}{105}\;\mbox{i}\;F_{\kappa\lambda}F_{\lambda\mu\nu}F_{\kappa\nu\mu} 
 + \;\frac{11}{420}\;F_{\kappa\lambda}F_{\mu\nu}F_{\lambda\kappa}F_{\nu\mu} 
 \nonumber 
\ear 
 
\bear
 O_5 &=& V^5 
 + \;2\;VV_{\kappa}VV_{\kappa} 
 + \;3\; V^2V_{\kappa}V_{\kappa} 
 + VV_{\kappa\lambda}V_{\lambda\kappa} 
 + V^2F_{\kappa\lambda}VF_{\lambda\kappa}
 \nonumber \\ 
  &&   
 -\;\frac{1}{3}\;\mbox{i}\;VF_{\kappa\lambda}V_{\lambda}V_{\kappa} 
 + \;\frac{2}{3}\;V^3F_{\kappa\lambda}F_{\lambda\kappa} 
 + \;\frac{5}{3}\;V_{\kappa}V_{\kappa\lambda}V_{\lambda} 
 -\;\frac{5}{3}\;\mbox{i}\;VV_{\kappa}F_{\kappa\lambda}V_{\lambda} 
 \nonumber \\ 
  &&  
 - \;2\;\mbox{i}\;VV_{\kappa}V_{\lambda}F_{\lambda\kappa} 
 + \;\frac{1}{14}\;V_{\kappa\lambda\mu}V_{\mu\lambda\kappa} 
 + \;\frac{1}{21}\;VF_{\kappa\lambda}F_{\lambda\mu}V_{\mu\kappa} 
 \nonumber \\ 
  &&  
 + \;\frac{1}{21}\;VF_{\kappa\lambda}V_{\lambda\mu}F_{\mu\kappa} 
 + \;\frac{2}{7}\;V^2F_{\kappa\lambda\mu}F_{\kappa\mu\lambda} 
 + \;\frac{2}{21}\;F_{\kappa\lambda}F_{\lambda\mu}V_{\kappa}V_{\mu} 
 \nonumber \\ 
  &&  
 + \;\frac{3}{7}\;F_{\kappa\lambda}F_{\lambda\kappa}V_{\mu}V_{\mu} 
 + \;\frac{3}{7}\;VF_{\kappa\lambda}F_{\mu\lambda\kappa}V_{\mu} 
 + \;\frac{3}{7}\;VV_{\kappa}F_{\kappa\lambda\mu}F_{\mu\lambda} 
 \nonumber \\ 
  &&  
 -\;\frac{3}{7}\;\mbox{i}\;F_{\kappa\lambda\mu}V_{\mu\kappa}V_{\lambda} 
 -\;\frac{3}{7}\;\mbox{i}\;F_{\kappa\lambda\mu}V_{\mu}V_{\lambda\kappa} 
 -\;\frac{3}{7}\;\mbox{i}\;VF_{\kappa\lambda}VF_{\lambda\mu}F_{\mu\kappa} 
 \nonumber \\ 
  &&  
 + \;\frac{3}{14}\;VF_{\kappa\lambda\mu}VF_{\kappa\mu\lambda} 
 -\;\frac{5}{14}\;F_{\kappa\lambda}V_{\lambda}F_{\kappa\mu}V_{\mu} 
 -\;\frac{5}{14}\;F_{\kappa\lambda}V_{\mu}F_{\mu\lambda}V_{\kappa} 
 \nonumber \\ 
  &&  
 + \;\frac{13}{21}\;VF_{\kappa\lambda\mu}F_{\mu\lambda}V_{\kappa} 
 + \;\frac{13}{21}\;VF_{\kappa\lambda\mu}V_{\kappa}F_{\mu\lambda} 
 + \;\frac{13}{21}\;VF_{\kappa\lambda}V_{\mu}F_{\mu\lambda\kappa} 
 \nonumber \\ 
  &&  
 + \;\frac{13}{21}\;VV_{\kappa}F_{\lambda\mu}F_{\kappa\mu\lambda} 
 -\;\frac{16}{21}\;VV_{\kappa\lambda}F_{\lambda\mu}F_{\mu\kappa} 
 -\;\frac{17}{21}\;F_{\kappa\lambda}F_{\lambda\mu}V_{\mu}V_{\kappa} 
 \nonumber \\ 
  &&  
 -\;\frac{17}{21}\;\mbox{i}\;F_{\kappa\lambda}V_{\lambda\mu}V_{\mu\kappa} 
 + \;\frac{17}{42}\;F_{\kappa\lambda}V_{\mu}F_{\lambda\kappa}V_{\mu} 
 -\;\frac{19}{21}\;\mbox{i}\;V^2F_{\kappa\lambda}F_{\lambda\mu}F_{\mu\kappa} 
 \nonumber \\ 
  &&  
 + \;\frac{1}{3}\;VF_{\kappa\lambda}F_{\lambda\mu}F_{\kappa\nu}F_{\nu\mu} 
 -\;\frac{1}{3}\;\mbox{i}\;F_{\kappa\lambda}F_{\mu\lambda\kappa}V_{\nu}F_{\nu\mu} 
 + \;\frac{1}{6}\;F_{\kappa\lambda\mu}F_{\kappa\nu\mu\lambda}V_{\nu} 
 \nonumber \\ 
  &&  
 + \;\frac{1}{6}\;F_{\kappa\lambda\mu}V_{\nu}F_{\nu\kappa\mu\lambda} 
 + \;\frac{1}{6}\;F_{\kappa\lambda}F_{\mu\nu\lambda\kappa}V_{\nu\mu} 
 + \;\frac{1}{6}\;F_{\kappa\lambda}V_{\mu\nu}F_{\nu\mu\lambda\kappa} 
 \nonumber \\ 
  &&  
 -\;\frac{1}{6}\;\mbox{i}\;F_{\kappa\lambda}F_{\mu\nu}V_{\nu}F_{\mu\lambda\kappa} 
 -\;\frac{1}{6}\;\mbox{i}\;VF_{\kappa\lambda\mu}F_{\kappa\nu}F_{\nu\mu\lambda} 
 + \;\frac{1}{12}\;VF_{\kappa\lambda}F_{\lambda\kappa}F_{\mu\nu}F_{\nu\mu} 
 \nonumber \\ 
  &&  
 + \;\frac{1}{14}\;VF_{\kappa\lambda\mu\nu}F_{\lambda\kappa\nu\mu} 
 -\;\frac{1}{14}\;\mbox{i}\;F_{\kappa\lambda}F_{\lambda\mu\nu}V_{\kappa}F_{\nu\mu} 
 + \;\frac{1}{42}\;\mbox{i}\;VF_{\kappa\lambda}F_{\lambda\mu\nu}F_{\kappa\nu\mu} 
 \nonumber \\ 
  &&  
 + \;\frac{2}{7}\;\mbox{i}\;F_{\kappa\lambda}F_{\lambda\mu}F_{\mu\nu}V_{\nu\kappa} 
 -\;\frac{2}{21}\;F_{\kappa\lambda\mu}F_{\kappa\mu\nu}V_{\nu\lambda} 
 -\;\frac{2}{21}\;F_{\kappa\lambda}F_{\mu\lambda\nu}V_{\nu\mu\kappa} 
 \nonumber \\ 
  &&  
 -\;\frac{2}{21}\;F_{\kappa\lambda}V_{\lambda\mu\nu}F_{\nu\mu\kappa} 
 + \;\frac{2}{21}\;\mbox{i}\;F_{\kappa\lambda}F_{\lambda\mu}F_{\mu\kappa\nu}V_{\nu} 
 -\;\frac{3}{7}\;\mbox{i}\;F_{\kappa\lambda}F_{\lambda\mu}F_{\nu\mu\kappa}V_{\nu} 
 \nonumber \\ 
  &&  
 -\;\frac{4}{7}\;\mbox{i}\;F_{\kappa\lambda}F_{\lambda\mu\nu}F_{\nu\kappa}V_{\mu} 
 + \;\frac{4}{21}\;\mbox{i}\;F_{\kappa\lambda}F_{\lambda\mu}V_{\nu}F_{\kappa\nu\mu} 
 -\;\frac{5}{21}\;VF_{\kappa\lambda}F_{\lambda\mu}F_{\mu\nu}F_{\nu\kappa} 
 \nonumber \\ 
  &&  
 + \;\frac{5}{21}\;VF_{\kappa\lambda}F_{\mu\nu}F_{\nu\lambda}F_{\mu\kappa} 
 -\;\frac{5}{21}\;\mbox{i}\;F_{\kappa\lambda}F_{\lambda\mu}V_{\nu}F_{\nu\mu\kappa} 
 -\;\frac{5}{21}\;\mbox{i}\;F_{\kappa\lambda}F_{\mu\nu}V_{\lambda}F_{\kappa\nu\mu} 
 \nonumber \\ 
  &&  
 -\;\frac{5}{21}\;\mbox{i}\;VF_{\kappa\lambda\mu}F_{\kappa\mu\nu}F_{\nu\lambda} 
 -\;\frac{5}{21}\;\mbox{i}\;VF_{\kappa\lambda\mu}F_{\nu\mu\lambda}F_{\nu\kappa} 
 -\;\frac{5}{21}\;\mbox{i}\;VF_{\kappa\lambda}F_{\mu\lambda\nu}F_{\mu\nu\kappa} 
 \nonumber \\ 
  &&  
 -\;\frac{8}{21}\;\mbox{i}\;VF_{\kappa\lambda\mu}F_{\mu\nu}F_{\kappa\nu\lambda} 
 -\;\frac{10}{21}\;\mbox{i}\;F_{\kappa\lambda}F_{\mu\lambda\nu}F_{\mu\kappa}V_{\nu} 
 -\;\frac{11}{42}\;\mbox{i}\;F_{\kappa\lambda}F_{\lambda\mu\nu}F_{\nu\mu}V_{\kappa} 
 \nonumber \\ 
  &&  
 -\;\frac{11}{42}\;\mbox{i}\;F_{\kappa\lambda}F_{\mu\lambda\kappa}F_{\mu\nu}V_{\nu} 
 + \;\frac{11}{84}\;VF_{\kappa\lambda}F_{\mu\nu}F_{\lambda\kappa}F_{\nu\mu} 
 + \;\frac{13}{42}\;F_{\kappa\lambda\mu}F_{\nu\mu\lambda}V_{\nu\kappa} 
 \nonumber \\ 
  &&  
 + \;\frac{17}{84}\;VF_{\kappa\lambda}F_{\mu\nu}F_{\nu\mu}F_{\lambda\kappa} 
 + \;\frac{1}{9}\;F_{\kappa\lambda}F_{\lambda\mu}F_{\nu\kappa\rho}F_{\nu\rho\mu} 
 -\;\frac{1}{18}\;\mbox{i}\;F_{\kappa\lambda}F_{\lambda\mu\nu\rho}F_{\mu\kappa\rho\nu} 
 \nonumber \\ 
  &&  
 -\;\frac{1}{189}\;F_{\kappa\lambda}F_{\lambda\mu}F_{\kappa\nu\rho}F_{\mu\rho\nu} 
 + \;\frac{1}{252}\;F_{\kappa\lambda\mu\nu\rho}F_{\mu\lambda\kappa\rho\nu} 
 + \;\frac{1}{378}\;F_{\kappa\lambda}F_{\lambda\mu\nu\rho}F_{\mu\kappa}F_{\rho\nu} 
 \nonumber \\ 
  &&  
 -\;\frac{2}{21}\;\mbox{i}\;F_{\kappa\lambda\mu}F_{\kappa\nu\mu\rho}F_{\nu\rho\lambda} 
 + \;\frac{2}{63}\;F_{\kappa\lambda}F_{\mu\nu\rho}F_{\lambda\kappa}F_{\mu\rho\nu} 
 + \;\frac{2}{945}\;\mbox{i}\;F_{\kappa\lambda}F_{\lambda\mu}F_{\mu\nu}F_{\nu\rho}F_{\rho\kappa} 
 \nonumber \\ 
  &&  
 -\;\frac{4}{63}\;\mbox{i}\;F_{\kappa\lambda}F_{\mu\nu\lambda\rho}F_{\nu\mu\rho\kappa} 
 -\;\frac{5}{63}\;F_{\kappa\lambda}F_{\lambda\mu}F_{\nu\mu\rho}F_{\nu\rho\kappa} 
 + \;\frac{5}{63}\;F_{\kappa\lambda}F_{\mu\lambda\nu}F_{\kappa\rho}F_{\mu\rho\nu} 
 \nonumber \\ 
  &&  
 + \;\frac{5}{63}\;F_{\kappa\lambda}F_{\mu\nu\rho}F_{\rho\lambda}F_{\mu\nu\kappa} 
 -\;\frac{5}{126}\;\mbox{i}\;F_{\kappa\lambda\mu}F_{\mu\kappa\nu\rho}F_{\lambda\rho\nu} 
 -\;\frac{5}{126}\;\mbox{i}\;F_{\kappa\lambda\mu}F_{\mu\nu\rho}F_{\lambda\kappa\rho\nu} 
 \nonumber \\ 
  &&  
 + \;\frac{8}{63}\;\mbox{i}\;F_{\kappa\lambda}F_{\lambda\mu}F_{\kappa\nu}F_{\mu\rho}F_{\rho\nu} 
 -\;\frac{8}{189}\;F_{\kappa\lambda}F_{\lambda\mu\nu}F_{\nu\kappa\rho}F_{\rho\mu} 
 -\;\frac{10}{189}\;F_{\kappa\lambda}F_{\lambda\mu\nu}F_{\nu\rho}F_{\rho\mu\kappa} 
 \nonumber \\ 
  &&  
 -\;\frac{10}{189}\;F_{\kappa\lambda}F_{\mu\lambda\nu}F_{\mu\rho}F_{\kappa\rho\nu} 
 + \;\frac{11}{189}\;F_{\kappa\lambda}F_{\mu\lambda\kappa}F_{\nu\rho}F_{\mu\rho\nu} 
 + \;\frac{11}{189}\;F_{\kappa\lambda}F_{\mu\nu\rho}F_{\rho\nu}F_{\mu\lambda\kappa} 
 \nonumber \\ 
  &&  
 -\;\frac{11}{378}\;F_{\kappa\lambda}F_{\lambda\mu}F_{\mu\kappa\nu\rho}F_{\rho\nu} 
 + \;\frac{13}{252}\;F_{\kappa\lambda}F_{\lambda\kappa}F_{\mu\nu\rho}F_{\mu\rho\nu} 
 -\;\frac{16}{63}\;F_{\kappa\lambda}F_{\lambda\mu\nu}F_{\nu\rho}F_{\kappa\rho\mu} 
 \nonumber \\ 
  &&  
 -\;\frac{16}{189}\;F_{\kappa\lambda}F_{\lambda\mu}F_{\nu\rho}F_{\mu\kappa\rho\nu} 
 -\;\frac{16}{945}\;\mbox{i}\;F_{\kappa\lambda}F_{\mu\nu}F_{\lambda\rho}F_{\nu\kappa}F_{\rho\mu} 
 -\;\frac{19}{756}\;F_{\kappa\lambda}F_{\lambda\mu\nu}F_{\kappa\rho}F_{\rho\nu\mu} 
 \nonumber \\ 
  &&  
 -\;\frac{19}{756}\;F_{\kappa\lambda}F_{\mu\nu\rho}F_{\mu\lambda}F_{\kappa\rho\nu} 
 -\;\frac{22}{189}\;\mbox{i}\;F_{\kappa\lambda}F_{\lambda\mu}F_{\kappa\nu}F_{\nu\rho}F_{\rho\mu} 
 + \;\frac{25}{189}\;F_{\kappa\lambda}F_{\mu\lambda\nu}F_{\mu\kappa\rho}F_{\rho\nu} 
 \nonumber \\ 
  &&  
 -\;\frac{26}{189}\;F_{\kappa\lambda}F_{\mu\lambda\nu}F_{\rho\mu\kappa}F_{\rho\nu} 
 -\;\frac{31}{378}\;\mbox{i}\;F_{\kappa\lambda}F_{\lambda\mu}F_{\nu\rho}F_{\mu\kappa}F_{\rho\nu} 
 -\;\frac{34}{189}\;F_{\kappa\lambda}F_{\lambda\mu\nu}F_{\rho\nu\kappa}F_{\rho\mu} 
 \nonumber \\ 
  &&  
 -\;\frac{41}{378}\;F_{\kappa\lambda}F_{\lambda\mu}F_{\mu\nu\rho}F_{\kappa\rho\nu} 
 -\;\frac{53}{378}\;\mbox{i}\;F_{\kappa\lambda}F_{\lambda\kappa}F_{\mu\nu}F_{\nu\rho}F_{\rho\mu} 
 + \;\frac{61}{756}\;F_{\kappa\lambda}F_{\mu\lambda\kappa}F_{\mu\nu\rho}F_{\rho\nu} 
 \nonumber \\ 
  &&  
 + \;\frac{61}{756}\;F_{\kappa\lambda}F_{\mu\nu}F_{\rho\lambda\kappa}F_{\rho\nu\mu} 
 \nonumber
\ear

The coefficient $O_6$ is quite lengthy and can be found in Appendix C. 
Even for the general case it is in principle possible to compute still 
higher coefficients of the inverse mass expansion. However,
the calculation is practically limited by the basis reduction, since the
implementation of the general rules given in Appendix A has to be extended
for every new order under consideration. 

\vskip.8cm
{\bf 4. Comparison with other methods}
\vskip.5cm

Finally, let us compare with other algorithms which
are available for the computation of the same expansion.

Generally, methods of calculating the higher derivative
expansion are either based on heat kernel 
 \cite{carson,carmcl,abc,bles,onofri,zuk1,zuk2} 
or Feynman diagram techniques
 \cite{vdv,kikkawa,mwz,fraser}.

In the heat kernel approach, one evaluates the operator 
trace eq.~(\ref{ime}) in $x$--space:
 
\begin{equation}
\Gamma [A,V] 
 = {\dps\int_0^{\infty}}{dT\over T}
{\rm tr}\,\int d^d x \,\,\langle x\mid 
\exp
\Bigl[-T
\bigl(-D^2 + m^2 + V(x)\bigr)\Bigr] 
\mid x\rangle \,.
\label{x-trace}
\end{equation}

\no
From the interacting heat kernel

\begin{equation}
\langle x\mid K(T) \mid y\rangle =
\langle x\mid
\exp
\Bigl[-T
\bigl(-D^2 + m^2 + V(x)\bigr)\Bigr]
\mid y\rangle
\end{equation}

\no
one separates off the known free one,

\begin{equation}
\langle x\mid K_0(T) \mid y\rangle =
\langle x\mid
\exp
\Bigl[-T
\bigl(-{\partial}^2 + m^2\bigr)\Bigr]
\mid y\rangle
=
{(4\pi T)}^{-d/2}
\exp (-{(x-y)^2\over 4T}) \,,
\end{equation}

\no
writing

\begin{equation}
\langle x\mid K(T) \mid y\rangle
=
{(4\pi T)}^{-d/2}
\exp (-{(x-y)^2\over 4T})
H(x,y;T) \,.
\label{sep}
\end{equation}

\no
H is then expanded in powers of T,

\begin{equation}
H(x,y;T) = \sum_{k=0}^{\infty}\,a_k(x,y)\,T^k 
\label{expandH}
\end{equation}

\no
and the heat kernel coefficients $a_k$ (which are 
functionals of the background fields) are calculated 
on the diagonal $x=y$.

For the calculation of those coefficients, a large 
variety of algorithms have been invented.
Roughly, they fall into three categories:

\begin{enumerate}
\item
Recursive $x$-space algorithms  \cite{bles,lanyov,abc}.
In our view these are too cumbersome for doing higher order 
calculations in gauge theory and will not be discussed here.

\item
The method of Zuk  \cite{zuk1,zuk2}, based on Onofri's
graphical representation of the heat kernel
coefficients  \cite{onofri}.

\item
Nonrecursive algorithms based on the insertion
of a momentum basis  \cite{nepo,chan}.

\end{enumerate}

\noindent
Let us first consider Zuk's method, which is manifestly 
gauge invariant, and also the one most closely related
to our work. 

\vskip.8cm
{\bf 4.1 Zuk's Method}
\vskip.5cm
 
\noindent
In Onofri's work  \cite{onofri}, the Baker-Campbell-Hausdorff 
formula was employed to represent the coefficients for the
pure scalar case by Feynman diagrams in a one-dimensional 
auxiliary field theory. Those Feynman diagrams are calculated using
the Green function

\begin{equation}
G^{(0)}(\tau_1,\tau_2) = \; \mid \tau_1 - \tau_2\mid - 
(\tau_1 + \tau_2 ) + {2\over T}\tau_1\tau_2\quad .
\label{defG0}     
\end{equation}

This Green function was also used by Zuk 
to calculate the effective Lagrangian for the pure
scalar case up to terms with four derivatives
 \cite{zuk1}. He then generalized the method to the 
gauge field case, and also used Fock-Schwinger gauge
to enforce manifest gauge invariance  \cite{zuk2}.
The `Quantum Mechanical Path Integral Method' \cite{mckeon},
which may be considered as an extension of the Onofri-Zuk 
formalism, also uses the Green function given above.

\begin{figure}[h]
\begin{center}  
\psfig{file=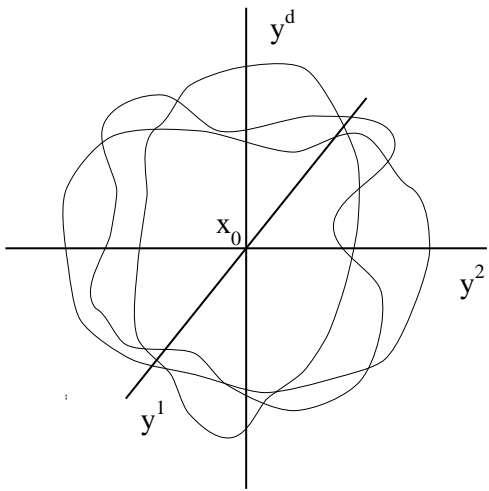}
\caption{\label{symmfig} The path integration for a uniformly distributed
background charge $\rho(\tau)=1/T$.}
\end{center}
\end{figure}

As explained in detail in Appendix B, the split in the path
integral eq. (\ref{pathsplit}), which is necessary to extract the zero
modes, is not unique. One still has the freedom of choosing a background 
charge $\rho(\tau)$ on the worldline, which parametrizes the boundary
conditions on the functions $y^\mu(\tau)$. 
This gives us some insight into the connection of the worldline 
path integral approach with the Onofri-Zuk formalism: 
If one uses a constant background charge,
\begin{equation}
\rho(\tau)=\frac{1}{T}\;,
\label{rhoconst}
\end{equation}
one obtains the Green function 
\begin{equation}
G^{(c)}(\tau_1,\tau_2) = | \tau_1 - \tau_2| -\;
\frac{(\tau_1 - \tau_2)^2}{T}-\frac{T}{6}\;,
\end{equation}
which agrees -- up to an (irrelevant) constant -- with the one used 
in our approach, eq.~(\ref{defG}).
The effective Lagrangian  ${\cal L} (x_0)$ is obtained as a path 
integral over the space of all loops having $x_0$ as their common 
center of mass (Fig.~\ref{symmfig}). 

\vskip.5cm
\vskip.5cm

\begin{figure}[h]
\begin{center}  
\psfig{file=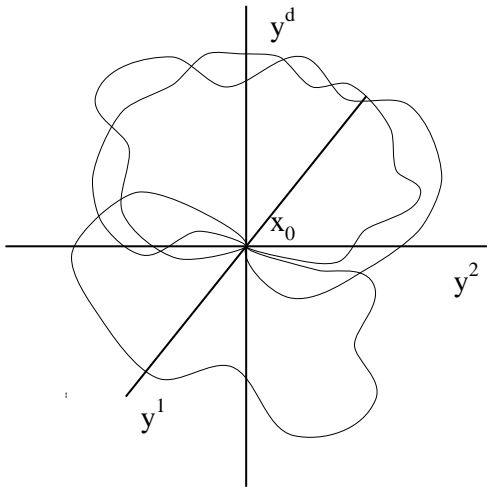}
\caption{\label{asymmfig} The path integration for a background charge
$\rho(\tau)=\delta(\tau)$.}
\end{center}
\end{figure}

If one uses the background charge
\begin{equation}
\rho(\tau)=\delta(\tau)
\label{rhodelta}
\end{equation}
instead, the resulting Green function turns out to be exactly the one 
used by Onofri, eq. (\ref{defG0}). In this case, the boundary condition 
reads
\begin{equation}
 y(0)=y(T)=0\;,
\end{equation}
and the effective Lagrangian  ${\cal L} (x_0)$ is given as path 
integral over the space of all loops intersecting in 
$x_0$ (Fig.~\ref{asymmfig}). 

\vskip.5cm
\vskip.5cm

The constant background charge has the special property that it is 
the only translationally invariant choice and therefore
$G^{(c)}$ (and equivalently $G$) depend only on $\tau_1-\tau_2$.
As a consequence, cyclically equivalent terms always come with
the same numerical coefficient, a fact which facilitates the 
cyclic identification process considerably. This does not hold true 
if one uses the Green function eq.~(\ref{defG0}); for example, 
of the three cyclically equivalent terms $V_{\mu}V_{\mu}V$, 
$VV_{\mu}V_{\mu}$ and $V_{\mu}VV_{\mu}$ appearing in the scalar 
effective action at $\O(T^4)$ the first two get assigned the same 
coefficient, while the coefficient of the third one is different. 
\par
Of all translationally invariant Green functions (which differ
only by constants), $G$ has the further advantage 
that it has vanishing diagonal terms, i.e.\ $G(\tau,\tau)=0$. 
This together with the symmetry of $G$ can be used to perform 
in the exponent the replacement
\begin{equation}
-\frac{1}{2}\sum_{i,j=1}^nG(\tau_i,\tau_j)
\partial_{(i)}\partial_{(j)}\quad\longrightarrow\quad
-\sum_{i<j}G(\tau_i,\tau_j)\partial_{(i)}\partial_{(j)} \,.
\end{equation}
For the pure scalar case the resulting effective Lagrangian is 
automatically in the minimal basis, i.e.~it consists only of terms 
without box operators. In general, in the process of identifying 
equivalent terms one never has to integrate by parts. On the other
hand one has to be cautious in comparing our results with the results
of standard local heat kernel methods, for the usage of the Green 
function eq. (\ref{defG}) amounts to implicit partial integrations.

The considerations above show that the redundancies arising in
Zuk's formalism can be avoided by using the translationally invariant
background charge and the corresponding Green function eq.\ (\ref{defG}).

\vskip.8cm
{\bf 4.2 A Modified Nonrecursive Heat Kernel Method}
\vskip.5cm

\def\unity{\mathord{1\hskip -1.5pt
    \vrule width .5pt height 7.75pt depth -.2pt \hskip -1.2pt
    \vrule width 2.5pt height .3pt depth -.05pt \hskip 1.5pt}}
\newcommand{\ehoch}[1]{{\rm e}^{#1}}
\newcommand{\Dl}[1]{\mbox{\raisebox{1pt}
{$\stackrel{\hspace*{4pt}\leftarrow}{D}_#1$}}}
\newcommand{\Dr}[1]{\mbox{\raisebox{1pt}
{$\stackrel{\hspace*{4pt}\rightarrow}{D}_#1$}}}

We have also explicitly tested our results using a modification 
of the nonrecursive heat kernel method mentioned above.
Here one evaluates the functional trace in a plane wave basis,
\begin{equation}
{\rm Tr}\;\ehoch{-TM}={\rm tr}\int\!\! d^d\!x \int\!\! 
\frac{d^dk}{(2\pi)^d}\;\ehoch{-ikx}\ehoch{-TM}\ehoch{ikx}\;.
\end{equation}

The net effect of commuting ${\rm e}^{ikx}$ to the left is the
substitution $D_\mu\rightarrow D_\mu+ik_\mu$ in $M$. After
rescaling the momenta $q_\mu=\sqrt{T}k_\mu$ one finds
\begin{equation}
{\rm Tr}\;\ehoch{-TM}=\frac{\ehoch{-m^2T}}{(4\pi T)^{d/2}}\;
{\rm tr}\int\!\! d^d\!x \int\!\! \frac{d^d\!q}{\pi^{d/2}}\;
\ehoch{-q^2}\exp\big(-T(-D^2+V)+2i\sqrt{T}q_\mu D_\mu\big)\;.
\label{qint}
\end{equation}
The last exponential is to be expanded in powers of $T$
(note that only even numbers of momenta $q_\mu$ contribute).
The $q$--integration produces totally symmetric combinations
of products of the metric tensor.
The intermediate result is a series in $T$, where every 
coefficient consists of a string of $V$'s and $D_\mu$'s and 
the latter are pairwise contracted. The covariant derivatives 
$D_\mu$ act to the right, thus a $D_\mu$ at the right end 
(acting on $\unity$) can be replaced with $iA_\mu$. This would 
however break covariance, which we want to avoid.
\par
The first of our modifications incorporates the `no double 
derivative' prescription discussed in Appendix A.
Of each contracted pair of derivatives one moves the first one 
to the left and the second one to the right,
\begin{equation}
\ldots \Dl{\mu} \ldots \Dr{\mu} \ldots\;,
\label{leftright}
\end{equation}
using the Leibniz rules
\begin{eqnarray}
\Dr{\mu} Y&=&D_\mu Y+Y\Dr{\mu}\;,\\
Y\Dl{\mu}&=&-D_\mu Y+\Dl{\mu} Y \;,
\end{eqnarray}
where $Y$ stands for any covariant structure. In this way one 
obtains terms of the generic form
\begin{equation}
\Dl{{\mu_1}}\ldots\Dl{{\mu_m}}X\Dr{{\nu_1}}\ldots\Dr{{\nu_n}}\;,
\end{equation}
where $X$ represents a string of covariant objects $V$, $DV$,
\ldots, $F$, $DF$, \ldots without active derivatives. Clearly 
the prescription (\ref{leftright}) avoids any self-contractions.
\par
The aim is now to reduce the number of active derivatives.
As an example consider terms where $X$ is antisymmetric in two 
of the indices $\nu_1\ldots\nu_n$. This leads then to terms where 
the number of right derivatives is reduced by two, at the price
of the appearance of a new field strength tensor $F$.
However, even if one includes Bianchi identities within $X$,
this turns out to be not enough.
\par
According to eq.\ (\ref{qint}) there is an overall trace and thus
a freedom of cyclic permutations. Cyclicity may however be exploited 
only after all derivatives have been executed. Nevertheless one can 
use this property already at this stage to `shuffle derivatives
to the right' in the subclass of terms which have only left
derivatives. In particular one can write
\begin{equation}
\Dl{\mu}X= ig A_\mu X=X ig A_\mu=X\Dr{\mu}\;,
\end{equation}
where the first equality comes from removing the total
derivative $\partial_\mu X$, the second makes use of the trace 
(no active derivatives!), and the last one follows from the
addition of $0=\partial_\mu\unity$. In a similar manner one can 
prove the identities
\begin{eqnarray}
\Dl{\mu}\Dl{\nu}X&=&-X\Dr{\nu}\Dr{\mu}+\Dl{\mu}X\Dr{\nu}
+\Dl{\nu}X\Dr{\mu}\;, \nonumber \\
\Dl{\lambda}\Dl{\mu}\Dl{\nu}X&=&+X\Dr{\nu}\Dr{\mu}\Dr{\lambda}
+\Dl{\lambda}\Dl{\mu}X\Dr{\nu}+\Dl{\lambda}\Dl{\nu}X\Dr{\mu}
+\Dl{\mu}\Dl{\nu}X\Dr{\lambda} \nonumber \\
&& -\Dl{\lambda}X\Dr{\nu}\Dr{\mu}-\Dl{\mu}X\Dr{\nu}\Dr{\lambda}
-\Dl{\nu}X\Dr{\mu}\Dr{\lambda}\;,
\end{eqnarray}
and more complicated ones for four or more left derivatives.
After each of these steps one has to investigate the resulting
terms again for their antisymmetry, possibly using Bianchi identities.
\par
The algorithm comes to an end if all terms have no more than
two derivatives acting at the same end. An important input at 
this point is the knowledge that the result can be written
in manifestly covariant form. This allows one to replace the
remaining derivatives according to
\begin{equation}
D_\mu\rightarrow 0\;,\quad\quad
D_\mu D_\nu\rightarrow \frac{ig}{2}F_{\mu\nu}\;.
\end{equation}
One can easily see that such a replacement rule does not exist
for three (or more) derivatives:
A structure $D_{\lambda}D_{\mu}D_{\nu}$ can originate from
$D_\lambda F_{\mu\nu}$, but equally well from $D_\nu F_{\mu\lambda}$,
and the difference of these possibilities is nonzero (it is just
$D_\mu F_{\lambda\nu}$ via Bianchi's identity).
The further treatment of the result, namely identifying cyclic 
equivalent terms and reduction to the minimal basis, proceeds along 
the same lines as described in the Appendix A. Using the method
described above we calculated all coefficients up to $O_5$ and 
all invariants in $O_6$ which contain two or more scalar potentials $V$.
By reduction into the minimal basis we found agreement with the
results of the worldline approach.

\vskip.8cm
{\bf 4.3 The Method of 't Hooft and van de Ven}
\vskip.5cm

Finally, let us comment on the calculation of the inverse mass 
expansion by Feynman diagrams  \cite{vdv,kikkawa,mwz,fraser}. 
Here, the version most suitable to gauge theory calculations 
appears to be the one invented by 't Hooft  \cite{tH}, and 
elaborated by van de Ven  \cite{vdv}.

\no
In this scheme, one first considers backgrounds obeying 

\begin{equation}
V \equiv - A_\mu A_\mu \,, \hspace{.8cm}
\partial_{\mu} A_{\nu} \equiv 0 \,.
\nonumber\\
\label{special}
\end{equation}

\no
For this special case, the effective Lagrangian can be
expanded in a basis consisting of strings
of $n$ $A_{\mu}$--matrices (denoted by ${\rm tr} J_j$),

\begin{equation}
{\cal L} = \sum_j a_j {\rm tr} J_j \quad \,.
\nonumber\\
\label{Lspecial}
\end{equation}

\no
The coefficients $a_j$ can be determined from the 
logarithmic divergence of the one-loop diagram with 
$n$ $A_{\mu}$-insertions in $d = 2n$ spacetime dimensions.
   
One then chooses a minimal basis of invariants for the 
general background, denoted ${\rm tr} I_j$, and subjects 
those invariants to the conditions (\ref{special}).
For any fixed order in $T$, they can then be written in 
terms of the basis ${\rm tr}J_j$,

\begin{equation}
{\rm tr} I_i = \sum_j P_{ij} {\rm tr} J_j \,, 
\label{defP}
\end{equation}

\no
with a certain numerical matrix $P$. If one restricts 
this equation to a fixed order in $T$, the numbers of 
invariants on both sides turn out to match, and the matrix
$P_{ij}$ to be invertible. After performing the inversion,
the effective Lagrangian for the general background
is obtained as

\begin{equation}
{\cal L} = \sum_{i,j} \; a_j \; P^{-1}_{ji} \; {\rm tr} \; I_i \,.
\nonumber\\
\label{Lfinal}
\end{equation}

\no
This method was used by van de Ven  \cite{vdv} to calculate 
the one-loop counterterms for Yang-Mills theory in even
dimensions $\le 10$, which is equivalent to calculating the
order $\O(T^5)$ in the inverse mass expansion. We have checked 
exact agreement with our result for the $\O(T^5)$ by explicitly 
performing the necessary partial integrations and basis reduction.
Since van de Ven considers the case of a real scalar field, this
also involves transposition symmetry of the coefficients, i.e.
the invariance (up to a sign) of the coefficients under inversion 
of the ordering of the simple factors, as explained in detail in 
\cite{muellerbasis}.

For a comparison of the efficiency of both methods, one would 
have to computerize this method, too, which has not been done 
yet. Obviously, the difficulty resides in the fact that the 
method requires the construction of a minimal basis of invariants 
{\sl a priori}, to ensure invertibility of the matrix $P$. 
Moreover, in order to compute higher coefficients one has to
find the inverse of matrices of the order of the length of the
minimal basis to get the prefactors of the coefficients. For the
calculation of $O_6$ in the case of a complex scalar field this
already amounts to a (symbolic) inversion of a 
$902\times 902$-matrix with rational elements.

\vskip.8cm
{\bf 5. Conclusions}
\vskip.5cm

We have applied the string-inspired method of evaluating
one-loop worldline path integrals to the calculation of 
inverse mass expansions of one-loop effective actions. 
Complete computerization of the method has allowed us to 
improve on existing results by one order for a background 
consisting of both a gauge field and a scalar potential, 
and by several orders for the case of only a scalar potential
\footnote[0]{As we have learned from A. van de Ven, he has 
recently also obtained the coefficient $O_6$ for the Yang-Mills
case, using a novel version of the recursive heat kernel 
method \cite{vandevennew}. This result has not yet been 
reduced to a minimal basis of invariants.}.
Comparing with the closely related algorithm used by Onofri, 
Fujiwara et al. and Zuk, we have traced the difference 
between both approaches to the different boundary conditions
imposed on the path integral. The results have been reduced 
to a minimal basis of invariants, and the reduction process 
was described in detail.
\par\medskip
As indicated in the introduction, the formalism can easily
be generalized to the case of a spin 1/2 particle in the loop
coupled to external gauge bosons. A systematic investigation 
of the one-loop effective action induced by a fermion in a scalar
background is currently being done \cite{mnss,fnms}.
Moreover a worldline path integral formulation of the one-loop 
effective action of a gluon circulating in the loop was 
described in \cite{rss}.
Finally, an extension of the present formalism to the two-loop 
case is under consideration, based on the construction of 
generalized worldline Green functions \cite {ss2}.

\newpage
{\sl Acknowledgements:}

We are grateful to E. D'Hoker,  U. M\"uller and A. E. van de Ven 
for various helpful discussions and informations. Two of us 
(D. Fliegner and C. Schubert) would like to thank Z. Bern 
and the UCLA Department of Physics for hospitality during 
part of this project.

\newpage

\vskip.8cm
{\bf Appendix A}
\vskip.5cm

As explained in Chapter 2 it is essential for practical purposes
to reduce the coefficients to a minimal basis. In the pure scalar
case the identification of cyclic equivalent invariants is sufficient.
No partial integrations have to be performed because of the 
existence of a minimal basis, which does not involve any box 
operators. In the gauged case the situation is more complicated.
Again there exists a basis without box operators, and partial 
integrations are not necessary, but cyclic permutations are
insufficient to reach a minimal basis. Additionally one has to use
several equalities, namely the Bianchi identities, the antisymmetry
of the field strength tensor and the exchange of covariant 
derivatives: 

\begin{equation}
D_\mu F_{\ka\la} + D_\ka F_{\la\mu} + D_\la F_{\mu\ka} = 0 \;,
\end{equation}

\begin{equation}
F_{\mu\nu} = - F_{\nu\mu} \;,
\end{equation}

\begin{equation}
D_\mu D_\nu X = D_\nu D_\mu X + ig [F_{\mu\nu}, X ] \,.
\end{equation}

In the following we describe the basis reduction algorithm 
proposed by M\"uller, the proof of minimality can be found in 
\cite{muellerbasis}. 

Before the basis reduction any invariant in our coefficients consists
of simple factors $X$ (a $V$, an $F$ or covariant derivatives thereof). 
Like in the pure scalar case there are no self-contractions within a 
simple factor:

\begin{equation}
 X \in \{ V, F_{\ka\la}, D_{\mu_1}D_{\mu_2}\ldots D_{\mu_n} V, 
 D_{\mu_1}D_{\mu_2}\ldots D_{\mu_n} F_{\mu_{n+1}\mu_{n+2}} 
 \,|\, \mu_i \neq \mu_j \}\,.
\end{equation}

The first step of the algorithm would be the elimination of 
self-contractions by partial integration, which is unnecessary
in our case. During the rest of the algorithm invariants with increasing
number of field strength tensors are produced due to exchange of
covariant derivatives. Therefore one has to start with the terms
containing the maximum number of covariant derivatives and collect 
the corrections to invariants with smaller number of derivatives before
the basis reduction. The remaining algorithm includes the following 
steps:

\begin{itemize}

\item Removal of derivatives of `middle' class

The Bianchi identity exchanges an index of a derivative and the indices
of a field strength tensor. This can be used for a reduction of single
contractions between different simple factors. Consider the following
example:

\begin{equation}
\mbox{tr} (D_\mu D_\nu D_\ro D_\si F_{\ka\la} \ldots
X_\nu \ldots F_{\si\la} \ldots X_\ro \ldots X_\ka \ldots X_\mu \ldots) 
\, .
\end{equation}

The contractions of the derivatives of $F_{\ka\la}$ belong to different
classes with respect to the contractions of $F_{\ka\la}$. 
This can be seen very easily in a diagrammatical picture, where
the (cyclic) function $\mbox{tr}$ is represented by a circle
(Fig. \ref{Sector}).

\begin{figure}[h]
\begin{center}
\psfig{file=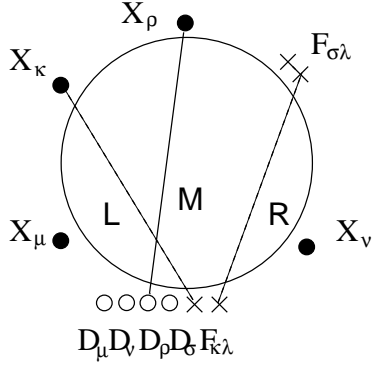}
\caption{\label{Sector}
Graphical representation of the sectors spanned by $F_{\kappa\lambda}$.
The indices of the field strength tensors are denoted by crosses, the 
indices of derivatives by dots, contractions are denoted by straight 
lines.}
\end{center}
\end{figure}

The loop is divided into a `right', a `middle' and a 
`left' sector by the contractions of $F_{\ka\la}$. 
Consequently the derivatives are members
of a `right' ($D_\nu$), a `middle' ($D_\ro$) and 
a `left' ($D_\mu$) class.
There are also derivatives that do not belong to any of 
these classes ($D_\si$). 

The Bianchi identity involves all classes of derivatives and can therefore
be used to eliminate one of them. It is useful to take the symmetric choice
and eliminate the `middle' derivative. In general the derivatives in a 
simple factor have to be exchanged, producing invariants with a higher number 
of field strength tensors, in order to apply the Bianchi identity:

\begin{equation}
D_{\mu} D_{\nu} D_{\ro} D_{\si} F_{\ka\la} = 
D_{\mu} D_{\nu} D_{\si} D_{\ro} F_{\ka\la} + \mbox{corrections} \,.
\end{equation}

Then the Bianchi identity is used to remove middle derivatives ($D_{\ro}$
in our example):

\begin{equation}
D_{\mu} D_{\nu} D_{\si} D_{\ro} F_{\ka\la} =
D_{\mu} D_{\nu} D_{\si} D_{\ka} F_{\ro\la} +
D_{\mu} D_{\nu} D_{\si} D_{\la} F_{\ka\ro} \,.
\end{equation}

The effect of using the Bianchi identity is a decrease of the `middle' sector
and correspondingly intersections between derivative contractions and this 
sector are removed. This procedure has to be done with all `middle' derivatives
in all simple factors, which corresponds to a minimization of the `middle'
sectors.

\item Reduction of multiple contractions between simple factors

In a next step multiple contractions between simple factors are considered.
The general aim of the reduction is to achieve that multiple contractions 
appear only between field strength tensors. To this end one applies the 
following rules:

\bear
\mbox{tr} ( \ldots F_{\mu\nu} \ldots D_\mu F_{\nu\ka} \ldots) &=& 
\frac{1}{2} \mbox{tr} ( \ldots F_{\mu\nu} \ldots D_\ka F_{\nu\mu} \ldots ) \,,
\\
\mbox{tr} ( \ldots D_\mu F_{\nu\ka} \ldots D_{\nu} F_{\mu\la} \ldots ) &=&
\frac{1}{2} \mbox{tr}( \ldots D_\ka F_{\nu\mu} \ldots D_{\la} F_{\mu\nu} \ldots ) 
\nonumber\\ 
& + & \mbox{tr} ( \ldots D_\mu F_{\nu\ka} \ldots D_{\mu} F_{\nu\la} \ldots ) \,, \\
\mbox{tr} ( \ldots F_{\mu\nu} \ldots D_\mu D_\nu X \ldots ) &=&
\frac{1}{2} \mbox{tr} ( \ldots F_{\mu\nu} \ldots ig [F_{\mu\nu}, X] ) \,.
\ear

The first and the second rule use the Bianchi identity combined with 
antisymmetry of the field strength tensor. The second one can be considered
as an exception from the first rule. In this case the reduction aim cannot 
be achieved completely. The third rule uses the exchange rule for 
derivatives and produces again invariants with a higher number of field 
strength tensors.

\item Arrangement of indices and simple factors 

After all the ordering has been done the indices can be fixed. 
Using our diagrammatical picture again we can always rearrange the indices 
in a simple factor in such a way that the contractions form the shortest 
possible connection on the circle. 
This can always be done using the exchange of derivatives
and the antisymmetry of the field strength tensors and will (in general) 
produce terms with a higher number of field strength tensors. 
We illustrate the rule in the following example:
\begin{eqnarray}
\mbox{tr}(\ldots D_{\ka} D_{\la} F_{\mu\nu} \ldots D_{\ka} D_{\la} F_{\mu\nu} 
\ldots ) & = &  
\mbox{tr}(\ldots D_{\ka} D_{\la} F_{\mu\nu} \ldots D_{\la} D_{\ka} F_{\nu\mu} 
\ldots ) + \ldots 
\end{eqnarray}
After arrangement of the indices cyclic equivalent invariants are reduced 
by simultaneously fixing the ordering of simple factors and relabelling the 
indices alphabetically. This completes the algorithm.

\end{itemize}

From the algorithm described above one can read off the properties of
the basis invariants:

\begin{itemize}

\setlength{\itemsep}{0pt}

\item The basis invariants are products of simple factors, which do
not contain any self-contractions. 

\item The basis invariants do not contain any `middle' derivatives.

\item In multiple contractions between simple factors, the field 
strength tensors are doubly contracted. There is one exception 
mentioned above.

\item The arrangement of indices is such that the contractions form 
the shortest possible connection on the circle.

\item The basis invariants are the lexically smallest ones among the
set of possible cyclic equivalent invariants.

\end{itemize}

The following table gives the number of basis invariants as a function
of the order in the proper-time parameter $T$:

\begin{table}[h]
\begin{center}
\begin{tabular}{|c|r|r|r|r|r|r|r|r|} 
\hline 
order & total & $v=0$ & $v=1$ & $v=2$ & $v=3$ & $v=4$ & $v=5$ & $v=6$\\
\hline
1 & 1	& 0	& 1	&  	&	&	&	&\\
2 & 2	& 1	& 0	& 1	&	&	&	&\\	
3 & 5	& 2	& 1	& 1	& 1	&	&	&\\
4 & 18	& 7	& 5	& 4	& 1	& 1	&	&\\
5 & 105	& 36	& 36	& 23	& 7	& 2	& 1	&\\ 
6 & 902	& 300	& 329	& 191	& 63	& 16	& 2	&1\\
\hline
\end{tabular}
\end{center}
\caption{\label{table} Number of basis invariants in different orders of 
the expansion. $v$ is the number of scalar background fields.}
\end{table}

\newpage

{\bf Appendix B}
\vskip.5cm

In this Appendix we investigate the connection of the split of the path 
integral with the freedom of choosing a Green function for the second
derivative operator.
Moreover we show that the final result (the effective action)
does not depend on the actual choice of the Green function.
\par
The Green function used in the worldline approach (eq.\ (\ref{defG}))
is in fact not the inverse of the second derivative acting on the complete
space of trajectories. Partial integration on the circle yields
\begin{equation}
\int_0^T d\tau_1\,G(\tau_1,\tau_2)\,
\frac{1}{2}\frac{\partial^2}{\partial\tau_1^2}x(\tau_1)=
x(\tau_2)-\frac{1}{T} \int_0^T d\tau_1 \,x(\tau_1)\;,
\end{equation}
where the second term should be absent. On the other hand we have
performed the path integration only over the relative coordinates
$y(\tau)\,$, which obey
\begin{equation}
\int_0^T d\tau\,y^\mu(\tau)=0\;.
\label{relcoord}
\end{equation}
This shows that the Green function (\ref{defG}) is the inverse of the 
second derivative on the space of relative coordinates $y$. 
\par
This can be generalized as follows:
An operator can be inverted only after extracting its zero modes.
For the Laplace operator on the circle, the zero modes are just
the constant functions. Therefore we perform the split
\begin{equation}
x(\tau)=x_0\,\unity +y(\tau)\;,\qquad
\int{\cal D}x = \int dx_0\int{\cal D} y\;.
\label{split}
\end{equation}
The value $x_0$ is determined as weighted average
\begin{equation}
x_0=\int_0^T d\tau\,\rho(\tau)x(\tau)\;,
\end{equation}
where the weight function $\rho$ -- the so called background charge 
-- is a periodic function on the circle with the normalization
\begin{equation}
\int_0^T d\tau\,\rho(\tau)=1\;.
\label{norm}
\end{equation}
For the path integration variables $y$ this leads to the general 
constraint
\begin{equation}
\int_0^T d\tau\,\rho(\tau)y(\tau)=0\;.
\label{constr}
\end{equation}
The defining equation for the Green function $G^{(\rho)}$ reads now
\begin{equation}
\P\frac{1}{2}\partial^2\P G^{(\rho)}\P=
\P G^{(\rho)}\P\frac{1}{2}\partial^2\P=\P\;,
\label{Pequ}
\end{equation}
where
\begin{equation}
\P=\unity-\frac{|\rho\rangle\langle\rho |}{\langle\rho |\rho\rangle}
\end{equation}
is the projector on the subspace of functions $y$ which obey the
constraint (\ref{constr}), $\P|y\rangle=|y\rangle$.
One can now show that for real Green functions the following conditions 
are sufficient to fulfill eq.~(\ref{Pequ}):
\begin{eqnarray}
G^{(\rho)}(\tau_1,\tau_2)&=&G^{(\rho)}(\tau_2,\tau_1)\;,\label{symm}\\
\rho(\tau_1)\int_0^T\rho(\tau)G^{(\rho)}(\tau,\tau_2)d\tau&=&
\rho(\tau_2)\int_0^T\rho(\tau)G^{(\rho)}(\tau_1,\tau)d\tau \;,
\label{intGro}\\
\frac{1}{2}\frac{\partial^2}{\partial\tau_1^2}
G^{(\rho)}(\tau_1,\tau_2)&=&\delta(\tau_1-\tau_2)-\rho(\tau_1)\;.
\end{eqnarray}
In order to solve the last equation, it is convenient to construct a 
generating functional $H$ as the solution to
\begin{equation}
\frac{1}{2}\frac{\partial^2}{\partial\tau_1^2}
H(\tau_1,\tau_2;\sigma)=\delta(\tau_1-\tau_2)-\delta(\tau_1-\sigma)\;.
\label{defH}
\end{equation}
$G^{(\rho)}$ is then obtained from a convolution of $H$ with $\rho$:
\begin{equation}
G^{(\rho)}(\tau_1,\tau_2) = \int_0^T d\sigma
\rho (\sigma )H(\tau_1,\tau_2;\sigma)\; .
\label{convol}
\end{equation}
The requirements of symmetry (\ref{symm}) and periodic boundary 
conditions determine $H$ uniquely as
\begin{equation}
H(\tau_1,\tau_2;\sigma) = |\tau_1-\tau_2|-|\tau_1-\sigma|
-|\tau_2-\sigma|+\frac{2}{T}(\tau_1-\sigma)(\tau_2-\sigma)+h(\sigma)
\label{solH}
\end{equation}
up to a function $h(\sigma)$. With the special choice
\begin{equation}
h(\sigma)=\int_0^T\rho(\tau)|\sigma-\tau|d\tau-\frac{2\sigma}{T}
\left(\sigma-\int_0^T\rho(\tau)\tau d\tau\right)
\end{equation}
we can achieve
\begin{equation}
\int_0^T\rho(\tau_1)G^{(\rho)}(\tau_1,\tau_2)d\tau_1=0\;,
\end{equation}
such that eq.~(\ref{intGro}) is trivially fulfilled.

Any $G^{(\rho)}$ constructed from (\ref{convol}) can be used as 
Green function for the evaluation of the path integral.
Different choices of $\rho$ lead to different effective
Lagrangians, but to the same effective action. Let us shortly 
verify this assertion for the scalar potential case, where
the integrand is just the universal exponential
\begin{equation}
\exp\biggl[-\frac{1}{2}\sum_{i,j=1}^nG^{(\rho)}(\tau_i,\tau_j)
\partial_{(i)}\partial_{(j)}\biggr]
\end{equation}
(note that for a general $G^{(\rho)}$ diagonal terms have to 
be included). Using the observation
\begin{equation}
H(\tau_1,\tau_2;\sigma)=G(\tau_1,\tau_2)-
G(\tau_1,\sigma)-G(\tau_2,\sigma)+h(\sigma)\;,
\end{equation}
where $G$ is our original Green function eq.~(\ref{defG}),
we may rewrite the exponent as
\par
\hfill\parbox{14.4cm}{
\begin{eqnarray*}
-\frac{1}{2}\sum_{i,j=1}^nG^{(\rho)}(\tau_i,\tau_j)
\partial_{(i)}\partial_{(j)}&=&-\frac{1}{2}\sum_{i,j=1}^n
G(\tau_i,\tau_j)\partial_{(i)}\partial_{(j)}\nonumber\\
&+&\sum_{i=1}^n\partial_{(i)}\int_0^Td\sigma\rho(\sigma)
\sum_{j=1}^nG(\tau_j,\sigma)\partial_{(j)} \nonumber\\
&-&\frac{1}{2}\sum_{i,j=1}^n\partial_{(i)}\partial_{(j)}
\int_0^Td\sigma\rho(\sigma)h(\sigma)\;.\nonumber
\end{eqnarray*} }
\hfill\parbox{0.8cm}{\begin{eqnarray}\label{exp} \end{eqnarray} }
\par\noindent
This shows that all $\rho$-- and $h$--dependent terms in the 
effective Lagrangian carry at least one free factor of 
$\sum\partial_{(i)}$ and therefore are total derivatives. 
This argument can be easily generalized to the gauge theory case.
In particular, by performing the shift
$h(\sigma)\rightarrow h(\sigma)+c$ we see from eq.~(\ref{exp}) 
that two Green functions that differ only by a constant $c$
lead to effective Lagrangians that differ only by total 
derivatives and thus give the same effective action.

\newpage

\vskip.5cm
{\bf Appendix C}
\vskip.5cm

The coefficient $O_6$ written in the minimal basis reads

\bear
  O_6 &=& 
   V^6 
 + 6 V^2V_{\kappa}VV_{\kappa} 
 + 4 V^3V_{\kappa}V_{\kappa} 
 \nonumber \\ 
  &&  
 + \;\frac{2}{7}\;\mbox{i}\;V^2F_{\kappa\lambda}V_{\lambda}V_{\kappa} 
 + \;\frac{5}{7}\;V^4F_{\kappa\lambda}F_{\lambda\kappa} 
 + \;\frac{8}{7}\;V^3F_{\kappa\lambda}VF_{\lambda\kappa} 
 \nonumber \\ 
  &&  
 - \;\frac{8}{7}\;\mbox{i}\;VF_{\kappa\lambda}V_{\lambda}VV_{\kappa} 
 + \;\frac{9}{7}\;VV_{\kappa\lambda}VV_{\lambda\kappa} 
 + \;\frac{9}{7}\;V_{\kappa}V_{\kappa}V_{\lambda}V_{\lambda} 
 \nonumber \\ 
  &&  
 + \;\frac{9}{14}\;V^2F_{\kappa\lambda}V^2F_{\lambda\kappa} 
 - \;\frac{10}{7}\;\mbox{i}\;VF_{\kappa\lambda}VV_{\lambda}V_{\kappa} 
 + \;\frac{12}{7}\;V^2V_{\kappa\lambda}V_{\lambda\kappa} 
 \nonumber \\ 
  &&  
 + \;\frac{17}{14}\;V_{\kappa}V_{\lambda}V_{\kappa}V_{\lambda} 
 + \;\frac{18}{7}\;VV_{\kappa}V_{\kappa\lambda}V_{\lambda} 
 - \;\frac{18}{7}\;\mbox{i}\;V^2V_{\kappa}F_{\kappa\lambda}V_{\lambda} 
 \nonumber \\ 
  &&  
 - \;\frac{24}{7}\;\mbox{i}\;V^2V_{\kappa}V_{\lambda}F_{\lambda\kappa} 
 + \;\frac{26}{7}\;VV_{\kappa\lambda}V_{\lambda}V_{\kappa} 
 + \;\frac{26}{7}\;VV_{\kappa}V_{\lambda}V_{\lambda\kappa} 
 \nonumber \\ 
  &&  
 - \;\frac{26}{7}\;\mbox{i}\;VV_{\kappa}VV_{\lambda}F_{\lambda\kappa} 
 - VF_{\kappa\lambda}V_{\lambda}F_{\kappa\mu}V_{\mu} 
 - 2 VV_{\kappa}F_{\kappa\lambda}V_{\mu}F_{\mu\lambda} 
 \nonumber \\ 
  &&  
 + \; V_{\kappa}V_{\kappa\lambda\mu}V_{\mu\lambda} 
 + \; V_{\kappa}V_{\lambda\mu}V_{\mu\lambda\kappa} 
 + \;\frac{1}{2}\;VF_{\kappa\lambda}F_{\lambda\kappa}V_{\mu}V_{\mu} 
 \nonumber \\ 
  &&  
 + \;\frac{1}{2}\;VV_{\kappa}V_{\kappa}F_{\lambda\mu}F_{\mu\lambda} 
 + \;\frac{1}{2}\;V^2F_{\kappa\lambda}F_{\mu\lambda\kappa}V_{\mu} 
 + \;\frac{1}{2}\;V^2V_{\kappa}F_{\kappa\lambda\mu}F_{\mu\lambda} 
 \nonumber \\ 
  &&  
 - \;\frac{1}{7}\;VF_{\kappa\lambda}V_{\mu}F_{\mu\lambda}V_{\kappa} 
 + \;\frac{2}{7}\;VV_{\kappa}F_{\lambda\mu}F_{\mu\kappa}V_{\lambda} 
 - \;\frac{2}{7}\;VV_{\kappa}V_{\lambda}F_{\kappa\mu}F_{\mu\lambda} 
 \nonumber \\ 
  &&  
 - \;\frac{2}{7}\;\mbox{i}\;VF_{\kappa\lambda\mu}V_{\mu\kappa}V_{\lambda} 
 - \;\frac{2}{7}\;\mbox{i}\;VF_{\kappa\lambda\mu}V_{\mu}V_{\lambda\kappa} 
 - \;\frac{2}{21}\;\mbox{i}\;VF_{\kappa\lambda}VF_{\lambda\mu}VF_{\mu\kappa} 
 \nonumber \\ 
  &&  
 + \;\frac{3}{7}\;VV_{\kappa\lambda\mu}V_{\mu\lambda\kappa} 
 + \;\frac{3}{7}\;V^2F_{\kappa\lambda}F_{\lambda\mu}V_{\mu\kappa} 
 + \;\frac{3}{7}\;V^2F_{\kappa\lambda}V_{\lambda\mu}F_{\mu\kappa} 
 \nonumber \\ 
  &&  
 - \;\frac{3}{7}\;\mbox{i}\;F_{\kappa\lambda}V_{\lambda\mu}V_{\kappa}V_{\mu} 
 - \;\frac{4}{7}\;VF_{\kappa\lambda}F_{\lambda\mu}VV_{\mu\kappa} 
 - \;\frac{4}{7}\;VF_{\kappa\lambda}F_{\lambda\mu}V_{\mu}V_{\kappa} 
 \nonumber \\ 
  &&  
 - \;\frac{4}{7}\;VF_{\kappa\lambda}VV_{\lambda\mu}F_{\mu\kappa} 
 - \;\frac{4}{7}\;VF_{\kappa\lambda}V_{\lambda}V_{\mu}F_{\mu\kappa} 
 - \;\frac{4}{7}\;VF_{\kappa\lambda}V_{\mu}V_{\lambda}F_{\mu\kappa} 
 \nonumber \\ 
  &&  
 - \;\frac{4}{7}\;\mbox{i}\;VF_{\kappa\lambda}V_{\lambda\mu}V_{\mu\kappa} 
 - \;\frac{4}{7}\;\mbox{i}\;V^2F_{\kappa\lambda}F_{\lambda\mu}VF_{\mu\kappa} 
 - \;\frac{4}{7}\;\mbox{i}\;V^2F_{\kappa\lambda}VF_{\lambda\mu}F_{\mu\kappa} 
 \nonumber \\ 
  &&  
 + \;\frac{5}{14}\;V^3F_{\kappa\lambda\mu}F_{\kappa\mu\lambda} 
 + \;\frac{6}{7}\;V^2F_{\kappa\lambda\mu}F_{\mu\lambda}V_{\kappa} 
 + \;\frac{6}{7}\;V^2F_{\kappa\lambda\mu}V_{\kappa}F_{\mu\lambda} 
 \nonumber \\ 
  &&  
 + \;\frac{6}{7}\;V^2F_{\kappa\lambda}V_{\mu}F_{\mu\lambda\kappa} 
 + \;\frac{6}{7}\;V^2V_{\kappa}F_{\lambda\mu}F_{\kappa\mu\lambda} 
 + \;\frac{8}{7}\;VF_{\kappa\lambda}F_{\lambda\mu}V_{\kappa}V_{\mu} 
 \nonumber \\ 
  &&  
 - \;\frac{8}{7}\;VV_{\kappa}F_{\lambda\mu}V_{\mu}F_{\lambda\kappa} 
 - \;\frac{9}{7}\;\mbox{i}\;VV_{\kappa\lambda}V_{\mu}F_{\lambda\mu\kappa} 
 - \;\frac{9}{7}\;\mbox{i}\;VV_{\kappa}V_{\lambda\mu}F_{\mu\lambda\kappa} 
 \nonumber \\
  &&  
 + \;\frac{9}{14}\;V^2F_{\kappa\lambda\mu}VF_{\kappa\mu\lambda} 
 - \;\frac{10}{7}\;\mbox{i}\;V^3F_{\kappa\lambda}F_{\lambda\mu}F_{\mu\kappa} 
 - \;\frac{11}{7}\;VV_{\kappa}F_{\kappa\lambda}F_{\lambda\mu}V_{\mu} 
 \nonumber \\ 
  &&  
 - \;\frac{11}{7}\;\mbox{i}\;F_{\kappa\lambda}V_{\lambda\mu}V_{\mu}V_{\kappa} 
 - \;\frac{11}{7}\;\mbox{i}\;F_{\kappa\lambda}V_{\lambda}V_{\mu}V_{\mu\kappa} 
 + \;\frac{11}{14}\;VF_{\kappa\lambda\mu}F_{\mu\lambda}VV_{\kappa} 
 \nonumber \\ 
  &&  
 + \;\frac{11}{14}\;VF_{\kappa\lambda}F_{\mu\lambda\kappa}VV_{\mu} 
 + \;\frac{11}{14}\;VF_{\kappa\lambda}VF_{\mu\lambda\kappa}V_{\mu} 
 + \;\frac{11}{14}\;VF_{\kappa\lambda}VV_{\mu}F_{\mu\lambda\kappa} 
 \nonumber \\ 
  &&  
 + \;\frac{11}{14}\;VF_{\kappa\lambda}V_{\mu}V_{\mu}F_{\lambda\kappa} 
 + \;\frac{11}{14}\;VV_{\kappa}F_{\lambda\mu}F_{\mu\lambda}V_{\kappa} 
 + \;\frac{11}{21}\;V_{\kappa\lambda}V_{\lambda\mu}V_{\mu\kappa} 
 \nonumber 
\ear

\newpage

\bear
  &&
 - \;\frac{12}{7}\;V^2V_{\kappa\lambda}F_{\lambda\mu}F_{\mu\kappa} 
 - \;\frac{15}{7}\;VV_{\kappa}V_{\lambda}F_{\lambda\mu}F_{\mu\kappa} 
 - \;\frac{15}{7}\;\mbox{i}\;VV_{\kappa\lambda}F_{\lambda\mu}V_{\mu\kappa} 
 \nonumber \\ 
  &&  
 - \;\frac{15}{7}\;\mbox{i}\;VV_{\kappa\lambda}V_{\lambda\mu}F_{\mu\kappa} 
 - \;\frac{17}{7}\;\mbox{i}\;F_{\kappa\lambda}V_{\mu}V_{\lambda}V_{\mu\kappa} 
 + \;\frac{17}{14}\;VF_{\kappa\lambda\mu}VV_{\kappa}F_{\mu\lambda} 
 \nonumber \\ 
  &&  
 + \;\frac{17}{14}\;VF_{\kappa\lambda}V_{\mu}F_{\lambda\kappa}V_{\mu} 
 + \;\frac{17}{14}\;VF_{\kappa\lambda}V_{\mu}VF_{\mu\lambda\kappa} 
 + \;\frac{17}{14}\;VV_{\kappa}F_{\lambda\mu}V_{\kappa}F_{\mu\lambda} 
 \nonumber \\ 
  &&  
 - \; 2 \; \mbox{i}\;F_{\kappa\lambda\mu}V_{\kappa}V_{\mu}V_{\lambda} 
 - \; 2 \; \mbox{i}\;F_{\kappa\lambda\mu}V_{\mu}V_{\lambda}V_{\kappa} 
 - \; 2 \; \mbox{i}\;F_{\kappa\lambda}V_{\mu}V_{\mu\lambda}V_{\kappa} 
 \nonumber \\ 
  &&  
 - \; \mbox{i}\;VV_{\kappa\lambda}F_{\lambda\kappa\mu}V_{\mu} 
 - \; \mbox{i}\;VV_{\kappa}F_{\lambda\kappa\mu}V_{\mu\lambda} 
 + \;\frac{1}{2}\;VF_{\kappa\lambda\mu}V_{\kappa\nu}F_{\nu\mu\lambda} 
 \nonumber \\ 
  &&  
 + \;\frac{1}{3}\;V^2F_{\kappa\lambda}F_{\mu\nu}F_{\nu\lambda}F_{\mu\kappa} 
 - \;\frac{1}{3}\;\mbox{i}\;V^2F_{\kappa\lambda\mu}F_{\nu\mu\lambda}F_{\nu\kappa} 
 - \;\frac{1}{6}\;\mbox{i}\;VF_{\kappa\lambda}VF_{\lambda\mu\nu}F_{\kappa\nu\mu} 
 \nonumber \\ 
  &&  
 - \;\frac{1}{6}\;\mbox{i}\;V^2F_{\kappa\lambda\mu}F_{\kappa\nu}F_{\nu\mu\lambda} 
 - \;\frac{1}{7}\;VF_{\kappa\lambda}VF_{\lambda\mu}F_{\kappa\nu}F_{\nu\mu} 
 - \;\frac{1}{9}\;\mbox{i}\;VF_{\kappa\lambda}V_{\mu\nu}F_{\nu\lambda}F_{\mu\kappa} 
 \nonumber \\ 
  &&  
 - \;\frac{1}{14}\;\mbox{i}\;VF_{\kappa\lambda}F_{\lambda\mu\nu}VF_{\kappa\nu\mu} 
 + \;\frac{1}{18}\;\mbox{i}\;VF_{\kappa\lambda}F_{\mu\nu}F_{\nu\lambda\kappa}V_{\mu} 
 - \;\frac{1}{21}\;VF_{\kappa\lambda\mu}F_{\kappa\mu\nu}V_{\nu\lambda} 
 \nonumber \\ 
  &&  
 + \;\frac{1}{21}\;VF_{\kappa\lambda\mu}F_{\mu\nu}V_{\nu\lambda\kappa} 
 - \;\frac{1}{21}\;VF_{\kappa\lambda\mu}V_{\mu\kappa\nu}F_{\nu\lambda} 
 + \;\frac{1}{21}\;VF_{\kappa\lambda\mu}V_{\mu\nu}F_{\kappa\nu\lambda} 
 \nonumber \\ 
  &&  
 - \;\frac{1}{21}\;VF_{\kappa\lambda}F_{\mu\lambda\nu}V_{\nu\mu\kappa} 
 - \;\frac{1}{21}\;VF_{\kappa\lambda}V_{\lambda\mu\nu}F_{\nu\mu\kappa} 
 + \;\frac{1}{21}\;\mbox{i}\;VF_{\kappa\lambda}F_{\lambda\mu\nu}V_{\kappa}F_{\nu\mu} 
 \nonumber \\ 
  &&  
 + \;\frac{1}{21}\;\mbox{i}\;VF_{\kappa\lambda}F_{\lambda\mu\nu}V_{\nu}F_{\mu\kappa} 
 + \;\frac{1}{21}\;\mbox{i}\;VV_{\kappa}F_{\kappa\lambda\mu}F_{\mu\nu}F_{\nu\lambda} 
 - \;\frac{1}{21}\;\mbox{i}\;VV_{\kappa}F_{\lambda\mu}F_{\mu\nu}F_{\nu\lambda\kappa} 
 \nonumber \\ 
  &&  
 + \;\frac{1}{42}\;V_{\kappa\lambda\mu\nu}V_{\nu\mu\lambda\kappa} 
 - \;\frac{2}{7}\;\mbox{i}\;F_{\kappa\lambda\mu\nu}V_{\nu\lambda}V_{\mu\kappa} 
 - \;\frac{2}{7}\;\mbox{i}\;VF_{\kappa\lambda\mu}V_{\nu}F_{\mu\lambda}F_{\nu\kappa} 
 \nonumber \\ 
  &&  
 - \;\frac{2}{9}\;\mbox{i}\;VF_{\kappa\lambda}F_{\lambda\mu}V_{\nu}F_{\nu\mu\kappa} 
 - \;\frac{2}{9}\;\mbox{i}\;VF_{\kappa\lambda}V_{\mu}F_{\mu\lambda\nu}F_{\nu\kappa} 
 + \;\frac{2}{21}\;F_{\kappa\lambda}F_{\mu\lambda\nu}V_{\kappa}V_{\nu\mu} 
 \nonumber \\ 
  &&  
 + \;\frac{2}{21}\;F_{\kappa\lambda}V_{\mu\nu}V_{\lambda}F_{\nu\mu\kappa} 
 + \;\frac{2}{21}\;VF_{\kappa\lambda\mu\nu}VF_{\lambda\kappa\nu\mu} 
 + \;\frac{2}{21}\;\mbox{i}\;VF_{\kappa\lambda}F_{\lambda\mu}V_{\nu}F_{\kappa\nu\mu} 
 \nonumber \\ 
  &&  
 + \;\frac{2}{21}\;\mbox{i}\;VF_{\kappa\lambda}V_{\mu}F_{\nu\mu\lambda}F_{\nu\kappa} 
 - \;\frac{2}{63}\;\mbox{i}\;VF_{\kappa\lambda}F_{\mu\nu}F_{\nu\lambda}V_{\mu\kappa} 
 - \;\frac{3}{7}\;F_{\kappa\lambda}V_{\lambda}F_{\mu\kappa\nu}V_{\nu\mu} 
 \nonumber \\ 
  &&  
 - \;\frac{3}{7}\;F_{\kappa\lambda}V_{\mu\nu}F_{\nu\mu\lambda}V_{\kappa} 
 - \;\frac{3}{7}\;\mbox{i}\;V^2F_{\kappa\lambda\mu}F_{\kappa\mu\nu}F_{\nu\lambda} 
 - \;\frac{3}{7}\;\mbox{i}\;V^2F_{\kappa\lambda}F_{\mu\lambda\nu}F_{\mu\nu\kappa} 
 \nonumber \\ 
  &&  
 - \;\frac{3}{14}\;\mbox{i}\;F_{\kappa\lambda}F_{\mu\nu}V_{\nu}F_{\lambda\kappa}V_{\mu} 
 - \;\frac{4}{7}\;VV_{\kappa\lambda\mu}F_{\mu\lambda\nu}F_{\nu\kappa} 
 - \;\frac{4}{7}\;VV_{\kappa\lambda}F_{\mu\lambda\nu}F_{\mu\nu\kappa} 
 \nonumber \\ 
  &&  
 - \;\frac{4}{9}\;F_{\kappa\lambda}V_{\lambda\mu}F_{\mu\nu}V_{\nu\kappa} 
 - \;\frac{4}{21}\;\mbox{i}\;F_{\kappa\lambda\mu\nu}V_{\nu\lambda\kappa}V_{\mu} 
 - \;\frac{4}{21}\;\mbox{i}\;F_{\kappa\lambda\mu\nu}V_{\nu}V_{\mu\lambda\kappa} 
 \nonumber \\ 
  &&  
 - \;\frac{4}{21}\;\mbox{i}\;VF_{\kappa\lambda}VF_{\mu\lambda\nu}F_{\mu\nu\kappa} 
 - \;\frac{4}{21}\;\mbox{i}\;VV_{\kappa\lambda}F_{\mu\nu}F_{\nu\lambda}F_{\mu\kappa} 
 + \;\frac{4}{21}\;\mbox{i}\;V^2F_{\kappa\lambda}F_{\lambda\mu\nu}F_{\kappa\nu\mu} 
 \nonumber \\ 
  &&  
 + \;\frac{4}{63}\;F_{\kappa\lambda}F_{\lambda\mu\nu}V_{\nu\kappa}V_{\mu} 
 + \;\frac{4}{63}\;F_{\kappa\lambda}F_{\mu\lambda\nu}V_{\mu\kappa}V_{\nu} 
 + \;\frac{4}{63}\;F_{\kappa\lambda}V_{\lambda\mu\nu}F_{\nu\kappa}V_{\mu} 
 \nonumber \\ 
  &&  
 - \;\frac{4}{63}\;\mbox{i}\;F_{\kappa\lambda}F_{\lambda\mu}V_{\nu}F_{\nu\kappa}V_{\mu} 
 + \;\frac{4}{63}\;\mbox{i}\;VF_{\kappa\lambda\mu}F_{\kappa\nu}F_{\nu\mu}V_{\lambda} 
 - \;\frac{5}{7}\;\mbox{i}\;VF_{\kappa\lambda\mu}V_{\nu}F_{\nu\kappa}F_{\mu\lambda} 
 \nonumber \\ 
  &&  
 + \;\frac{5}{9}\;F_{\kappa\lambda}F_{\mu\lambda\kappa}V_{\nu}V_{\nu\mu} 
 + \;\frac{5}{9}\;F_{\kappa\lambda}F_{\mu\nu\lambda\kappa}V_{\nu}V_{\mu} 
 + \;\frac{5}{9}\;F_{\kappa\lambda}V_{\mu\nu}V_{\nu}F_{\mu\lambda\kappa} 
 \nonumber \\ 
  &&  
 + \;\frac{5}{9}\;F_{\kappa\lambda}V_{\mu}F_{\mu\nu\lambda\kappa}V_{\nu} 
 + \;\frac{5}{9}\;F_{\kappa\lambda}V_{\mu}V_{\nu}F_{\nu\mu\lambda\kappa} 
 - \;\frac{5}{9}\;\mbox{i}\;VF_{\kappa\lambda}V_{\mu}F_{\lambda\nu}F_{\nu\mu\kappa} 
 \nonumber \\ 
  &&  
 - \;\frac{5}{9}\;\mbox{i}\;VV_{\kappa}F_{\kappa\lambda}F_{\mu\nu}F_{\lambda\nu\mu} 
 + \;\frac{5}{14}\;VF_{\kappa\lambda\mu\nu}F_{\lambda\nu\mu}V_{\kappa} 
 + \;\frac{5}{14}\;VF_{\kappa\lambda\mu\nu}V_{\lambda\kappa}F_{\nu\mu} 
 \nonumber \\ 
  &&  
 + \;\frac{5}{14}\;VF_{\kappa\lambda}V_{\mu\nu}F_{\nu\mu\lambda\kappa} 
 + \;\frac{5}{14}\;VV_{\kappa}F_{\lambda\mu\nu}F_{\lambda\kappa\nu\mu} 
 - \;\frac{5}{18}\;\mbox{i}\;F_{\kappa\lambda}F_{\mu\nu}F_{\lambda\kappa}V_{\nu}V_{\mu} 
 \nonumber \\ 
  &&  
 + \;\frac{5}{21}\;F_{\kappa\lambda\mu}F_{\kappa\mu\lambda}V_{\nu}V_{\nu} 
 + \;\frac{5}{21}\;F_{\kappa\lambda}F_{\lambda\kappa}V_{\mu\nu}V_{\nu\mu} 
 + \;\frac{5}{21}\;VF_{\kappa\lambda\mu}F_{\kappa\nu\mu\lambda}V_{\nu} 
 \nonumber \\ 
  &&  
 + \;\frac{5}{21}\;VF_{\kappa\lambda}F_{\mu\nu\lambda\kappa}V_{\nu\mu} 
 + \;\frac{5}{21}\;VV_{\kappa\lambda}F_{\lambda\kappa\mu\nu}F_{\nu\mu} 
 + \;\frac{5}{21}\;VV_{\kappa}F_{\kappa\lambda\mu\nu}F_{\lambda\nu\mu} 
 \nonumber \\ 
  &&  
 + \;\frac{5}{21}\;\mbox{i}\;F_{\kappa\lambda}F_{\lambda\mu}F_{\kappa\nu}V_{\mu}V_{\nu} 
 + \;\frac{5}{42}\;V^2F_{\kappa\lambda\mu\nu}F_{\lambda\kappa\nu\mu} 
 - \;\frac{5}{63}\;VF_{\kappa\lambda}F_{\lambda\mu}VF_{\mu\nu}F_{\nu\kappa} 
 \nonumber \\ 
  &&  
 + \;\frac{5}{63}\;V^2F_{\kappa\lambda}F_{\lambda\kappa}F_{\mu\nu}F_{\nu\mu} 
 + \;\frac{5}{63}\;\mbox{i}\;F_{\kappa\lambda}F_{\lambda\kappa}F_{\mu\nu}V_{\nu}V_{\mu} 
 + \;\frac{5}{63}\;\mbox{i}\;VF_{\kappa\lambda}F_{\mu\nu}V_{\lambda}F_{\kappa\nu\mu} 
 \nonumber \\ 
  &&  
 + \;\frac{5}{63}\;\mbox{i}\;VF_{\kappa\lambda}V_{\lambda}F_{\kappa\mu\nu}F_{\nu\mu} 
 + \;\frac{5}{126}\;VF_{\kappa\lambda}F_{\lambda\kappa}VF_{\mu\nu}F_{\nu\mu} 
 - \;\frac{7}{9}\;\mbox{i}\;VF_{\kappa\lambda}F_{\mu\nu}V_{\nu\lambda}F_{\mu\kappa} 
 \nonumber \\ 
  &&  
 - \;\frac{8}{9}\;F_{\kappa\lambda}F_{\lambda\mu}V_{\mu\nu}V_{\nu\kappa} 
 - \;\frac{8}{9}\;F_{\kappa\lambda}F_{\lambda\mu}V_{\nu}V_{\nu\mu\kappa} 
 - \;\frac{8}{9}\;F_{\kappa\lambda}F_{\mu\lambda\nu}V_{\mu}V_{\nu\kappa} 
 \nonumber \\ 
  &&  
 - \;\frac{8}{9}\;F_{\kappa\lambda}V_{\lambda\mu}V_{\nu}F_{\nu\mu\kappa} 
 - \;\frac{8}{9}\;F_{\kappa\lambda}V_{\mu}F_{\mu\lambda\nu}V_{\nu\kappa} 
 + \;\frac{8}{9}\;\mbox{i}\;F_{\kappa\lambda}F_{\lambda\mu}F_{\mu\nu}V_{\nu}V_{\kappa} 
 \nonumber \\ 
  &&  
 + \;\frac{8}{9}\;\mbox{i}\;VV_{\kappa}F_{\lambda\kappa\mu}F_{\lambda\nu}F_{\nu\mu} 
 - \;\frac{8}{21}\;\mbox{i}\;VF_{\kappa\lambda\mu}VF_{\kappa\mu\nu}F_{\nu\lambda} 
 - \;\frac{8}{21}\;\mbox{i}\;VF_{\kappa\lambda}F_{\mu\lambda\nu}VF_{\mu\nu\kappa} 
 \nonumber \\ 
  &&  
 - \;\frac{8}{63}\;\mbox{i}\;F_{\kappa\lambda}F_{\lambda\mu}V_{\nu}V_{\kappa}F_{\nu\mu} 
 - \;\frac{10}{21}\;VV_{\kappa\lambda\mu}F_{\mu\nu}F_{\lambda\nu\kappa} 
 + \;\frac{10}{21}\;V^2F_{\kappa\lambda}F_{\lambda\mu}F_{\kappa\nu}F_{\nu\mu} 
 \nonumber \\ 
  &&  
 - \;\frac{10}{21}\;V^2F_{\kappa\lambda}F_{\lambda\mu}F_{\mu\nu}F_{\nu\kappa} 
 - \;\frac{10}{21}\;\mbox{i}\;F_{\kappa\lambda}F_{\lambda\kappa}V_{\mu}V_{\nu}F_{\nu\mu} 
 - \;\frac{10}{21}\;\mbox{i}\;VF_{\kappa\lambda}F_{\mu\lambda\kappa}V_{\nu}F_{\nu\mu} 
 \nonumber \\ 
  &&  
 + \;\frac{10}{63}\;F_{\kappa\lambda}F_{\lambda\mu}V_{\kappa\nu}V_{\nu\mu} 
 + \;\frac{10}{63}\;F_{\kappa\lambda}F_{\lambda\mu}V_{\mu\kappa\nu}V_{\nu} 
 + \;\frac{10}{63}\;F_{\kappa\lambda}V_{\lambda\mu}F_{\nu\mu\kappa}V_{\nu} 
 \nonumber \\ 
  &&  
 - \;\frac{10}{63}\;VF_{\kappa\lambda}VF_{\lambda\mu}F_{\mu\nu}F_{\nu\kappa} 
 - \;\frac{10}{63}\;\mbox{i}\;F_{\kappa\lambda}F_{\lambda\mu}F_{\kappa\nu}V_{\nu}V_{\mu} 
 - \;\frac{10}{63}\;\mbox{i}\;F_{\kappa\lambda}F_{\lambda\mu}V_{\kappa}V_{\nu}F_{\nu\mu} 
 \nonumber \\ 
  &&  
 + \;\frac{10}{63}\;\mbox{i}\;VF_{\kappa\lambda}F_{\lambda\mu}F_{\kappa\nu}V_{\nu\mu} 
 - \;\frac{10}{63}\;\mbox{i}\;VF_{\kappa\lambda}F_{\lambda\mu}F_{\mu\kappa\nu}V_{\nu} 
 - \;\frac{11}{21}\;F_{\kappa\lambda\mu}F_{\kappa\mu\nu}V_{\nu}V_{\lambda} 
 \nonumber \\ 
  &&  
 - \;\frac{11}{21}\;F_{\kappa\lambda}F_{\mu\lambda\nu}V_{\nu\mu}V_{\kappa} 
 - \;\frac{11}{21}\;F_{\kappa\lambda}V_{\lambda\mu}F_{\mu\kappa\nu}V_{\nu} 
 - \;\frac{11}{21}\;F_{\kappa\lambda}V_{\lambda}V_{\mu\nu}F_{\nu\mu\kappa} 
 \nonumber \\ 
  &&  
 - \;\frac{11}{21}\;\mbox{i}\;F_{\kappa\lambda\mu}V_{\mu\kappa\nu}V_{\nu\lambda} 
 - \;\frac{11}{21}\;\mbox{i}\;F_{\kappa\lambda\mu}V_{\mu\nu}V_{\nu\lambda\kappa} 
 - \;\frac{11}{21}\;\mbox{i}\;F_{\kappa\lambda}V_{\lambda\mu\nu}V_{\nu\mu\kappa} 
 \nonumber \\ 
  &&  
 - \;\frac{11}{21}\;\mbox{i}\;VF_{\kappa\lambda\mu}F_{\kappa\nu}V_{\nu}F_{\mu\lambda} 
 - \;\frac{11}{21}\;\mbox{i}\;VF_{\kappa\lambda\mu}F_{\mu\lambda}V_{\nu}F_{\nu\kappa} 
 - \;\frac{11}{21}\;\mbox{i}\;VF_{\kappa\lambda}V_{\mu}F_{\mu\nu}F_{\nu\lambda\kappa} 
 \nonumber \\ 
  &&  
 - \;\frac{11}{21}\;\mbox{i}\;VV_{\kappa}F_{\lambda\mu}F_{\nu\mu\lambda}F_{\nu\kappa} 
 - \;\frac{11}{42}\;F_{\kappa\lambda\mu}V_{\mu}F_{\kappa\lambda\nu}V_{\nu} 
 + \;\frac{11}{42}\;F_{\kappa\lambda\mu}V_{\nu}F_{\kappa\mu\lambda}V_{\nu} 
 \nonumber \\ 
  &&  
 - \;\frac{11}{42}\;F_{\kappa\lambda\mu}V_{\nu}F_{\kappa\nu\mu}V_{\lambda} 
 + \;\frac{11}{42}\;F_{\kappa\lambda}V_{\mu\nu}F_{\lambda\kappa}V_{\nu\mu} 
 + \;\frac{11}{63}\;VF_{\kappa\lambda}VF_{\mu\nu}F_{\nu\lambda}F_{\mu\kappa} 
 \nonumber \\ 
  &&  
 + \;\frac{11}{84}\;VF_{\kappa\lambda}VF_{\lambda\kappa}F_{\mu\nu}F_{\nu\mu} 
 + \;\frac{11}{84}\;VF_{\kappa\lambda}VF_{\mu\nu}F_{\nu\mu}F_{\lambda\kappa} 
 + \;\frac{11}{252}\;VF_{\kappa\lambda}F_{\mu\nu}VF_{\lambda\kappa}F_{\nu\mu} 
 \nonumber \\ 
  &&  
 - \;\frac{13}{18}\;\mbox{i}\;VF_{\kappa\lambda}V_{\mu}F_{\nu\lambda\kappa}F_{\nu\mu} 
 - \;\frac{13}{42}\;\mbox{i}\;VF_{\kappa\lambda}F_{\mu\nu}V_{\nu}F_{\mu\lambda\kappa} 
 + \;\frac{13}{63}\;\mbox{i}\;VF_{\kappa\lambda}F_{\lambda\mu}F_{\mu\nu}V_{\nu\kappa} 
 \nonumber \\ 
  &&  
 + \;\frac{13}{63}\;\mbox{i}\;VF_{\kappa\lambda}F_{\lambda\mu}V_{\mu\nu}F_{\nu\kappa} 
 + \;\frac{13}{63}\;\mbox{i}\;VF_{\kappa\lambda}V_{\lambda\mu}F_{\mu\nu}F_{\nu\kappa} 
 + \;\frac{13}{126}\;V^2F_{\kappa\lambda}F_{\mu\nu}F_{\lambda\kappa}F_{\nu\mu} 
 \nonumber \\ 
  &&  
 + \;\frac{16}{21}\;\mbox{i}\;F_{\kappa\lambda}F_{\mu\nu}V_{\nu}F_{\mu\lambda}V_{\kappa} 
 - \;\frac{16}{21}\;\mbox{i}\;V^2F_{\kappa\lambda\mu}F_{\mu\nu}F_{\kappa\nu\lambda} 
 - \;\frac{16}{63}\;\mbox{i}\;VF_{\kappa\lambda}F_{\mu\nu}F_{\lambda\nu\mu}V_{\kappa} 
 \nonumber \\ 
  &&  
 + \;\frac{17}{21}\;VV_{\kappa\lambda}F_{\lambda\mu\nu}F_{\kappa\nu\mu} 
 + \;\frac{17}{21}\;\mbox{i}\;VF_{\kappa\lambda}V_{\lambda\mu}F_{\kappa\nu}F_{\nu\mu} 
 + \;\frac{17}{42}\;VF_{\kappa\lambda\mu\nu}F_{\nu\mu}V_{\lambda\kappa} 
 \nonumber \\ 
  &&  
 + \;\frac{17}{42}\;VF_{\kappa\lambda\mu\nu}V_{\lambda}F_{\kappa\nu\mu} 
 + \;\frac{17}{42}\;VF_{\kappa\lambda\mu}V_{\nu}F_{\nu\kappa\mu\lambda} 
 + \;\frac{17}{42}\;VV_{\kappa\lambda}F_{\mu\nu}F_{\lambda\kappa\nu\mu} 
 \nonumber \\ 
  &&  
 - \;\frac{17}{63}\;\mbox{i}\;VF_{\kappa\lambda}F_{\lambda\mu}V_{\kappa\nu}F_{\nu\mu} 
 + \;\frac{17}{84}\;VF_{\kappa\lambda}F_{\mu\nu}VF_{\nu\mu}F_{\lambda\kappa} 
 - \;\frac{19}{21}\;\mbox{i}\;VF_{\kappa\lambda}F_{\lambda\mu\nu}F_{\nu\kappa}V_{\mu} 
 \nonumber \\ 
  &&  
 - \;\frac{19}{42}\;\mbox{i}\;VV_{\kappa}F_{\lambda\mu\nu}F_{\nu\mu}F_{\lambda\kappa} 
 + \;\frac{19}{63}\;\mbox{i}\;F_{\kappa\lambda}F_{\mu\nu}V_{\lambda}F_{\nu\kappa}V_{\mu} 
 - \;\frac{19}{63}\;\mbox{i}\;VV_{\kappa}F_{\lambda\mu\nu}F_{\lambda\kappa}F_{\nu\mu} 
 \nonumber \\ 
  &&  
 - \;\frac{20}{63}\;\mbox{i}\;VF_{\kappa\lambda\mu}F_{\mu\lambda}F_{\kappa\nu}V_{\nu} 
 - \;\frac{20}{63}\;\mbox{i}\;VF_{\kappa\lambda\mu}V_{\kappa}F_{\mu\nu}F_{\nu\lambda} 
 - \;\frac{20}{63}\;\mbox{i}\;VF_{\kappa\lambda}F_{\mu\lambda\nu}V_{\mu}F_{\nu\kappa} 
 \nonumber \\ 
  &&  
 - \;\frac{22}{21}\;F_{\kappa\lambda}F_{\lambda\mu\nu}V_{\nu}V_{\mu\kappa} 
 - \;\frac{22}{21}\;F_{\kappa\lambda}V_{\lambda\mu}V_{\nu}F_{\kappa\nu\mu} 
 - \;\frac{22}{21}\;F_{\kappa\lambda}V_{\mu}F_{\nu\mu\lambda}V_{\nu\kappa} 
 \nonumber \\ 
  &&  
 + \;\frac{22}{21}\;\mbox{i}\;F_{\kappa\lambda}F_{\lambda\mu}V_{\nu}F_{\nu\mu}V_{\kappa} 
 + \;\frac{23}{21}\;\mbox{i}\;VV_{\kappa\lambda}F_{\lambda\mu}F_{\mu\nu}F_{\nu\kappa} 
 - \;\frac{23}{21}\;\mbox{i}\;VV_{\kappa}F_{\lambda\mu}F_{\mu\nu}F_{\kappa\nu\lambda} 
 \nonumber \\ 
  &&  
 + \;\frac{23}{42}\;VF_{\kappa\lambda\mu}F_{\nu\mu\lambda}V_{\nu\kappa} 
 + \;\frac{23}{63}\;VF_{\kappa\lambda}F_{\mu\nu}VF_{\nu\lambda}F_{\mu\kappa} 
 - \;\frac{23}{63}\;\mbox{i}\;F_{\kappa\lambda}F_{\lambda\mu}V_{\kappa}F_{\mu\nu}V_{\nu} 
 \nonumber \\ 
  &&  
 + \;\frac{23}{84}\;V^2F_{\kappa\lambda}F_{\mu\nu}F_{\nu\mu}F_{\lambda\kappa} 
 - \;\frac{23}{126}\;\mbox{i}\;VF_{\kappa\lambda}F_{\lambda\mu\nu}F_{\nu\mu}V_{\kappa} 
 - \;\frac{23}{126}\;\mbox{i}\;VF_{\kappa\lambda}V_{\lambda}F_{\mu\nu}F_{\kappa\nu\mu} 
 \nonumber \\ 
  &&  
 - \;\frac{25}{42}\;\mbox{i}\;VF_{\kappa\lambda\mu}VF_{\nu\mu\lambda}F_{\nu\kappa} 
 - \;\frac{26}{63}\;\mbox{i}\;VF_{\kappa\lambda\mu}F_{\kappa\nu}V_{\mu}F_{\nu\lambda} 
 - \;\frac{26}{63}\;\mbox{i}\;VF_{\kappa\lambda}V_{\mu}F_{\lambda\nu}F_{\kappa\nu\mu} 
 \nonumber \\ 
  &&  
 + \;\frac{26}{63}\;\mbox{i}\;VV_{\kappa\lambda}F_{\lambda\mu}F_{\kappa\nu}F_{\nu\mu} 
 - \;\frac{29}{63}\;F_{\kappa\lambda}V_{\lambda\mu}F_{\kappa\nu}V_{\nu\mu} 
 - \;\frac{29}{63}\;F_{\kappa\lambda}V_{\mu\nu}F_{\nu\lambda}V_{\mu\kappa} 
 \nonumber \\ 
  &&  
 - \;\frac{29}{63}\;F_{\kappa\lambda}V_{\mu}V_{\lambda\nu}F_{\nu\mu\kappa} 
 - \;\frac{31}{63}\;\mbox{i}\;F_{\kappa\lambda}F_{\lambda\mu}V_{\nu}F_{\mu\kappa}V_{\nu} 
 + \;\frac{31}{126}\;VF_{\kappa\lambda}VF_{\mu\nu}F_{\lambda\kappa}F_{\nu\mu} 
 \nonumber \\ 
  &&  
 - \;\frac{31}{126}\;\mbox{i}\;VV_{\kappa}F_{\lambda\mu}F_{\kappa\nu}F_{\nu\mu\lambda} 
 + \;\frac{32}{63}\;VF_{\kappa\lambda}F_{\lambda\mu}VF_{\kappa\nu}F_{\nu\mu} 
 - \;\frac{34}{63}\;F_{\kappa\lambda}F_{\mu\nu}V_{\nu\lambda}V_{\mu\kappa} 
 \nonumber \\ 
  &&  
 + \;\frac{37}{252}\;F_{\kappa\lambda\mu}V_{\kappa}F_{\nu\mu\lambda}V_{\nu} 
 + \;\frac{37}{252}\;F_{\kappa\lambda\mu}V_{\nu}F_{\nu\mu\lambda}V_{\kappa} 
 + \;\frac{38}{63}\;F_{\kappa\lambda\mu}F_{\nu\mu\lambda}V_{\nu}V_{\kappa} 
 \nonumber \\ 
  &&  
 + \;\frac{43}{63}\;\mbox{i}\;F_{\kappa\lambda}F_{\lambda\mu}V_{\mu}F_{\kappa\nu}V_{\nu} 
 + \;\frac{43}{126}\;F_{\kappa\lambda}F_{\mu\lambda\kappa}V_{\mu\nu}V_{\nu} 
 + \;\frac{43}{126}\;F_{\kappa\lambda}V_{\mu}V_{\mu\nu}F_{\nu\lambda\kappa} 
 \nonumber \\ 
  &&  
 - \;\frac{43}{126}\;\mbox{i}\;F_{\kappa\lambda}F_{\lambda\kappa}V_{\mu}F_{\mu\nu}V_{\nu} 
 - \;\frac{43}{126}\;\mbox{i}\;VF_{\kappa\lambda}F_{\mu\lambda\kappa}F_{\mu\nu}V_{\nu} 
 - \;\frac{43}{126}\;\mbox{i}\;VV_{\kappa}F_{\kappa\lambda}F_{\lambda\mu\nu}F_{\nu\mu} 
 \nonumber \\ 
  &&  
 + \;\frac{46}{63}\;\mbox{i}\;VF_{\kappa\lambda\mu}V_{\mu}F_{\lambda\nu}F_{\nu\kappa} 
 - \;\frac{50}{63}\;\mbox{i}\;VF_{\kappa\lambda\mu}F_{\mu\nu}V_{\lambda}F_{\nu\kappa} 
 - \;\frac{53}{63}\;\mbox{i}\;F_{\kappa\lambda}F_{\lambda\mu}F_{\mu\kappa}V_{\nu}V_{\nu} 
 \nonumber \\ 
  &&  
 - \;\frac{53}{63}\;\mbox{i}\;VF_{\kappa\lambda}F_{\lambda\mu}F_{\nu\mu\kappa}V_{\nu} 
 - \;\frac{61}{126}\;\mbox{i}\;VF_{\kappa\lambda\mu}F_{\kappa\nu}F_{\mu\lambda}V_{\nu} 
 - \;\frac{62}{63}\;F_{\kappa\lambda}V_{\mu}V_{\lambda\nu}F_{\kappa\nu\mu} 
 \nonumber \\ 
  &&  
 - \;\frac{64}{63}\;\mbox{i}\;VF_{\kappa\lambda}F_{\mu\lambda\nu}F_{\mu\kappa}V_{\nu} 
 - \;\frac{64}{63}\;\mbox{i}\;VV_{\kappa}F_{\lambda\mu}F_{\nu\mu\kappa}F_{\nu\lambda} 
 - \;\frac{65}{63}\;\mbox{i}\;VF_{\kappa\lambda\mu}F_{\mu\nu}F_{\nu\lambda}V_{\kappa} 
 \nonumber \\ 
  &&  
 - \;\frac{74}{63}\;\mbox{i}\;VV_{\kappa}F_{\lambda\mu}F_{\mu\kappa\nu}F_{\nu\lambda} 
 + \;\frac{97}{126}\;F_{\kappa\lambda\mu}F_{\nu\mu\lambda}V_{\kappa}V_{\nu} 
 + \;\frac{97}{126}\;F_{\kappa\lambda}V_{\mu\nu}F_{\nu\lambda\kappa}V_{\mu} 
 \nonumber \\ 
  &&  
 + \;\frac{97}{126}\;F_{\kappa\lambda}V_{\mu}F_{\nu\lambda\kappa}V_{\nu\mu} 
 - \;\frac{97}{126}\;\mbox{i}\;F_{\kappa\lambda}F_{\mu\nu}V_{\lambda}F_{\nu\mu}V_{\kappa} 
 - \;\frac{1}{3}\;\mbox{i}\;F_{\kappa\lambda}F_{\lambda\mu\nu\rho}F_{\mu\rho\kappa}V_{\nu} 
 \nonumber \\ 
  &&  
 + \;\frac{1}{5}\;F_{\kappa\lambda}F_{\lambda\mu}F_{\nu\kappa\rho}V_{\nu}F_{\rho\mu} 
 - \;\frac{1}{5}\;F_{\kappa\lambda}F_{\mu\nu}V_{\rho}F_{\rho\lambda}F_{\kappa\nu\mu} 
 + \;\frac{1}{5}\;VF_{\kappa\lambda\mu}F_{\kappa\mu\nu}F_{\lambda\rho}F_{\rho\nu} 
 \nonumber \\ 
  &&  
 - \;\frac{1}{5}\;\mbox{i}\;F_{\kappa\lambda}F_{\mu\nu\rho}F_{\mu\lambda\rho\nu}V_{\kappa} 
 - \;\frac{1}{5}\;\mbox{i}\;F_{\kappa\lambda}F_{\mu\nu\rho}V_{\mu\lambda}F_{\kappa\rho\nu} 
 + \;\frac{1}{7}\;F_{\kappa\lambda}F_{\lambda\mu}F_{\kappa\nu}V_{\rho}F_{\rho\nu\mu} 
 \nonumber \\ 
  &&  
 + \;\frac{1}{7}\;F_{\kappa\lambda}F_{\lambda\mu}F_{\nu\rho}V_{\rho}F_{\kappa\nu\mu} 
 + \;\frac{1}{7}\;VF_{\kappa\lambda}F_{\lambda\mu}F_{\kappa\nu\rho}F_{\mu\rho\nu} 
 + \;\frac{1}{7}\;VF_{\kappa\lambda}F_{\lambda\mu}F_{\nu\kappa\rho}F_{\nu\rho\mu} 
 \nonumber \\ 
  &&  
 + \;\frac{1}{7}\;\mbox{i}\;F_{\kappa\lambda}F_{\lambda\mu}V_{\nu\rho}F_{\rho\kappa\nu\mu} 
 + \;\frac{1}{7}\;\mbox{i}\;F_{\kappa\lambda}F_{\mu\lambda\nu}V_{\rho}F_{\mu\kappa\rho\nu} 
 - \;\frac{1}{7}\;\mbox{i}\;F_{\kappa\lambda}F_{\mu\nu\rho}V_{\rho}F_{\nu\mu\lambda\kappa} 
 \nonumber \\ 
  &&  
 - \;\frac{1}{7}\;\mbox{i}\;VF_{\kappa\lambda\mu\nu}F_{\lambda\rho}F_{\rho\kappa\nu\mu} 
 + \;\frac{1}{9}\;\mbox{i}\;F_{\kappa\lambda}F_{\lambda\mu\nu}F_{\nu\kappa\rho}V_{\rho\mu} 
 + \;\frac{1}{9}\;\mbox{i}\;F_{\kappa\lambda}F_{\lambda\mu\nu}V_{\nu\kappa\rho}F_{\rho\mu} 
 \nonumber \\ 
  &&  
 + \;\frac{1}{9}\;\mbox{i}\;F_{\kappa\lambda}F_{\mu\nu}V_{\nu\lambda\rho}F_{\kappa\rho\mu} 
 + \;\frac{1}{10}\;F_{\kappa\lambda\mu}F_{\kappa\nu\rho\mu\lambda}V_{\rho\nu} 
 + \;\frac{1}{10}\;F_{\kappa\lambda\mu}V_{\nu\rho}F_{\rho\nu\kappa\mu\lambda} 
 \nonumber \\ 
  &&  
 + \;\frac{1}{10}\;VF_{\kappa\lambda}F_{\lambda\mu\nu\rho}F_{\mu\kappa}F_{\rho\nu} 
 + \;\frac{1}{14}\;F_{\kappa\lambda\mu\nu}F_{\lambda\kappa\rho\nu\mu}V_{\rho} 
 + \;\frac{1}{14}\;F_{\kappa\lambda\mu\nu}V_{\rho}F_{\rho\lambda\kappa\nu\mu} 
 \nonumber \\ 
  &&  
 - \;\frac{1}{14}\;F_{\kappa\lambda}F_{\lambda\mu\nu}F_{\kappa\rho}V_{\rho}F_{\nu\mu} 
 + \;\frac{1}{14}\;F_{\kappa\lambda}F_{\mu\nu\rho\lambda\kappa}V_{\rho\nu\mu} 
 + \;\frac{1}{14}\;F_{\kappa\lambda}V_{\mu\nu\rho}F_{\rho\nu\mu\lambda\kappa} 
 \nonumber \\ 
  &&  
 - \;\frac{1}{14}\;VF_{\kappa\lambda}F_{\lambda\mu\nu}F_{\kappa\rho}F_{\rho\nu\mu} 
 - \;\frac{1}{15}\;VF_{\kappa\lambda}F_{\lambda\mu\nu}F_{\rho\nu\mu}F_{\rho\kappa} 
 - \;\frac{1}{15}\;VF_{\kappa\lambda}F_{\lambda\mu}F_{\mu\nu\rho}F_{\kappa\rho\nu} 
 \nonumber \\ 
  &&  
 + \;\frac{1}{15}\;\mbox{i}\;VF_{\kappa\lambda}F_{\mu\nu}F_{\nu\lambda}F_{\kappa\rho}F_{\rho\mu} 
 - \;\frac{1}{18}\;F_{\kappa\lambda}F_{\lambda\mu\nu}V_{\rho}F_{\rho\kappa}F_{\nu\mu} 
 - \;\frac{1}{21}\;F_{\kappa\lambda\mu\nu}F_{\lambda\kappa\nu\rho}V_{\rho\mu} 
 \nonumber \\ 
  &&  
 - \;\frac{1}{21}\;F_{\kappa\lambda}F_{\mu\nu\lambda\rho}V_{\rho\nu\mu\kappa} 
 - \;\frac{1}{21}\;F_{\kappa\lambda}V_{\lambda\mu\nu\rho}F_{\rho\nu\mu\kappa} 
 - \;\frac{1}{21}\;\mbox{i}\;F_{\kappa\lambda\mu}F_{\kappa\mu\nu}V_{\rho}F_{\rho\nu\lambda} 
 \nonumber \\ 
  &&  
 - \;\frac{1}{21}\;\mbox{i}\;F_{\kappa\lambda}F_{\lambda\mu\nu\rho}V_{\kappa}F_{\mu\rho\nu} 
 + \;\frac{1}{21}\;\mbox{i}\;F_{\kappa\lambda}F_{\lambda\mu}F_{\mu\nu\kappa\rho}V_{\rho\nu} 
 - \;\frac{1}{21}\;\mbox{i}\;F_{\kappa\lambda}V_{\mu}F_{\mu\nu\lambda\rho}F_{\nu\rho\kappa} 
 \nonumber \\ 
  &&  
 - \;\frac{1}{21}\;\mbox{i}\;VF_{\kappa\lambda}F_{\mu\nu}F_{\lambda\rho}F_{\nu\mu}F_{\rho\kappa} 
 - \;\frac{1}{35}\;F_{\kappa\lambda}F_{\lambda\mu}F_{\kappa\nu\rho}V_{\rho}F_{\nu\mu} 
 - \;\frac{1}{35}\;VF_{\kappa\lambda\mu}F_{\mu\nu}F_{\lambda\rho}F_{\rho\nu\kappa} 
 \nonumber \\ 
  &&  
 - \;\frac{1}{35}\;\mbox{i}\;F_{\kappa\lambda}F_{\lambda\mu}F_{\nu\kappa\rho}V_{\rho\nu\mu} 
 - \;\frac{1}{35}\;\mbox{i}\;F_{\kappa\lambda}F_{\lambda\mu}V_{\kappa\nu\rho}F_{\rho\nu\mu} 
 + \;\frac{1}{35}\;\mbox{i}\;F_{\kappa\lambda}F_{\mu\nu\rho}F_{\mu\rho\lambda}V_{\nu\kappa} 
 \nonumber \\ 
  &&  
 + \;\frac{1}{35}\;\mbox{i}\;F_{\kappa\lambda}V_{\lambda\mu}F_{\nu\kappa\rho}F_{\nu\rho\mu} 
 + \;\frac{1}{35}\;\mbox{i}\;VF_{\kappa\lambda\mu}F_{\mu\nu\rho}F_{\lambda\kappa\rho\nu} 
 + \;\frac{1}{35}\;\mbox{i}\;VF_{\kappa\lambda}F_{\mu\nu}F_{\nu\rho}F_{\rho\lambda}F_{\mu\kappa} 
 \nonumber \\ 
  &&  
 + \;\frac{1}{42}\;VF_{\kappa\lambda\mu\nu\rho}F_{\mu\lambda\kappa\rho\nu} 
 - \;\frac{1}{63}\;VF_{\kappa\lambda}F_{\mu\nu\rho}F_{\lambda\rho\nu}F_{\mu\kappa} 
 + \;\frac{1}{105}\;F_{\kappa\lambda}F_{\lambda\mu}F_{\nu\rho}V_{\rho}F_{\nu\mu\kappa} 
 \nonumber \\ 
  &&  
 - \;\frac{1}{105}\;\mbox{i}\;VF_{\kappa\lambda\mu}F_{\mu\kappa\nu\rho}F_{\lambda\rho\nu} 
 + \;\frac{1}{126}\;F_{\kappa\lambda}F_{\mu\nu}F_{\lambda\kappa}F_{\nu\rho}V_{\rho\mu} 
 - \;\frac{1}{126}\;VF_{\kappa\lambda}F_{\mu\nu}F_{\nu\rho\lambda\kappa}F_{\rho\mu} 
 \nonumber \\ 
  &&  
 - \;\frac{1}{315}\;F_{\kappa\lambda}F_{\lambda\mu}V_{\nu}F_{\mu\kappa\rho}F_{\rho\nu} 
 - \;\frac{2}{7}\;F_{\kappa\lambda}F_{\mu\lambda\kappa}F_{\nu\rho}V_{\rho}F_{\nu\mu} 
 + \;\frac{2}{7}\;VF_{\kappa\lambda\mu}F_{\mu\nu}F_{\kappa\lambda\rho}F_{\rho\nu} 
 \nonumber \\ 
  &&  
 + \;\frac{2}{7}\;VF_{\kappa\lambda}F_{\mu\lambda\nu}F_{\kappa\rho}F_{\mu\rho\nu} 
 - \;\frac{2}{7}\;\mbox{i}\;F_{\kappa\lambda}F_{\mu\nu\lambda\rho}F_{\nu\kappa}V_{\rho\mu} 
 - \;\frac{2}{9}\;\mbox{i}\;F_{\kappa\lambda\mu}F_{\kappa\nu\rho}V_{\mu}F_{\lambda\rho\nu} 
 \nonumber \\ 
  &&  
 - \;\frac{2}{15}\;VF_{\kappa\lambda}F_{\mu\nu}F_{\nu\lambda\rho}F_{\rho\mu\kappa} 
 - \;\frac{2}{15}\;\mbox{i}\;F_{\kappa\lambda}V_{\mu}F_{\nu\rho\lambda\kappa}F_{\rho\nu\mu} 
 + \;\frac{2}{21}\;F_{\kappa\lambda}F_{\lambda\mu}F_{\mu\kappa\nu}V_{\rho}F_{\rho\nu} 
 \nonumber \\ 
  &&  
 + \;\frac{2}{21}\;VF_{\kappa\lambda\mu}F_{\nu\rho}F_{\kappa\mu\lambda}F_{\rho\nu} 
 + \;\frac{2}{21}\;VF_{\kappa\lambda}F_{\mu\nu\rho}F_{\lambda\kappa}F_{\mu\rho\nu} 
 + \;\frac{2}{21}\;\mbox{i}\;F_{\kappa\lambda}F_{\lambda\mu\nu\rho}V_{\rho}F_{\mu\nu\kappa} 
 \nonumber \\ 
  &&  
 - \;\frac{2}{21}\;\mbox{i}\;VF_{\kappa\lambda\mu\nu}F_{\lambda\kappa\nu\rho}F_{\rho\mu} 
 - \;\frac{2}{21}\;\mbox{i}\;VF_{\kappa\lambda}F_{\mu\nu\lambda\rho}F_{\nu\mu\rho\kappa} 
 - \;\frac{2}{35}\;F_{\kappa\lambda}F_{\lambda\mu}F_{\kappa\nu}V_{\rho}F_{\mu\rho\nu} 
 \nonumber \\ 
  &&  
 - \;\frac{2}{35}\;\mbox{i}\;F_{\kappa\lambda}V_{\lambda}F_{\kappa\mu\nu\rho}F_{\mu\rho\nu} 
 - \;\frac{2}{63}\;VF_{\kappa\lambda}F_{\lambda\mu}F_{\nu\rho}F_{\mu\kappa\rho\nu} 
 - \;\frac{2}{63}\;VF_{\kappa\lambda}F_{\mu\nu}F_{\lambda\rho\nu\mu}F_{\rho\kappa} 
 \nonumber \\ 
  &&  
 + \;\frac{2}{63}\;VF_{\kappa\lambda}F_{\mu\nu}F_{\nu\lambda\rho}F_{\kappa\rho\mu} 
 - \;\frac{2}{105}\;\mbox{i}\;VF_{\kappa\lambda}F_{\lambda\mu\nu\rho}F_{\mu\kappa\rho\nu} 
 - \;\frac{2}{315}\;VF_{\kappa\lambda}F_{\mu\nu}F_{\rho\nu\lambda}F_{\kappa\rho\mu} 
 \nonumber \\ 
  &&  
 + \;\frac{3}{7}\;F_{\kappa\lambda}F_{\lambda\mu}V_{\nu}F_{\kappa\rho}F_{\rho\nu\mu} 
 - \;\frac{3}{14}\;\mbox{i}\;F_{\kappa\lambda}F_{\mu\lambda\kappa}V_{\nu\rho}F_{\rho\nu\mu} 
 - \;\frac{3}{14}\;\mbox{i}\;F_{\kappa\lambda}F_{\mu\nu\lambda\kappa}V_{\rho}F_{\nu\rho\mu} 
 \nonumber \\ 
  &&  
 - \;\frac{3}{14}\;\mbox{i}\;F_{\kappa\lambda}V_{\lambda\mu}F_{\kappa\nu\rho}F_{\mu\rho\nu} 
 + \;\frac{3}{28}\;F_{\kappa\lambda}F_{\lambda\kappa}F_{\mu\nu\rho}F_{\rho\nu}V_{\mu} 
 + \;\frac{3}{28}\;F_{\kappa\lambda}F_{\lambda\kappa}V_{\mu}F_{\nu\rho}F_{\mu\rho\nu} 
 \nonumber \\ 
  &&  
 + \;\frac{3}{28}\;VF_{\kappa\lambda\mu}F_{\mu\lambda}F_{\kappa\nu\rho}F_{\rho\nu} 
 + \;\frac{3}{28}\;VF_{\kappa\lambda}F_{\mu\lambda\kappa}F_{\nu\rho}F_{\mu\rho\nu} 
 - \;\frac{3}{35}\;F_{\kappa\lambda\mu}F_{\kappa\nu\mu\rho}V_{\rho\nu\lambda} 
 \nonumber \\ 
  &&  
 - \;\frac{3}{35}\;F_{\kappa\lambda\mu}F_{\nu\mu\rho}V_{\rho\nu\lambda\kappa} 
 - \;\frac{3}{35}\;F_{\kappa\lambda\mu}V_{\mu\nu\rho}F_{\rho\kappa\nu\lambda} 
 + \;\frac{3}{35}\;F_{\kappa\lambda}F_{\lambda\mu}F_{\kappa\nu}V_{\mu\rho}F_{\rho\nu} 
 \nonumber \\ 
  &&  
 + \;\frac{3}{35}\;F_{\kappa\lambda}F_{\lambda\mu}F_{\nu\rho}V_{\rho\kappa}F_{\nu\mu} 
 - \;\frac{3}{35}\;F_{\kappa\lambda}F_{\lambda\mu}V_{\nu}F_{\nu\rho}F_{\rho\mu\kappa} 
 - \;\frac{3}{35}\;VF_{\kappa\lambda\mu}F_{\kappa\mu\nu}F_{\nu\rho}F_{\rho\lambda} 
 \nonumber \\ 
  &&  
 - \;\frac{3}{35}\;VF_{\kappa\lambda}F_{\lambda\mu}F_{\nu\mu\rho}F_{\nu\rho\kappa} 
 - \;\frac{3}{35}\;VF_{\kappa\lambda}F_{\mu\lambda\nu}F_{\mu\nu\rho}F_{\rho\kappa} 
 + \;\frac{3}{35}\;\mbox{i}\;F_{\kappa\lambda\mu}F_{\kappa\mu\nu}F_{\nu\lambda\rho}V_{\rho} 
 \nonumber \\ 
  &&  
 + \;\frac{3}{35}\;\mbox{i}\;F_{\kappa\lambda}F_{\mu\lambda\nu}F_{\nu\mu\kappa\rho}V_{\rho} 
 - \;\frac{3}{35}\;\mbox{i}\;F_{\kappa\lambda}F_{\mu\lambda\nu}V_{\mu\rho}F_{\rho\nu\kappa} 
 + \;\frac{3}{35}\;\mbox{i}\;VF_{\kappa\lambda}F_{\lambda\mu}F_{\kappa\nu}F_{\mu\rho}F_{\rho\nu} 
 \nonumber \\ 
  &&  
 + \;\frac{3}{70}\;F_{\kappa\lambda}F_{\lambda\mu}F_{\kappa\nu\rho}F_{\rho\nu}V_{\mu} 
 + \;\frac{3}{70}\;F_{\kappa\lambda}F_{\lambda\mu}V_{\kappa}F_{\nu\rho}F_{\mu\rho\nu} 
 + \;\frac{3}{70}\;VF_{\kappa\lambda\mu}F_{\kappa\mu\lambda}F_{\nu\rho}F_{\rho\nu} 
 \nonumber \\ 
  &&  
 + \;\frac{3}{70}\;VF_{\kappa\lambda}F_{\lambda\kappa}F_{\mu\nu\rho}F_{\mu\rho\nu} 
 - \;\frac{4}{7}\;\mbox{i}\;F_{\kappa\lambda}F_{\lambda\mu\nu}F_{\rho\nu\kappa}V_{\rho\mu} 
 - \;\frac{4}{15}\;F_{\kappa\lambda}F_{\lambda\mu}F_{\nu\rho}V_{\mu}F_{\kappa\rho\nu} 
 \nonumber \\ 
  &&  
 + \;\frac{4}{15}\;F_{\kappa\lambda}F_{\mu\lambda\nu}F_{\kappa\rho}V_{\mu}F_{\rho\nu} 
 - \;\frac{4}{21}\;VF_{\kappa\lambda\mu}F_{\nu\kappa\rho}F_{\nu\mu}F_{\rho\lambda} 
 + \;\frac{4}{21}\;VF_{\kappa\lambda\mu}F_{\nu\rho}F_{\kappa\rho\mu}F_{\nu\lambda} 
 \nonumber \\ 
  &&  
 + \;\frac{4}{21}\;VF_{\kappa\lambda}F_{\mu\nu\rho}F_{\rho\lambda}F_{\mu\nu\kappa} 
 - \;\frac{4}{21}\;\mbox{i}\;F_{\kappa\lambda}F_{\mu\nu\rho}V_{\lambda}F_{\mu\kappa\rho\nu} 
 - \;\frac{4}{21}\;\mbox{i}\;VF_{\kappa\lambda\mu\nu}F_{\nu\rho}F_{\lambda\kappa\rho\mu} 
 \nonumber \\ 
  &&  
 - \;\frac{4}{35}\;\mbox{i}\;F_{\kappa\lambda}F_{\lambda\mu}V_{\nu\rho}F_{\rho\nu\mu\kappa} 
 - \;\frac{4}{35}\;\mbox{i}\;F_{\kappa\lambda}F_{\mu\nu\rho}V_{\rho\mu}F_{\nu\lambda\kappa} 
 - \;\frac{4}{35}\;\mbox{i}\;VF_{\kappa\lambda\mu}F_{\kappa\nu\mu\rho}F_{\nu\rho\lambda} 
 \nonumber \\ 
  &&  
 + \;\frac{4}{35}\;\mbox{i}\;VF_{\kappa\lambda}F_{\mu\nu}F_{\nu\rho}F_{\mu\lambda}F_{\rho\kappa} 
 + \;\frac{4}{105}\;F_{\kappa\lambda}F_{\lambda\mu}F_{\kappa\nu}F_{\mu\rho}V_{\rho\nu} 
 + \;\frac{4}{105}\;F_{\kappa\lambda}F_{\lambda\mu}V_{\nu\rho}F_{\rho\kappa}F_{\nu\mu} 
 \nonumber \\ 
  &&  
 - \;\frac{4}{315}\;VF_{\kappa\lambda}F_{\lambda\mu\nu\rho}F_{\rho\nu}F_{\mu\kappa} 
 + \;\frac{4}{315}\;\mbox{i}\;F_{\kappa\lambda}F_{\mu\nu}V_{\nu\lambda\rho}F_{\rho\mu\kappa} 
 + \;\frac{4}{315}\;\mbox{i}\;VF_{\kappa\lambda}F_{\lambda\mu}F_{\mu\nu}F_{\nu\rho}F_{\rho\kappa} 
 \nonumber \\ 
  &&  
 - \;\frac{5}{21}\;\mbox{i}\;F_{\kappa\lambda}F_{\mu\lambda\nu}V_{\rho}F_{\rho\mu\nu\kappa} 
 + \;\frac{5}{21}\;\mbox{i}\;VF_{\kappa\lambda}F_{\lambda\mu}F_{\nu\rho}F_{\rho\kappa}F_{\nu\mu} 
 + \;\frac{5}{28}\;F_{\kappa\lambda}F_{\lambda\kappa}F_{\mu\nu\rho}V_{\mu}F_{\rho\nu} 
 \nonumber \\ 
  &&  
 + \;\frac{5}{28}\;F_{\kappa\lambda}F_{\lambda\kappa}F_{\mu\nu}V_{\rho}F_{\rho\nu\mu} 
 - \;\frac{5}{42}\;F_{\kappa\lambda}F_{\lambda\mu}F_{\nu\rho}V_{\mu\kappa}F_{\rho\nu} 
 - \;\frac{5}{42}\;F_{\kappa\lambda}F_{\mu\nu}F_{\lambda\kappa}V_{\nu\rho}F_{\rho\mu} 
 \nonumber \\ 
  &&  
 + \;\frac{5}{42}\;VF_{\kappa\lambda\mu}F_{\nu\rho}F_{\rho\nu}F_{\kappa\mu\lambda} 
 + \;\frac{5}{63}\;F_{\kappa\lambda}F_{\lambda\kappa}F_{\mu\nu}F_{\nu\rho}V_{\rho\mu} 
 + \;\frac{5}{63}\;F_{\kappa\lambda}F_{\lambda\kappa}F_{\mu\nu}V_{\nu\rho}F_{\rho\mu} 
 \nonumber \\ 
  &&  
 - \;\frac{5}{63}\;VF_{\kappa\lambda\mu}F_{\nu\rho}F_{\rho\mu\lambda}F_{\nu\kappa} 
 + \;\frac{5}{63}\;VF_{\kappa\lambda}F_{\lambda\mu}F_{\mu\kappa\nu\rho}F_{\rho\nu} 
 - \;\frac{5}{63}\;\mbox{i}\;F_{\kappa\lambda}F_{\lambda\mu\nu\rho}V_{\mu\kappa}F_{\rho\nu} 
 \nonumber \\ 
  &&  
 - \;\frac{5}{63}\;\mbox{i}\;VF_{\kappa\lambda}F_{\lambda\mu}F_{\nu\rho}F_{\rho\nu}F_{\mu\kappa} 
 - \;\frac{5}{63}\;\mbox{i}\;VF_{\kappa\lambda}F_{\mu\nu}F_{\nu\mu}F_{\lambda\rho}F_{\rho\kappa} 
 + \;\frac{5}{126}\;VF_{\kappa\lambda}F_{\mu\nu}F_{\lambda\rho}F_{\rho\kappa\nu\mu} 
 \nonumber \\ 
  &&  
 - \;\frac{5}{126}\;VF_{\kappa\lambda}F_{\mu\nu}F_{\nu\rho}F_{\rho\mu\lambda\kappa} 
 + \;\frac{5}{126}\;\mbox{i}\;F_{\kappa\lambda}F_{\lambda\mu\nu\rho}F_{\kappa\rho\nu}V_{\mu} 
 - \;\frac{6}{35}\;VF_{\kappa\lambda\mu}F_{\kappa\nu}F_{\rho\mu\lambda}F_{\rho\nu} 
 \nonumber \\ 
  &&  
 + \;\frac{6}{35}\;\mbox{i}\;F_{\kappa\lambda\mu}F_{\kappa\mu\nu}V_{\rho}F_{\lambda\rho\nu} 
 + \;\frac{6}{35}\;\mbox{i}\;F_{\kappa\lambda}V_{\mu}F_{\nu\rho\mu\lambda}F_{\rho\nu\kappa} 
 - \;\frac{6}{35}\;\mbox{i}\;VF_{\kappa\lambda\mu\nu}F_{\lambda\rho\nu\mu}F_{\rho\kappa} 
 \nonumber \\ 
  &&  
 - \;\frac{7}{18}\;\mbox{i}\;F_{\kappa\lambda}F_{\mu\nu}F_{\lambda\rho\nu\mu}V_{\rho\kappa} 
 - \;\frac{8}{21}\;\mbox{i}\;F_{\kappa\lambda}F_{\lambda\mu\nu\rho}F_{\rho\kappa}V_{\nu\mu} 
 - \;\frac{8}{21}\;\mbox{i}\;F_{\kappa\lambda}F_{\mu\nu\lambda\rho}V_{\nu}F_{\mu\rho\kappa} 
 \nonumber \\ 
  &&  
 - \;\frac{8}{35}\;F_{\kappa\lambda}F_{\lambda\mu\nu}F_{\nu\kappa}V_{\rho}F_{\rho\mu} 
 - \;\frac{8}{35}\;\mbox{i}\;F_{\kappa\lambda}V_{\mu\nu}F_{\nu\lambda\rho}F_{\kappa\rho\mu} 
 - \;\frac{8}{35}\;\mbox{i}\;VF_{\kappa\lambda\mu\nu}F_{\lambda\nu\rho}F_{\kappa\rho\mu} 
 \nonumber \\ 
  &&  
 - \;\frac{8}{35}\;\mbox{i}\;VF_{\kappa\lambda\mu}F_{\nu\mu\rho}F_{\nu\kappa\rho\lambda} 
 - \;\frac{9}{35}\;\mbox{i}\;F_{\kappa\lambda}F_{\lambda\mu}F_{\nu\rho\mu\kappa}V_{\rho\nu} 
 - \;\frac{9}{35}\;\mbox{i}\;F_{\kappa\lambda}F_{\mu\nu\lambda\rho}F_{\nu\mu\kappa}V_{\rho} 
 \nonumber \\ 
  &&  
 - \;\frac{9}{35}\;\mbox{i}\;F_{\kappa\lambda}V_{\mu}F_{\nu\lambda\rho}F_{\nu\kappa\rho\mu} 
 + \;\frac{9}{35}\;\mbox{i}\;VF_{\kappa\lambda}F_{\mu\nu}F_{\lambda\rho}F_{\rho\nu}F_{\mu\kappa} 
 - \;\frac{9}{70}\;F_{\kappa\lambda}F_{\mu\nu}F_{\lambda\rho}V_{\rho}F_{\kappa\nu\mu} 
 \nonumber \\ 
  &&  
 - \;\frac{10}{63}\;VF_{\kappa\lambda\mu}F_{\kappa\nu}F_{\rho\nu\mu}F_{\rho\lambda} 
 - \;\frac{10}{63}\;VF_{\kappa\lambda\mu}F_{\mu\nu}F_{\nu\lambda\rho}F_{\rho\kappa} 
 - \;\frac{10}{63}\;VF_{\kappa\lambda}F_{\lambda\mu\nu}F_{\nu\rho}F_{\rho\mu\kappa} 
 \nonumber \\ 
  &&  
 - \;\frac{10}{63}\;VF_{\kappa\lambda}F_{\mu\lambda\nu}F_{\mu\rho}F_{\kappa\rho\nu} 
 - \;\frac{10}{63}\;\mbox{i}\;VF_{\kappa\lambda}F_{\lambda\mu}F_{\kappa\nu}F_{\nu\rho}F_{\rho\mu} 
 - \;\frac{11}{35}\;VF_{\kappa\lambda}F_{\lambda\mu\nu}F_{\rho\nu\kappa}F_{\rho\mu} 
 \nonumber \\ 
  &&  
 - \;\frac{11}{35}\;\mbox{i}\;F_{\kappa\lambda}V_{\mu}F_{\nu\lambda\rho}F_{\rho\nu\mu\kappa} 
 - \;\frac{11}{42}\;F_{\kappa\lambda}F_{\lambda\kappa}V_{\mu\nu}F_{\nu\rho}F_{\rho\mu} 
 - \;\frac{11}{42}\;VF_{\kappa\lambda}F_{\mu\nu\lambda\kappa}F_{\nu\rho}F_{\rho\mu} 
 \nonumber \\ 
  &&  
 - \;\frac{11}{45}\;F_{\kappa\lambda}F_{\lambda\mu}F_{\nu\mu\rho}F_{\nu\kappa}V_{\rho} 
 - \;\frac{11}{45}\;F_{\kappa\lambda}F_{\lambda\mu}V_{\nu}F_{\mu\rho}F_{\kappa\rho\nu} 
 - \;\frac{11}{70}\;F_{\kappa\lambda}F_{\mu\nu}V_{\rho}F_{\rho\nu}F_{\mu\lambda\kappa} 
 \nonumber \\ 
  &&  
 + \;\frac{11}{105}\;VF_{\kappa\lambda}F_{\mu\nu\rho}F_{\mu\rho\lambda}F_{\nu\kappa} 
 + \;\frac{11}{105}\;VF_{\kappa\lambda}F_{\mu\nu\rho}F_{\mu\rho\nu}F_{\lambda\kappa} 
 - \;\frac{11}{105}\;\mbox{i}\;VF_{\kappa\lambda\mu\nu}F_{\lambda\kappa\rho}F_{\rho\nu\mu} 
 \nonumber \\ 
  &&  
 - \;\frac{11}{105}\;\mbox{i}\;VF_{\kappa\lambda\mu}F_{\nu\kappa\rho}F_{\rho\nu\mu\lambda} 
 - \;\frac{11}{210}\;F_{\kappa\lambda}F_{\mu\nu}F_{\rho\lambda\kappa}F_{\rho\nu}V_{\mu} 
 + \;\frac{12}{35}\;F_{\kappa\lambda}F_{\lambda\mu}F_{\kappa\nu\rho}F_{\rho\mu}V_{\nu} 
 \nonumber \\ 
  &&  
 + \;\frac{13}{35}\;F_{\kappa\lambda}F_{\lambda\mu}F_{\nu\kappa\rho}F_{\nu\mu}V_{\rho} 
 - \;\frac{13}{35}\;F_{\kappa\lambda}F_{\lambda\mu}F_{\nu\mu\kappa}V_{\rho}F_{\rho\nu} 
 - \;\frac{13}{35}\;F_{\kappa\lambda}F_{\lambda\mu}F_{\nu\rho}F_{\rho\kappa}V_{\nu\mu} 
 \nonumber \\ 
  &&  
 + \;\frac{13}{35}\;F_{\kappa\lambda}F_{\lambda\mu}V_{\nu}F_{\kappa\rho}F_{\mu\rho\nu} 
 + \;\frac{13}{63}\;F_{\kappa\lambda}F_{\lambda\mu}F_{\mu\nu}V_{\rho}F_{\rho\nu\kappa} 
 - \;\frac{13}{70}\;VF_{\kappa\lambda\mu}F_{\nu\mu\lambda}F_{\kappa\rho}F_{\rho\nu} 
 \nonumber \\ 
  &&  
 - \;\frac{13}{105}\;\mbox{i}\;VF_{\kappa\lambda\mu\nu}F_{\rho\nu\mu}F_{\lambda\rho\kappa} 
 - \;\frac{13}{126}\;\mbox{i}\;F_{\kappa\lambda}F_{\lambda\mu\nu\rho}V_{\mu}F_{\kappa\rho\nu} 
 - \;\frac{13}{210}\;\mbox{i}\;F_{\kappa\lambda}F_{\mu\nu\rho}F_{\rho\lambda\kappa}V_{\nu\mu} 
 \nonumber \\ 
  &&  
 - \;\frac{13}{210}\;\mbox{i}\;F_{\kappa\lambda}F_{\mu\nu\rho}F_{\rho\mu\lambda\kappa}V_{\nu} 
 + \;\frac{13}{630}\;VF_{\kappa\lambda}F_{\mu\nu\rho}F_{\mu\lambda}F_{\kappa\rho\nu} 
 - \;\frac{16}{63}\;F_{\kappa\lambda}F_{\mu\lambda\nu}F_{\mu\rho}V_{\kappa}F_{\rho\nu} 
 \nonumber \\ 
  &&  
 + \;\frac{16}{63}\;F_{\kappa\lambda}F_{\mu\nu}V_{\rho}F_{\nu\lambda}F_{\rho\mu\kappa} 
 - \;\frac{16}{105}\;\mbox{i}\;F_{\kappa\lambda}F_{\lambda\mu\nu\rho}F_{\mu\rho\nu}V_{\kappa} 
 - \;\frac{16}{105}\;\mbox{i}\;F_{\kappa\lambda}F_{\mu\lambda\kappa}F_{\nu\mu\rho}V_{\rho\nu} 
 \nonumber \\ 
  &&  
 - \;\frac{16}{105}\;\mbox{i}\;F_{\kappa\lambda}F_{\mu\nu\lambda\kappa}F_{\nu\mu\rho}V_{\rho} 
 - \;\frac{16}{105}\;\mbox{i}\;F_{\kappa\lambda}V_{\lambda}F_{\mu\nu\rho}F_{\mu\kappa\rho\nu} 
 - \;\frac{16}{105}\;\mbox{i}\;F_{\kappa\lambda}V_{\mu\nu}F_{\nu\mu\rho}F_{\rho\lambda\kappa} 
 \nonumber \\ 
  &&  
 - \;\frac{16}{105}\;\mbox{i}\;F_{\kappa\lambda}V_{\mu}F_{\nu\mu\rho}F_{\rho\nu\lambda\kappa} 
 + \;\frac{16}{315}\;F_{\kappa\lambda}F_{\mu\nu}F_{\lambda\rho}V_{\kappa}F_{\rho\nu\mu} 
 + \;\frac{16}{315}\;VF_{\kappa\lambda}F_{\mu\nu}F_{\rho\nu\lambda}F_{\rho\mu\kappa} 
 \nonumber \\ 
  &&  
 - \;\frac{17}{35}\;F_{\kappa\lambda}F_{\lambda\mu\nu}F_{\nu\rho}V_{\kappa}F_{\rho\mu} 
 + \;\frac{17}{63}\;VF_{\kappa\lambda}F_{\mu\lambda\nu}F_{\mu\kappa\rho}F_{\rho\nu} 
 - \;\frac{17}{90}\;F_{\kappa\lambda}F_{\lambda\mu}V_{\mu}F_{\kappa\nu\rho}F_{\rho\nu} 
 \nonumber \\ 
  &&  
 - \;\frac{17}{105}\;\mbox{i}\;F_{\kappa\lambda}V_{\mu\nu}F_{\rho\lambda\kappa}F_{\nu\rho\mu} 
 - \;\frac{17}{105}\;\mbox{i}\;VF_{\kappa\lambda\mu}F_{\nu\rho\mu\lambda}F_{\rho\nu\kappa} 
 + \;\frac{17}{315}\;F_{\kappa\lambda}F_{\mu\lambda\nu}V_{\rho}F_{\mu\kappa}F_{\rho\nu} 
 \nonumber \\ 
  &&  
 + \;\frac{17}{630}\;VF_{\kappa\lambda\mu}F_{\nu\rho}F_{\rho\kappa}F_{\nu\mu\lambda} 
 - \;\frac{19}{63}\;VF_{\kappa\lambda\mu}F_{\mu\nu\rho}F_{\rho\lambda}F_{\nu\kappa} 
 - \;\frac{19}{126}\;\mbox{i}\;F_{\kappa\lambda}F_{\lambda\mu\nu}V_{\rho}F_{\rho\kappa\nu\mu} 
 \nonumber \\ 
  &&  
 - \;\frac{19}{210}\;F_{\kappa\lambda}F_{\lambda\mu}F_{\nu\rho}F_{\kappa\rho\nu}V_{\mu} 
 + \;\frac{19}{315}\;F_{\kappa\lambda}F_{\lambda\mu}F_{\kappa\nu\rho}V_{\mu}F_{\rho\nu} 
 + \;\frac{19}{315}\;F_{\kappa\lambda}F_{\lambda\mu}V_{\kappa}F_{\mu\nu\rho}F_{\rho\nu} 
 \nonumber \\ 
  &&  
 - \;\frac{20}{63}\;F_{\kappa\lambda}F_{\lambda\mu}V_{\nu}F_{\mu\rho}F_{\rho\nu\kappa} 
 + \;\frac{20}{63}\;F_{\kappa\lambda}F_{\lambda\mu}V_{\nu}F_{\nu\kappa\rho}F_{\rho\mu} 
 - \;\frac{20}{63}\;VF_{\kappa\lambda\mu}F_{\kappa\nu}F_{\mu\rho}F_{\rho\nu\lambda} 
 \nonumber \\ 
  &&  
 - \;\frac{20}{63}\;VF_{\kappa\lambda\mu}F_{\kappa\nu}F_{\nu\mu\rho}F_{\rho\lambda} 
 - \;\frac{20}{63}\;VF_{\kappa\lambda}F_{\mu\lambda\nu}F_{\mu\rho}F_{\rho\nu\kappa} 
 - \;\frac{20}{63}\;\mbox{i}\;F_{\kappa\lambda\mu}F_{\mu\nu\rho}F_{\lambda\rho\nu}V_{\kappa} 
 \nonumber \\ 
  &&  
 - \;\frac{20}{63}\;\mbox{i}\;F_{\kappa\lambda}V_{\mu}F_{\mu\lambda\nu\rho}F_{\kappa\rho\nu} 
 + \;\frac{22}{105}\;VF_{\kappa\lambda}F_{\lambda\mu\nu}F_{\nu\kappa\rho}F_{\rho\mu} 
 - \;\frac{23}{35}\;F_{\kappa\lambda}F_{\lambda\mu}F_{\nu\rho}F_{\rho\mu\kappa}V_{\nu} 
 \nonumber \\ 
  &&  
 - \;\frac{23}{45}\;\mbox{i}\;F_{\kappa\lambda}F_{\mu\lambda\nu}F_{\rho\mu\kappa}V_{\rho\nu} 
 + \;\frac{23}{63}\;F_{\kappa\lambda}F_{\lambda\mu}V_{\nu}F_{\nu\rho}F_{\kappa\rho\mu} 
 + \;\frac{23}{63}\;\mbox{i}\;F_{\kappa\lambda}F_{\lambda\mu\nu}F_{\nu\rho}V_{\rho\mu\kappa} 
 \nonumber \\ 
  &&  
 + \;\frac{23}{63}\;\mbox{i}\;F_{\kappa\lambda}F_{\lambda\mu\nu}V_{\nu\rho}F_{\kappa\rho\mu} 
 + \;\frac{23}{63}\;\mbox{i}\;F_{\kappa\lambda}F_{\lambda\mu\nu}V_{\nu\rho}F_{\rho\mu\kappa} 
 + \;\frac{23}{63}\;\mbox{i}\;F_{\kappa\lambda}F_{\mu\lambda\nu}F_{\mu\rho}V_{\rho\nu\kappa} 
 \nonumber \\ 
  &&  
 + \;\frac{23}{63}\;\mbox{i}\;F_{\kappa\lambda}F_{\mu\lambda\nu}V_{\mu\rho}F_{\kappa\rho\nu} 
 - \;\frac{23}{90}\;\mbox{i}\;F_{\kappa\lambda}F_{\mu\nu}V_{\lambda\rho}F_{\rho\kappa\nu\mu} 
 + \;\frac{23}{105}\;F_{\kappa\lambda}F_{\lambda\mu}F_{\mu\nu}F_{\nu\kappa\rho}V_{\rho} 
 \nonumber \\ 
  &&  
 - \;\frac{23}{105}\;F_{\kappa\lambda}F_{\mu\nu}F_{\lambda\rho}V_{\nu}F_{\rho\mu\kappa} 
 - \;\frac{23}{105}\;VF_{\kappa\lambda\mu}F_{\mu\nu}F_{\nu\rho}F_{\kappa\rho\lambda} 
 + \;\frac{23}{105}\;VF_{\kappa\lambda\mu}F_{\nu\rho}F_{\rho\mu}F_{\kappa\nu\lambda} 
 \nonumber \\ 
  &&  
 - \;\frac{23}{105}\;VF_{\kappa\lambda}F_{\lambda\mu\nu}F_{\nu\rho}F_{\kappa\rho\mu} 
 - \;\frac{23}{105}\;\mbox{i}\;F_{\kappa\lambda\mu}F_{\kappa\mu\nu}F_{\rho\nu\lambda}V_{\rho} 
 + \;\frac{23}{105}\;\mbox{i}\;F_{\kappa\lambda}F_{\lambda\mu}F_{\nu\mu\rho}V_{\rho\nu\kappa} 
 \nonumber \\ 
  &&  
 + \;\frac{23}{105}\;\mbox{i}\;F_{\kappa\lambda}F_{\lambda\mu}V_{\mu\nu\rho}F_{\rho\nu\kappa} 
 + \;\frac{23}{105}\;\mbox{i}\;F_{\kappa\lambda}F_{\mu\lambda\nu}F_{\mu\nu\rho}V_{\rho\kappa} 
 - \;\frac{23}{105}\;\mbox{i}\;F_{\kappa\lambda}F_{\mu\lambda\nu}F_{\mu\rho\nu\kappa}V_{\rho} 
 \nonumber \\ 
  &&  
 + \;\frac{23}{105}\;\mbox{i}\;F_{\kappa\lambda}V_{\lambda\mu}F_{\nu\mu\rho}F_{\nu\rho\kappa} 
 - \;\frac{23}{126}\;VF_{\kappa\lambda\mu\nu}F_{\lambda\rho}F_{\rho\kappa}F_{\nu\mu} 
 - \;\frac{23}{315}\;F_{\kappa\lambda}F_{\lambda\mu\nu}F_{\kappa\rho}V_{\nu}F_{\rho\mu} 
 \nonumber \\ 
  &&  
 + \;\frac{23}{315}\;F_{\kappa\lambda}F_{\lambda\mu}F_{\kappa\nu}F_{\nu\mu\rho}V_{\rho} 
 - \;\frac{23}{315}\;F_{\kappa\lambda}F_{\lambda\mu}F_{\kappa\nu}F_{\nu\rho}V_{\rho\mu} 
 - \;\frac{23}{315}\;F_{\kappa\lambda}F_{\lambda\mu}F_{\nu\rho}F_{\rho\mu}V_{\nu\kappa} 
 \nonumber \\ 
  &&  
 + \;\frac{23}{315}\;\mbox{i}\;F_{\kappa\lambda}F_{\lambda\mu\nu}F_{\kappa\rho\nu\mu}V_{\rho} 
 - \;\frac{25}{63}\;F_{\kappa\lambda}F_{\lambda\mu}F_{\nu\rho}F_{\kappa\rho\mu}V_{\nu} 
 - \;\frac{26}{63}\;VF_{\kappa\lambda}F_{\mu\lambda\nu}F_{\rho\mu\kappa}F_{\rho\nu} 
 \nonumber \\ 
  &&  
 + \;\frac{26}{315}\;F_{\kappa\lambda}F_{\mu\nu}F_{\lambda\rho}F_{\kappa\nu\mu}V_{\rho} 
 - \;\frac{29}{70}\;\mbox{i}\;F_{\kappa\lambda}F_{\mu\nu\rho}F_{\lambda\rho\nu}V_{\mu\kappa} 
 + \;\frac{31}{252}\;VF_{\kappa\lambda\mu}F_{\mu\lambda}F_{\nu\rho}F_{\kappa\rho\nu} 
 \nonumber \\ 
  &&  
 + \;\frac{31}{315}\;F_{\kappa\lambda}F_{\lambda\mu\nu}F_{\nu\rho}F_{\mu\kappa}V_{\rho} 
 + \;\frac{31}{315}\;F_{\kappa\lambda}F_{\lambda\mu}V_{\nu}F_{\rho\nu\kappa}F_{\rho\mu} 
 - \;\frac{31}{315}\;F_{\kappa\lambda}F_{\mu\nu}V_{\rho}F_{\nu\lambda}F_{\kappa\rho\mu} 
 \nonumber \\ 
  &&  
 + \;\frac{31}{315}\;\mbox{i}\;F_{\kappa\lambda}F_{\mu\lambda\nu}F_{\mu\kappa\rho}V_{\rho\nu} 
 + \;\frac{31}{315}\;\mbox{i}\;F_{\kappa\lambda}F_{\mu\lambda\nu}V_{\mu\kappa\rho}F_{\rho\nu} 
 - \;\frac{32}{105}\;F_{\kappa\lambda}F_{\lambda\mu\nu}F_{\nu\kappa}F_{\mu\rho}V_{\rho} 
 \nonumber \\ 
  &&  
 + \;\frac{32}{105}\;F_{\kappa\lambda}F_{\lambda\mu}F_{\kappa\nu}F_{\rho\nu\mu}V_{\rho} 
 - \;\frac{32}{105}\;F_{\kappa\lambda}F_{\mu\lambda\kappa}F_{\mu\nu}V_{\rho}F_{\rho\nu} 
 - \;\frac{32}{315}\;\mbox{i}\;VF_{\kappa\lambda}F_{\mu\nu}F_{\lambda\rho}F_{\nu\kappa}F_{\rho\mu} 
 \nonumber \\ 
  &&  
 + \;\frac{34}{105}\;F_{\kappa\lambda}F_{\lambda\mu\nu}V_{\rho}F_{\nu\kappa}F_{\rho\mu} 
 - \;\frac{34}{315}\;\mbox{i}\;VF_{\kappa\lambda}F_{\lambda\mu}F_{\nu\rho}F_{\rho\mu}F_{\nu\kappa} 
 - \;\frac{34}{315}\;\mbox{i}\;VF_{\kappa\lambda}F_{\mu\nu}F_{\nu\lambda}F_{\mu\rho}F_{\rho\kappa} 
 \nonumber \\ 
  &&  
 - \;\frac{37}{105}\;\mbox{i}\;VF_{\kappa\lambda}F_{\lambda\mu}F_{\mu\nu}F_{\kappa\rho}F_{\rho\nu} 
 + \;\frac{37}{210}\;F_{\kappa\lambda\mu\nu}F_{\lambda\rho\nu\mu}V_{\rho\kappa} 
 + \;\frac{37}{252}\;VF_{\kappa\lambda\mu}F_{\nu\rho}F_{\mu\lambda}F_{\kappa\rho\nu} 
 \nonumber \\ 
  &&  
 - \;\frac{37}{630}\;F_{\kappa\lambda}F_{\lambda\mu}F_{\nu\rho}V_{\kappa}F_{\mu\rho\nu} 
 - \;\frac{38}{315}\;F_{\kappa\lambda}F_{\lambda\mu}F_{\kappa\nu}V_{\nu\rho}F_{\rho\mu} 
 - \;\frac{41}{105}\;F_{\kappa\lambda}F_{\lambda\mu}F_{\mu\nu\rho}F_{\rho\kappa}V_{\nu} 
 \nonumber \\ 
  &&  
 - \;\frac{41}{105}\;\mbox{i}\;F_{\kappa\lambda\mu}F_{\kappa\nu\rho}F_{\mu\rho\nu}V_{\lambda} 
 + \;\frac{41}{180}\;F_{\kappa\lambda}F_{\mu\lambda\kappa}F_{\nu\rho}V_{\mu}F_{\rho\nu} 
 + \;\frac{41}{180}\;F_{\kappa\lambda}F_{\mu\nu}V_{\rho}F_{\nu\mu}F_{\rho\lambda\kappa} 
 \nonumber \\ 
  &&  
 + \;\frac{41}{210}\;F_{\kappa\lambda\mu}F_{\nu\rho\mu\lambda}V_{\rho\nu\kappa} 
 + \;\frac{41}{210}\;F_{\kappa\lambda\mu}V_{\kappa\nu\rho}F_{\rho\nu\mu\lambda} 
 - \;\frac{41}{210}\;\mbox{i}\;VF_{\kappa\lambda}F_{\lambda\kappa}F_{\mu\nu}F_{\nu\rho}F_{\rho\mu} 
 \nonumber \\ 
  &&  
 - \;\frac{41}{210}\;\mbox{i}\;VF_{\kappa\lambda}F_{\lambda\mu}F_{\mu\kappa}F_{\nu\rho}F_{\rho\nu} 
 - \;\frac{43}{126}\;F_{\kappa\lambda}F_{\lambda\mu}F_{\nu\rho}F_{\mu\rho\nu}V_{\kappa} 
 - \;\frac{43}{126}\;\mbox{i}\;F_{\kappa\lambda}F_{\lambda\mu\nu\rho}F_{\rho\nu}V_{\mu\kappa} 
 \nonumber \\ 
  &&  
 - \;\frac{43}{126}\;\mbox{i}\;F_{\kappa\lambda}F_{\mu\nu\lambda\kappa}F_{\nu\rho}V_{\rho\mu} 
 + \;\frac{43}{315}\;VF_{\kappa\lambda\mu}F_{\kappa\nu\rho}F_{\rho\mu}F_{\nu\lambda} 
 - \;\frac{44}{105}\;\mbox{i}\;F_{\kappa\lambda\mu}F_{\nu\kappa\rho}V_{\mu}F_{\nu\rho\lambda} 
 \nonumber \\ 
  &&  
 + \;\frac{46}{105}\;F_{\kappa\lambda}F_{\lambda\mu}F_{\mu\nu}F_{\nu\rho}V_{\rho\kappa} 
 + \;\frac{46}{105}\;F_{\kappa\lambda}F_{\lambda\mu}F_{\mu\nu}V_{\rho}F_{\kappa\rho\nu} 
 + \;\frac{47}{420}\;VF_{\kappa\lambda\mu}F_{\kappa\nu\rho}F_{\mu\lambda}F_{\rho\nu} 
 \nonumber \\ 
  &&  
 + \;\frac{47}{420}\;VF_{\kappa\lambda}F_{\mu\nu}F_{\rho\lambda\kappa}F_{\rho\nu\mu} 
 - \;\frac{53}{105}\;\mbox{i}\;F_{\kappa\lambda\mu}F_{\mu\nu\rho}F_{\kappa\rho\lambda}V_{\nu} 
 + \;\frac{53}{315}\;F_{\kappa\lambda}F_{\lambda\mu}F_{\mu\nu}F_{\kappa\rho}V_{\rho\nu} 
 \nonumber \\ 
  &&  
 + \;\frac{53}{315}\;F_{\kappa\lambda}F_{\lambda\mu}F_{\mu\nu}V_{\kappa\rho}F_{\rho\nu} 
 - \;\frac{58}{315}\;F_{\kappa\lambda}F_{\mu\nu}F_{\lambda\rho}V_{\nu}F_{\kappa\rho\mu} 
 + \;\frac{58}{315}\;\mbox{i}\;F_{\kappa\lambda}V_{\mu\nu}F_{\rho\nu\lambda}F_{\rho\mu\kappa} 
 \nonumber \\ 
  &&  
 - \;\frac{59}{315}\;F_{\kappa\lambda}F_{\lambda\mu}F_{\mu\nu\rho}V_{\kappa}F_{\rho\nu} 
 + \;\frac{61}{252}\;VF_{\kappa\lambda\mu}F_{\nu\rho}F_{\kappa\rho\nu}F_{\mu\lambda} 
 + \;\frac{61}{252}\;VF_{\kappa\lambda}F_{\mu\nu\rho}F_{\rho\nu}F_{\mu\lambda\kappa} 
 \nonumber \\ 
  &&  
 - \;\frac{67}{315}\;\mbox{i}\;F_{\kappa\lambda}F_{\mu\nu}V_{\nu\rho}F_{\rho\mu\lambda\kappa} 
 - \;\frac{67}{630}\;\mbox{i}\;VF_{\kappa\lambda}F_{\lambda\mu}F_{\nu\rho}F_{\mu\kappa}F_{\rho\nu} 
 - \;\frac{67}{630}\;\mbox{i}\;VF_{\kappa\lambda}F_{\mu\nu}F_{\lambda\kappa}F_{\nu\rho}F_{\rho\mu} 
 \nonumber \\ 
  &&  
 - \;\frac{68}{315}\;VF_{\kappa\lambda\mu}F_{\kappa\nu}F_{\mu\rho}F_{\lambda\rho\nu} 
 - \;\frac{71}{315}\;F_{\kappa\lambda}F_{\lambda\mu\nu}F_{\nu\mu}F_{\kappa\rho}V_{\rho} 
 - \;\frac{71}{315}\;F_{\kappa\lambda}F_{\lambda\mu}F_{\mu\nu\rho}F_{\rho\nu}V_{\kappa} 
 \nonumber \\ 
  &&  
 - \;\frac{71}{315}\;F_{\kappa\lambda}F_{\lambda\mu}V_{\mu}F_{\nu\rho}F_{\kappa\rho\nu} 
 - \;\frac{71}{315}\;VF_{\kappa\lambda\mu}F_{\kappa\nu}F_{\nu\rho}F_{\rho\mu\lambda} 
 - \;\frac{73}{210}\;\mbox{i}\;F_{\kappa\lambda\mu}F_{\kappa\nu\rho}V_{\rho}F_{\nu\mu\lambda} 
 \nonumber \\ 
  &&  
 - \;\frac{73}{315}\;F_{\kappa\lambda}F_{\lambda\mu}F_{\mu\nu}F_{\rho\nu\kappa}V_{\rho} 
 - \;\frac{73}{315}\;\mbox{i}\;F_{\kappa\lambda\mu}F_{\nu\kappa\rho}V_{\nu}F_{\rho\mu\lambda} 
 - \;\frac{73}{630}\;VF_{\kappa\lambda\mu\nu}F_{\lambda\rho}F_{\nu\mu}F_{\rho\kappa} 
 \nonumber \\ 
  &&  
 - \;\frac{73}{630}\;\mbox{i}\;VF_{\kappa\lambda}F_{\mu\nu}F_{\lambda\rho}F_{\rho\kappa}F_{\nu\mu} 
 - \;\frac{73}{630}\;\mbox{i}\;VF_{\kappa\lambda}F_{\mu\nu}F_{\nu\rho}F_{\lambda\kappa}F_{\rho\mu} 
 + \;\frac{73}{1260}\;F_{\kappa\lambda}F_{\lambda\kappa}F_{\mu\nu}F_{\rho\nu\mu}V_{\rho} 
 \nonumber \\ 
  &&  
 + \;\frac{73}{1260}\;F_{\kappa\lambda}F_{\lambda\kappa}V_{\mu}F_{\mu\nu\rho}F_{\rho\nu} 
 + \;\frac{73}{1260}\;VF_{\kappa\lambda}F_{\mu\lambda\kappa}F_{\mu\nu\rho}F_{\rho\nu} 
 - \;\frac{82}{315}\;F_{\kappa\lambda}F_{\lambda\mu}V_{\nu}F_{\rho\mu\kappa}F_{\rho\nu} 
 \nonumber \\ 
  &&  
 - \;\frac{83}{630}\;F_{\kappa\lambda}F_{\lambda\mu\nu}F_{\kappa\rho}F_{\nu\mu}V_{\rho} 
 - \;\frac{88}{315}\;F_{\kappa\lambda}F_{\mu\lambda\nu}F_{\mu\rho}F_{\nu\kappa}V_{\rho} 
 + \;\frac{89}{315}\;\mbox{i}\;F_{\kappa\lambda}V_{\mu\nu}F_{\rho\nu\lambda}F_{\kappa\rho\mu} 
 \nonumber \\ 
  &&  
 - \;\frac{92}{315}\;F_{\kappa\lambda}F_{\mu\lambda\nu}F_{\mu\kappa}F_{\nu\rho}V_{\rho} 
 - \;\frac{92}{315}\;VF_{\kappa\lambda\mu}F_{\nu\mu\lambda}F_{\nu\rho}F_{\rho\kappa} 
 - \;\frac{92}{315}\;\mbox{i}\;VF_{\kappa\lambda}F_{\mu\nu}F_{\nu\rho}F_{\rho\mu}F_{\lambda\kappa} 
 \nonumber \\ 
  &&  
 + \;\frac{106}{315}\;VF_{\kappa\lambda\mu}F_{\mu\nu}F_{\lambda\rho}F_{\kappa\rho\nu} 
 - \;\frac{107}{315}\;F_{\kappa\lambda}F_{\mu\lambda\nu}F_{\mu\kappa}V_{\rho}F_{\rho\nu} 
 - \;\frac{107}{315}\;\mbox{i}\;F_{\kappa\lambda}V_{\mu\nu}F_{\nu\lambda\rho}F_{\rho\mu\kappa} 
 \nonumber \\ 
  &&  
 + \;\frac{107}{1260}\;F_{\kappa\lambda}F_{\mu\nu}F_{\lambda\kappa}F_{\rho\nu\mu}V_{\rho} 
 + \;\frac{107}{1260}\;F_{\kappa\lambda}F_{\mu\nu}F_{\lambda\kappa}V_{\rho}F_{\rho\nu\mu} 
 - \;\frac{113}{630}\;\mbox{i}\;F_{\kappa\lambda\mu}F_{\kappa\nu\rho}F_{\rho\mu\lambda}V_{\nu} 
 \nonumber \\ 
  &&  
 + \;\frac{127}{315}\;F_{\kappa\lambda}F_{\mu\nu}F_{\lambda\rho}F_{\nu\kappa}V_{\rho\mu} 
 - \;\frac{127}{630}\;VF_{\kappa\lambda\mu\nu}F_{\nu\mu}F_{\lambda\rho}F_{\rho\kappa} 
 - \;\frac{142}{315}\;F_{\kappa\lambda}F_{\lambda\mu}F_{\nu\mu\kappa}F_{\nu\rho}V_{\rho} 
 \nonumber \\ 
  &&  
 + \;\frac{143}{1260}\;VF_{\kappa\lambda}F_{\mu\nu\rho}F_{\mu\lambda\kappa}F_{\rho\nu} 
 - \;\frac{151}{315}\;VF_{\kappa\lambda\mu}F_{\nu\mu\rho}F_{\nu\lambda}F_{\rho\kappa} 
 - \;\frac{191}{630}\;\mbox{i}\;F_{\kappa\lambda}V_{\mu}F_{\lambda\nu\rho}F_{\mu\kappa\rho\nu} 
 \nonumber \\ 
  &&  
 + \;\frac{191}{1260}\;VF_{\kappa\lambda\mu}F_{\kappa\nu\rho}F_{\rho\nu}F_{\mu\lambda} 
 + \;\frac{191}{1260}\;VF_{\kappa\lambda}F_{\mu\nu}F_{\rho\nu\mu}F_{\rho\lambda\kappa} 
 - \;\frac{211}{315}\;VF_{\kappa\lambda\mu}F_{\mu\nu}F_{\rho\nu\lambda}F_{\rho\kappa} 
 \nonumber \\ 
  &&  
 - \;\frac{211}{630}\;\mbox{i}\;F_{\kappa\lambda}F_{\lambda\mu\nu}F_{\rho\nu\mu}V_{\rho\kappa} 
 - \;\frac{211}{630}\;\mbox{i}\;F_{\kappa\lambda}V_{\lambda\mu}F_{\mu\nu\rho}F_{\kappa\rho\nu} 
 + \;\frac{223}{1260}\;F_{\kappa\lambda}F_{\mu\nu}F_{\rho\lambda\kappa}F_{\nu\mu}V_{\rho} 
 \nonumber \\ 
  &&  
 + \;\frac{223}{1260}\;F_{\kappa\lambda}F_{\mu\nu}V_{\rho}F_{\lambda\kappa}F_{\rho\nu\mu} 
 + \;\frac{1}{21}\;F_{\kappa\lambda}F_{\lambda\mu}F_{\nu\rho\kappa\sigma}F_{\rho\nu\sigma\mu} 
 - \;\frac{1}{22}\;\mbox{i}\;F_{\kappa\lambda}F_{\lambda\mu\nu}F_{\rho\sigma}F_{\kappa\sigma\rho}F_{\nu\mu} 
 \nonumber \\ 
  &&  
 + \;\frac{1}{33}\;F_{\kappa\lambda\mu}F_{\kappa\mu\nu}F_{\rho\lambda\sigma}F_{\rho\sigma\nu} 
 - \;\frac{1}{33}\;F_{\kappa\lambda}F_{\lambda\mu}F_{\nu\rho\mu\sigma}F_{\rho\nu\sigma\kappa} 
 + \;\frac{1}{33}\;\mbox{i}\;F_{\kappa\lambda}F_{\lambda\mu}F_{\nu\mu\rho}F_{\nu\sigma}F_{\sigma\rho\kappa} 
 \nonumber \\ 
  &&  
 - \;\frac{1}{55}\;\mbox{i}\;F_{\kappa\lambda\mu\nu}F_{\lambda\rho\nu\sigma}F_{\rho\kappa\sigma\mu} 
 + \;\frac{1}{66}\;F_{\kappa\lambda}F_{\mu\nu\rho\sigma}F_{\lambda\kappa}F_{\nu\mu\sigma\rho} 
 - \;\frac{1}{99}\;\mbox{i}\;F_{\kappa\lambda}F_{\lambda\mu}F_{\kappa\nu\rho\sigma}F_{\sigma\rho}F_{\nu\mu} 
 \nonumber \\ 
  &&  
 - \;\frac{1}{154}\;\mbox{i}\;F_{\kappa\lambda}F_{\lambda\mu\nu}F_{\rho\sigma}F_{\kappa\nu\mu}F_{\sigma\rho} 
 - \;\frac{1}{154}\;\mbox{i}\;F_{\kappa\lambda}F_{\lambda\mu}F_{\kappa\nu\rho\sigma}F_{\nu\mu}F_{\sigma\rho} 
 + \;\frac{1}{198}\;\mbox{i}\;F_{\kappa\lambda}F_{\lambda\mu}F_{\nu\rho}F_{\mu\sigma}F_{\sigma\kappa\rho\nu} 
 \nonumber \\ 
  &&  
 + \;\frac{1}{231}\;F_{\kappa\lambda}F_{\lambda\mu}F_{\nu\rho}F_{\mu\sigma}F_{\sigma\kappa}F_{\rho\nu} 
 + \;\frac{1}{252}\;\mbox{i}\;F_{\kappa\lambda}F_{\lambda\mu\nu}F_{\kappa\rho\sigma}F_{\sigma\rho}F_{\nu\mu} 
 - \;\frac{1}{315}\;\mbox{i}\;F_{\kappa\lambda}F_{\lambda\mu}F_{\kappa\nu\rho}F_{\rho\mu\sigma}F_{\sigma\nu} 
 \nonumber \\ 
  &&  
 - \;\frac{1}{495}\;\mbox{i}\;F_{\kappa\lambda}F_{\lambda\mu}F_{\nu\rho}F_{\rho\kappa\sigma}F_{\sigma\nu\mu} 
 + \;\frac{1}{924}\;F_{\kappa\lambda\mu\nu\rho\sigma}F_{\nu\mu\lambda\kappa\sigma\rho} 
 + \;\frac{1}{990}\;\mbox{i}\;F_{\kappa\lambda}F_{\mu\nu}F_{\lambda\rho\sigma}F_{\kappa\nu\mu}F_{\sigma\rho} 
 \nonumber \\ 
  &&  
 + \;\frac{1}{1386}\;F_{\kappa\lambda}F_{\lambda\mu\nu\rho\sigma}F_{\nu\mu\kappa}F_{\sigma\rho} 
 - \;\frac{1}{3465}\;F_{\kappa\lambda}F_{\lambda\mu\nu}F_{\nu\rho\kappa\sigma}F_{\rho\sigma\mu} 
 + \;\frac{2}{63}\;F_{\kappa\lambda}F_{\lambda\mu}F_{\kappa\nu}F_{\rho\sigma}F_{\sigma\nu}F_{\rho\mu} 
 \nonumber \\ 
  &&  
 + \;\frac{2}{315}\;\mbox{i}\;F_{\kappa\lambda}F_{\lambda\mu}F_{\nu\rho\sigma}F_{\kappa\sigma\rho}F_{\nu\mu} 
 + \;\frac{2}{385}\;\mbox{i}\;F_{\kappa\lambda}F_{\lambda\mu}F_{\nu\rho\sigma}F_{\nu\kappa}F_{\mu\sigma\rho} 
 + \;\frac{2}{693}\;F_{\kappa\lambda}F_{\lambda\mu\nu\rho\sigma}F_{\nu\kappa}F_{\mu\sigma\rho} 
 \nonumber \\ 
  &&  
 - \;\frac{3}{35}\;\mbox{i}\;F_{\kappa\lambda}F_{\lambda\mu}F_{\kappa\nu\rho}F_{\rho\sigma}F_{\sigma\nu\mu} 
 - \;\frac{3}{35}\;\mbox{i}\;F_{\kappa\lambda}F_{\lambda\mu}F_{\nu\kappa\rho}F_{\nu\sigma}F_{\mu\sigma\rho} 
 + \;\frac{3}{35}\;\mbox{i}\;F_{\kappa\lambda}F_{\lambda\mu}F_{\nu\rho}F_{\kappa\sigma\rho\nu}F_{\sigma\mu} 
 \nonumber \\ 
  &&  
 + \;\frac{3}{35}\;\mbox{i}\;F_{\kappa\lambda}F_{\mu\lambda\nu}F_{\mu\rho\sigma}F_{\sigma\kappa}F_{\rho\nu} 
 - \;\frac{3}{55}\;\mbox{i}\;F_{\kappa\lambda}F_{\mu\nu}F_{\lambda\rho}F_{\sigma\nu\kappa}F_{\mu\sigma\rho} 
 + \;\frac{3}{77}\;F_{\kappa\lambda}F_{\mu\nu\lambda\rho}F_{\kappa\sigma}F_{\nu\mu\sigma\rho} 
 \nonumber \\ 
  &&  
 + \;\frac{3}{77}\;F_{\kappa\lambda}F_{\mu\nu\rho\sigma}F_{\sigma\lambda}F_{\nu\mu\rho\kappa} 
 - \;\frac{3}{77}\;\mbox{i}\;F_{\kappa\lambda}F_{\mu\nu}F_{\lambda\kappa}F_{\nu\rho\sigma}F_{\mu\sigma\rho} 
 + \;\frac{4}{63}\;\mbox{i}\;F_{\kappa\lambda}F_{\lambda\mu\nu}F_{\kappa\rho}F_{\rho\nu\sigma}F_{\sigma\mu} 
 \nonumber \\ 
  &&  
 + \;\frac{4}{63}\;\mbox{i}\;F_{\kappa\lambda}F_{\lambda\mu}F_{\mu\nu}F_{\kappa\rho\sigma}F_{\nu\sigma\rho} 
 - \;\frac{4}{99}\;F_{\kappa\lambda}F_{\lambda\mu\nu\rho}F_{\kappa\sigma\rho\nu}F_{\sigma\mu} 
 - \;\frac{4}{231}\;\mbox{i}\;F_{\kappa\lambda\mu}F_{\nu\rho\kappa\sigma}F_{\sigma\rho\nu\mu\lambda} 
 \nonumber \\ 
  &&  
 - \;\frac{4}{231}\;\mbox{i}\;F_{\kappa\lambda\mu}F_{\nu\rho\sigma\mu\lambda}F_{\sigma\rho\nu\kappa} 
 - \;\frac{4}{231}\;\mbox{i}\;F_{\kappa\lambda}F_{\lambda\mu}F_{\kappa\nu}F_{\rho\sigma}F_{\nu\mu\sigma\rho} 
 - \;\frac{4}{315}\;F_{\kappa\lambda}F_{\lambda\mu}F_{\kappa\nu}F_{\nu\rho}F_{\rho\sigma}F_{\sigma\mu} 
 \nonumber \\ 
  &&  
 - \;\frac{4}{693}\;F_{\kappa\lambda}F_{\mu\lambda\nu}F_{\kappa\rho\sigma}F_{\nu\mu\sigma\rho} 
 - \;\frac{4}{693}\;F_{\kappa\lambda}F_{\mu\nu\rho\sigma}F_{\lambda\sigma\rho}F_{\nu\mu\kappa} 
 - \;\frac{5}{63}\;F_{\kappa\lambda}F_{\lambda\mu\nu}F_{\rho\sigma\nu\kappa}F_{\sigma\rho\mu} 
 \nonumber \\ 
  &&  
 + \;\frac{5}{77}\;\mbox{i}\;F_{\kappa\lambda}F_{\lambda\mu}F_{\kappa\nu}F_{\rho\mu\sigma}F_{\rho\sigma\nu} 
 + \;\frac{5}{77}\;\mbox{i}\;F_{\kappa\lambda}F_{\lambda\mu}F_{\nu\rho\sigma}F_{\nu\sigma\kappa}F_{\rho\mu} 
 - \;\frac{5}{154}\;F_{\kappa\lambda}F_{\lambda\mu\nu\rho}F_{\mu\kappa\sigma}F_{\sigma\rho\nu} 
 \nonumber \\ 
  &&  
 - \;\frac{5}{198}\;F_{\kappa\lambda}F_{\lambda\mu\nu\rho}F_{\kappa\sigma}F_{\sigma\mu\rho\nu} 
 - \;\frac{5}{198}\;F_{\kappa\lambda}F_{\mu\nu\rho\sigma}F_{\nu\lambda}F_{\mu\kappa\sigma\rho} 
 - \;\frac{5}{231}\;\mbox{i}\;F_{\kappa\lambda}F_{\mu\nu\rho\lambda\sigma}F_{\rho\nu\mu\sigma\kappa} 
 \nonumber \\ 
  &&  
 + \;\frac{5}{693}\;F_{\kappa\lambda}F_{\lambda\mu}F_{\kappa\nu\rho\sigma}F_{\nu\mu\sigma\rho} 
 + \;\frac{6}{55}\;\mbox{i}\;F_{\kappa\lambda}F_{\lambda\mu}F_{\nu\kappa\rho}F_{\nu\mu\sigma}F_{\sigma\rho} 
 + \;\frac{7}{99}\;\mbox{i}\;F_{\kappa\lambda}F_{\lambda\mu\nu}F_{\kappa\rho}F_{\nu\sigma}F_{\sigma\rho\mu} 
 \nonumber \\ 
  &&  
 - \;\frac{7}{198}\;\mbox{i}\;F_{\kappa\lambda}F_{\mu\nu}F_{\lambda\kappa}F_{\rho\nu\sigma}F_{\rho\sigma\mu} 
 - \;\frac{7}{396}\;F_{\kappa\lambda\mu}F_{\mu\nu\rho}F_{\kappa\lambda\sigma}F_{\sigma\rho\nu} 
 - \;\frac{7}{396}\;F_{\kappa\lambda\mu}F_{\nu\kappa\rho}F_{\sigma\mu\lambda}F_{\nu\sigma\rho} 
 \nonumber \\ 
  &&  
 - \;\frac{8}{77}\;\mbox{i}\;F_{\kappa\lambda}F_{\lambda\mu\nu}F_{\rho\sigma}F_{\nu\kappa}F_{\mu\sigma\rho} 
 - \;\frac{8}{165}\;\mbox{i}\;F_{\kappa\lambda}F_{\lambda\mu\nu}F_{\nu\mu}F_{\rho\sigma}F_{\kappa\sigma\rho} 
 - \;\frac{8}{231}\;F_{\kappa\lambda}F_{\lambda\mu}F_{\kappa\nu}F_{\rho\sigma}F_{\sigma\mu}F_{\rho\nu} 
 \nonumber \\ 
  &&  
 - \;\frac{8}{231}\;F_{\kappa\lambda}F_{\lambda\mu}F_{\nu\rho}F_{\kappa\sigma}F_{\sigma\rho}F_{\nu\mu} 
 + \;\frac{8}{231}\;\mbox{i}\;F_{\kappa\lambda}F_{\lambda\kappa}F_{\mu\nu}F_{\nu\rho\sigma}F_{\mu\sigma\rho} 
 + \;\frac{8}{693}\;F_{\kappa\lambda}F_{\lambda\mu\nu\rho}F_{\sigma\mu\kappa}F_{\sigma\rho\nu} 
 \nonumber \\ 
  &&  
 + \;\frac{9}{55}\;\mbox{i}\;F_{\kappa\lambda}F_{\lambda\mu}F_{\mu\nu\rho}F_{\sigma\rho\kappa}F_{\sigma\nu} 
 - \;\frac{10}{231}\;\mbox{i}\;F_{\kappa\lambda\mu}F_{\kappa\nu\rho\mu\sigma}F_{\rho\nu\sigma\lambda} 
 - \;\frac{10}{231}\;\mbox{i}\;F_{\kappa\lambda\mu}F_{\nu\rho\mu\sigma}F_{\rho\nu\kappa\sigma\lambda} 
 \nonumber \\ 
  &&  
 + \;\frac{10}{231}\;\mbox{i}\;F_{\kappa\lambda}F_{\lambda\mu}F_{\nu\rho}F_{\mu\kappa\sigma}F_{\sigma\rho\nu} 
 - \;\frac{10}{693}\;F_{\kappa\lambda\mu}F_{\kappa\mu\nu}F_{\lambda\rho\sigma}F_{\nu\sigma\rho} 
 - \;\frac{11}{63}\;\mbox{i}\;F_{\kappa\lambda}F_{\lambda\mu}F_{\nu\kappa\rho}F_{\nu\sigma}F_{\sigma\rho\mu} 
 \nonumber \\ 
  &&  
 - \;\frac{11}{315}\;\mbox{i}\;F_{\kappa\lambda}F_{\mu\nu}F_{\lambda\rho}F_{\sigma\nu\mu}F_{\kappa\sigma\rho} 
 + \;\frac{13}{165}\;\mbox{i}\;F_{\kappa\lambda}F_{\lambda\mu}F_{\nu\rho\sigma}F_{\nu\mu}F_{\kappa\sigma\rho} 
 - \;\frac{13}{420}\;\mbox{i}\;F_{\kappa\lambda}F_{\lambda\mu\nu}F_{\kappa\rho\sigma}F_{\nu\mu}F_{\sigma\rho} 
 \nonumber \\ 
  &&  
 - \;\frac{13}{462}\;\mbox{i}\;F_{\kappa\lambda}F_{\lambda\mu\nu\rho\sigma}F_{\nu\mu\kappa\sigma\rho} 
 - \;\frac{13}{693}\;F_{\kappa\lambda}F_{\lambda\mu\nu\rho}F_{\rho\sigma}F_{\sigma\mu\nu\kappa} 
 - \;\frac{13}{693}\;F_{\kappa\lambda}F_{\mu\nu\lambda\rho}F_{\nu\sigma}F_{\mu\kappa\sigma\rho} 
 \nonumber \\ 
  &&  
 + \;\frac{13}{924}\;F_{\kappa\lambda\mu}F_{\kappa\mu\lambda}F_{\nu\rho\sigma}F_{\nu\sigma\rho} 
 - \;\frac{13}{1155}\;\mbox{i}\;F_{\kappa\lambda}F_{\mu\nu}F_{\lambda\rho}F_{\nu\kappa\sigma}F_{\sigma\rho\mu} 
 - \;\frac{16}{99}\;F_{\kappa\lambda}F_{\mu\nu\rho}F_{\rho\lambda\sigma}F_{\mu\kappa\sigma\nu} 
 \nonumber \\ 
  &&  
 + \;\frac{16}{231}\;F_{\kappa\lambda\mu}F_{\nu\mu\rho}F_{\kappa\lambda\sigma}F_{\nu\sigma\rho} 
 - \;\frac{16}{315}\;\mbox{i}\;F_{\kappa\lambda}F_{\lambda\mu}F_{\kappa\nu}F_{\rho\nu\sigma}F_{\rho\sigma\mu} 
 - \;\frac{16}{315}\;\mbox{i}\;F_{\kappa\lambda}F_{\lambda\mu}F_{\nu\kappa\rho}F_{\nu\rho\sigma}F_{\sigma\mu} 
 \nonumber \\ 
  &&  
 - \;\frac{16}{495}\;\mbox{i}\;F_{\kappa\lambda}F_{\lambda\mu}F_{\nu\rho\sigma}F_{\nu\mu\kappa}F_{\sigma\rho} 
 - \;\frac{16}{1155}\;F_{\kappa\lambda\mu}F_{\kappa\mu\nu}F_{\rho\nu\sigma}F_{\rho\sigma\lambda} 
 - \;\frac{17}{693}\;F_{\kappa\lambda}F_{\lambda\mu\nu\rho}F_{\mu\sigma}F_{\sigma\kappa\rho\nu} 
 \nonumber \\ 
  &&  
 + \;\frac{17}{1386}\;F_{\kappa\lambda\mu}F_{\kappa\nu\rho}F_{\sigma\mu\lambda}F_{\sigma\rho\nu} 
 + \;\frac{17}{6930}\;F_{\kappa\lambda}F_{\mu\nu\rho}F_{\rho\sigma\lambda\kappa}F_{\mu\sigma\nu} 
 + \;\frac{18}{385}\;\mbox{i}\;F_{\kappa\lambda}F_{\lambda\mu}F_{\nu\rho}F_{\rho\kappa\sigma}F_{\mu\sigma\nu} 
 \nonumber \\ 
  &&  
 + \;\frac{19}{77}\;\mbox{i}\;F_{\kappa\lambda}F_{\lambda\mu}F_{\nu\rho}F_{\sigma\rho\mu}F_{\kappa\sigma\nu} 
 - \;\frac{19}{198}\;\mbox{i}\;F_{\kappa\lambda}F_{\mu\lambda\kappa}F_{\mu\nu}F_{\rho\sigma}F_{\nu\sigma\rho} 
 + \;\frac{19}{315}\;F_{\kappa\lambda}F_{\mu\nu\rho\sigma}F_{\nu\lambda\kappa}F_{\mu\sigma\rho} 
 \nonumber \\ 
  &&  
 + \;\frac{19}{315}\;F_{\kappa\lambda}F_{\mu\nu\rho}F_{\sigma\lambda\kappa}F_{\sigma\mu\rho\nu} 
 + \;\frac{19}{462}\;\mbox{i}\;F_{\kappa\lambda}F_{\lambda\mu}F_{\kappa\nu}F_{\mu\rho\sigma}F_{\nu\sigma\rho} 
 - \;\frac{19}{693}\;\mbox{i}\;F_{\kappa\lambda}F_{\lambda\mu}F_{\nu\rho}F_{\sigma\mu\kappa}F_{\sigma\rho\nu} 
 \nonumber \\ 
  &&  
 + \;\frac{19}{924}\;F_{\kappa\lambda}F_{\lambda\kappa}F_{\mu\nu\rho\sigma}F_{\nu\mu\sigma\rho} 
 - \;\frac{19}{1155}\;\mbox{i}\;F_{\kappa\lambda}F_{\mu\nu}F_{\lambda\rho}F_{\nu\kappa\sigma}F_{\mu\sigma\rho} 
 + \;\frac{19}{41580}\;F_{\kappa\lambda}F_{\lambda\kappa}F_{\mu\nu}F_{\nu\mu}F_{\rho\sigma}F_{\sigma\rho} 
 \nonumber \\ 
  &&  
 + \;\frac{20}{231}\;\mbox{i}\;F_{\kappa\lambda}F_{\lambda\mu}F_{\nu\rho}F_{\rho\mu\kappa\sigma}F_{\sigma\nu} 
 - \;\frac{20}{693}\;\mbox{i}\;F_{\kappa\lambda}F_{\lambda\mu\nu\rho}F_{\rho\sigma}F_{\nu\kappa}F_{\sigma\mu} 
 - \;\frac{23}{99}\;F_{\kappa\lambda}F_{\mu\nu}F_{\lambda\rho}F_{\kappa\sigma}F_{\rho\nu}F_{\sigma\mu} 
 \nonumber \\ 
  &&  
 - \;\frac{23}{126}\;\mbox{i}\;F_{\kappa\lambda}F_{\lambda\mu\nu}F_{\nu\kappa}F_{\rho\sigma}F_{\mu\sigma\rho} 
 - \;\frac{23}{198}\;\mbox{i}\;F_{\kappa\lambda}F_{\lambda\kappa}F_{\mu\nu\rho}F_{\rho\sigma}F_{\mu\sigma\nu} 
 + \;\frac{23}{231}\;F_{\kappa\lambda}F_{\mu\lambda\nu}F_{\rho\kappa\sigma}F_{\rho\mu\sigma\nu} 
 \nonumber \\ 
  &&  
 + \;\frac{23}{231}\;F_{\kappa\lambda}F_{\mu\nu\rho\sigma}F_{\nu\sigma\lambda}F_{\mu\rho\kappa} 
 - \;\frac{23}{252}\;\mbox{i}\;F_{\kappa\lambda}F_{\mu\nu}F_{\rho\lambda\kappa}F_{\sigma\nu\mu}F_{\sigma\rho} 
 + \;\frac{23}{315}\;\mbox{i}\;F_{\kappa\lambda}F_{\mu\lambda\nu}F_{\mu\rho}F_{\nu\sigma}F_{\kappa\sigma\rho} 
 \nonumber \\ 
  &&  
 + \;\frac{23}{462}\;F_{\kappa\lambda}F_{\lambda\mu}F_{\nu\rho}F_{\kappa\sigma}F_{\sigma\mu}F_{\rho\nu} 
 + \;\frac{23}{693}\;F_{\kappa\lambda}F_{\lambda\kappa}F_{\mu\nu}F_{\nu\rho}F_{\mu\sigma}F_{\sigma\rho} 
 - \;\frac{23}{693}\;F_{\kappa\lambda}F_{\mu\nu\rho}F_{\rho\lambda\sigma}F_{\sigma\mu\nu\kappa} 
 \nonumber \\ 
  &&  
 - \;\frac{23}{3465}\;F_{\kappa\lambda}F_{\lambda\mu}F_{\mu\kappa\nu\rho\sigma}F_{\nu\sigma\rho} 
 + \;\frac{25}{154}\;\mbox{i}\;F_{\kappa\lambda}F_{\lambda\mu}F_{\mu\nu}F_{\rho\sigma}F_{\nu\kappa\sigma\rho} 
 + \;\frac{25}{231}\;\mbox{i}\;F_{\kappa\lambda}F_{\lambda\mu}F_{\mu\nu\rho}F_{\rho\kappa\sigma}F_{\sigma\nu} 
 \nonumber \\ 
  &&  
 + \;\frac{25}{231}\;\mbox{i}\;F_{\kappa\lambda}F_{\lambda\mu}F_{\nu\kappa\rho}F_{\mu\sigma}F_{\nu\sigma\rho} 
 + \;\frac{25}{231}\;\mbox{i}\;F_{\kappa\lambda}F_{\lambda\mu}F_{\nu\rho\sigma}F_{\sigma\kappa}F_{\nu\rho\mu} 
 - \;\frac{25}{693}\;F_{\kappa\lambda}F_{\lambda\mu\nu}F_{\nu\kappa\rho\sigma}F_{\mu\sigma\rho} 
 \nonumber \\ 
  &&  
 - \;\frac{25}{1386}\;F_{\kappa\lambda}F_{\mu\nu}F_{\lambda\rho}F_{\kappa\sigma}F_{\nu\mu}F_{\sigma\rho} 
 - \;\frac{26}{385}\;F_{\kappa\lambda}F_{\mu\nu\rho}F_{\sigma\mu\lambda}F_{\sigma\kappa\rho\nu} 
 + \;\frac{26}{495}\;\mbox{i}\;F_{\kappa\lambda}F_{\lambda\mu}F_{\mu\kappa\nu\rho}F_{\rho\sigma}F_{\sigma\nu} 
 \nonumber \\ 
  &&  
 + \;\frac{26}{495}\;\mbox{i}\;F_{\kappa\lambda}F_{\lambda\mu}F_{\nu\rho\sigma}F_{\sigma\rho}F_{\kappa\nu\mu} 
 + \;\frac{26}{693}\;F_{\kappa\lambda}F_{\mu\nu\lambda\kappa}F_{\rho\sigma}F_{\nu\mu\sigma\rho} 
 + \;\frac{26}{693}\;F_{\kappa\lambda}F_{\mu\nu\rho\sigma}F_{\sigma\rho}F_{\nu\mu\lambda\kappa} 
 \nonumber \\ 
  &&  
 + \;\frac{26}{693}\;\mbox{i}\;F_{\kappa\lambda}F_{\lambda\mu}F_{\mu\nu\kappa\rho}F_{\nu\sigma}F_{\sigma\rho} 
 - \;\frac{27}{154}\;\mbox{i}\;F_{\kappa\lambda}F_{\lambda\mu}F_{\nu\mu\kappa}F_{\rho\sigma}F_{\nu\sigma\rho} 
 - \;\frac{27}{385}\;\mbox{i}\;F_{\kappa\lambda}F_{\lambda\mu\nu}F_{\nu\rho}F_{\sigma\mu\kappa}F_{\sigma\rho} 
 \nonumber \\ 
  &&  
 - \;\frac{27}{770}\;\mbox{i}\;F_{\kappa\lambda\mu}F_{\mu\kappa\nu\rho\sigma}F_{\nu\lambda\sigma\rho} 
 - \;\frac{27}{770}\;\mbox{i}\;F_{\kappa\lambda\mu}F_{\mu\nu\rho\sigma}F_{\nu\lambda\kappa\sigma\rho} 
 - \;\frac{29}{495}\;F_{\kappa\lambda}F_{\mu\nu\rho}F_{\mu\lambda\sigma}F_{\sigma\kappa\rho\nu} 
 \nonumber \\ 
  &&  
 - \;\frac{29}{693}\;F_{\kappa\lambda}F_{\lambda\mu\nu\rho}F_{\rho\kappa\sigma}F_{\mu\sigma\nu} 
 - \;\frac{29}{693}\;\mbox{i}\;F_{\kappa\lambda}F_{\lambda\kappa}F_{\mu\nu\rho}F_{\mu\sigma}F_{\sigma\rho\nu} 
 - \;\frac{31}{990}\;F_{\kappa\lambda}F_{\mu\lambda\nu}F_{\nu\mu\kappa\rho\sigma}F_{\sigma\rho} 
 \nonumber \\ 
  &&  
 - \;\frac{31}{990}\;F_{\kappa\lambda}F_{\mu\nu\rho}F_{\mu\rho\sigma}F_{\sigma\nu\lambda\kappa} 
 - \;\frac{32}{231}\;F_{\kappa\lambda}F_{\lambda\mu\nu\rho}F_{\mu\rho\sigma}F_{\kappa\sigma\nu} 
 - \;\frac{32}{231}\;F_{\kappa\lambda}F_{\lambda\mu\nu}F_{\rho\nu\sigma}F_{\rho\kappa\sigma\mu} 
 \nonumber \\ 
  &&  
 - \;\frac{32}{231}\;\mbox{i}\;F_{\kappa\lambda}F_{\lambda\mu\nu}F_{\nu\rho}F_{\mu\sigma}F_{\kappa\sigma\rho} 
 - \;\frac{32}{315}\;F_{\kappa\lambda}F_{\mu\nu}F_{\rho\sigma\nu\lambda}F_{\sigma\kappa\rho\mu} 
 - \;\frac{32}{495}\;\mbox{i}\;F_{\kappa\lambda}F_{\lambda\mu}F_{\mu\nu}F_{\rho\kappa\sigma}F_{\rho\sigma\nu} 
 \nonumber \\ 
  &&  
 - \;\frac{32}{1155}\;F_{\kappa\lambda}F_{\mu\lambda\nu}F_{\mu\rho\nu\sigma}F_{\rho\sigma\kappa} 
 + \;\frac{32}{1155}\;\mbox{i}\;F_{\kappa\lambda}F_{\lambda\mu}F_{\mu\kappa\nu}F_{\nu\rho\sigma}F_{\sigma\rho} 
 - \;\frac{32}{1155}\;\mbox{i}\;F_{\kappa\lambda}F_{\lambda\mu}F_{\nu\rho}F_{\sigma\rho\mu}F_{\sigma\nu\kappa} 
 \nonumber \\ 
  &&  
 + \;\frac{32}{1155}\;\mbox{i}\;F_{\kappa\lambda}F_{\mu\nu}F_{\rho\lambda\sigma}F_{\rho\nu}F_{\kappa\sigma\mu} 
 - \;\frac{34}{693}\;F_{\kappa\lambda}F_{\lambda\mu}F_{\mu\nu}F_{\rho\sigma}F_{\nu\kappa}F_{\sigma\rho} 
 - \;\frac{34}{1155}\;\mbox{i}\;F_{\kappa\lambda\mu\nu}F_{\lambda\rho\kappa\sigma}F_{\sigma\rho\nu\mu} 
 \nonumber \\ 
  &&  
 + \;\frac{37}{693}\;F_{\kappa\lambda}F_{\lambda\kappa}F_{\mu\nu}F_{\rho\sigma}F_{\sigma\nu}F_{\rho\mu} 
 - \;\frac{37}{1386}\;F_{\kappa\lambda}F_{\lambda\mu\nu}F_{\rho\kappa\sigma}F_{\sigma\rho\nu\mu} 
 - \;\frac{37}{1386}\;F_{\kappa\lambda}F_{\mu\nu\rho\sigma}F_{\nu\mu\lambda}F_{\kappa\sigma\rho} 
 \nonumber \\ 
  &&  
 - \;\frac{37}{1386}\;\mbox{i}\;F_{\kappa\lambda}F_{\lambda\mu}F_{\kappa\nu\rho}F_{\mu\sigma}F_{\sigma\rho\nu} 
 - \;\frac{37}{6930}\;\mbox{i}\;F_{\kappa\lambda}F_{\lambda\mu\nu}F_{\nu\rho\sigma}F_{\mu\kappa}F_{\sigma\rho} 
 - \;\frac{38}{495}\;F_{\kappa\lambda}F_{\lambda\mu}F_{\mu\nu\rho\sigma}F_{\nu\kappa\sigma\rho} 
 \nonumber \\ 
  &&  
 - \;\frac{38}{1485}\;F_{\kappa\lambda}F_{\lambda\mu}F_{\mu\nu}F_{\nu\rho}F_{\rho\sigma}F_{\sigma\kappa} 
 - \;\frac{40}{693}\;F_{\kappa\lambda}F_{\mu\lambda\nu}F_{\rho\sigma}F_{\nu\mu\kappa\sigma\rho} 
 + \;\frac{41}{385}\;\mbox{i}\;F_{\kappa\lambda}F_{\mu\lambda\nu}F_{\mu\rho}F_{\sigma\nu\kappa}F_{\sigma\rho} 
 \nonumber \\ 
  &&  
 - \;\frac{41}{1386}\;F_{\kappa\lambda}F_{\lambda\mu\nu\rho\sigma}F_{\sigma\rho}F_{\nu\mu\kappa} 
 + \;\frac{41}{1386}\;\mbox{i}\;F_{\kappa\lambda}F_{\lambda\mu}F_{\mu\kappa\nu}F_{\rho\sigma}F_{\nu\sigma\rho} 
 - \;\frac{43}{231}\;F_{\kappa\lambda}F_{\lambda\mu\nu\rho}F_{\rho\sigma}F_{\mu\kappa\sigma\nu} 
 \nonumber \\ 
  &&  
 - \;\frac{43}{495}\;F_{\kappa\lambda}F_{\mu\nu\rho}F_{\mu\sigma\rho\lambda}F_{\kappa\sigma\nu} 
 - \;\frac{43}{693}\;\mbox{i}\;F_{\kappa\lambda}F_{\lambda\mu\nu}F_{\rho\kappa\sigma}F_{\rho\nu}F_{\sigma\mu} 
 - \;\frac{43}{693}\;\mbox{i}\;F_{\kappa\lambda}F_{\mu\lambda\nu}F_{\kappa\rho\sigma}F_{\sigma\mu}F_{\rho\nu} 
 \nonumber \\ 
  &&  
 + \;\frac{43}{3465}\;F_{\kappa\lambda\mu}F_{\kappa\nu\rho}F_{\sigma\rho\mu}F_{\sigma\nu\lambda} 
 - \;\frac{43}{6930}\;F_{\kappa\lambda}F_{\mu\nu}F_{\rho\lambda\sigma}F_{\sigma\rho\kappa\nu\mu} 
 - \;\frac{46}{693}\;F_{\kappa\lambda}F_{\mu\nu\lambda\kappa}F_{\rho\nu\sigma}F_{\rho\sigma\mu} 
 \nonumber \\ 
  &&  
 - \;\frac{46}{693}\;F_{\kappa\lambda}F_{\mu\nu}F_{\lambda\rho\sigma\nu\mu}F_{\sigma\rho\kappa} 
 + \;\frac{46}{693}\;F_{\kappa\lambda}F_{\mu\nu}F_{\rho\sigma\nu\lambda}F_{\sigma\rho\mu\kappa} 
 - \;\frac{46}{693}\;\mbox{i}\;F_{\kappa\lambda}F_{\lambda\mu\nu}F_{\kappa\rho\sigma}F_{\sigma\nu}F_{\rho\mu} 
 \nonumber \\ 
  &&  
 + \;\frac{47}{990}\;\mbox{i}\;F_{\kappa\lambda}F_{\lambda\mu}F_{\kappa\nu}F_{\nu\mu\rho\sigma}F_{\sigma\rho} 
 - \;\frac{48}{385}\;\mbox{i}\;F_{\kappa\lambda}F_{\lambda\mu}F_{\nu\kappa\rho}F_{\sigma\nu\mu}F_{\sigma\rho} 
 + \;\frac{52}{693}\;F_{\kappa\lambda}F_{\lambda\mu}F_{\nu\rho}F_{\kappa\sigma}F_{\rho\nu}F_{\sigma\mu} 
 \nonumber \\ 
  &&  
 - \;\frac{52}{3465}\;\mbox{i}\;F_{\kappa\lambda}F_{\mu\lambda\nu}F_{\rho\mu\sigma}F_{\rho\kappa}F_{\sigma\nu} 
 - \;\frac{53}{693}\;F_{\kappa\lambda}F_{\mu\nu\lambda\rho}F_{\sigma\nu\kappa}F_{\mu\sigma\rho} 
 - \;\frac{53}{693}\;\mbox{i}\;F_{\kappa\lambda}F_{\mu\nu}F_{\lambda\rho\sigma}F_{\nu\mu}F_{\kappa\sigma\rho} 
 \nonumber \\ 
  &&  
 - \;\frac{53}{3465}\;F_{\kappa\lambda}F_{\lambda\mu}F_{\mu\nu}F_{\kappa\rho}F_{\rho\sigma}F_{\sigma\nu} 
 - \;\frac{58}{495}\;F_{\kappa\lambda}F_{\lambda\mu}F_{\mu\kappa}F_{\nu\rho}F_{\rho\sigma}F_{\sigma\nu} 
 + \;\frac{58}{3465}\;\mbox{i}\;F_{\kappa\lambda}F_{\lambda\mu}F_{\nu\rho\sigma}F_{\kappa\nu\mu}F_{\sigma\rho} 
 \nonumber \\ 
  &&  
 - \;\frac{61}{1155}\;\mbox{i}\;F_{\kappa\lambda}F_{\lambda\mu}F_{\mu\nu\rho}F_{\rho\sigma}F_{\sigma\nu\kappa} 
 - \;\frac{61}{1155}\;\mbox{i}\;F_{\kappa\lambda}F_{\lambda\mu}F_{\nu\mu\rho}F_{\nu\sigma}F_{\kappa\sigma\rho} 
 + \;\frac{61}{1386}\;\mbox{i}\;F_{\kappa\lambda}F_{\lambda\mu}F_{\mu\nu\rho}F_{\kappa\sigma}F_{\sigma\rho\nu} 
 \nonumber \\ 
  &&  
 - \;\frac{62}{385}\;\mbox{i}\;F_{\kappa\lambda}F_{\lambda\mu}F_{\kappa\nu\rho}F_{\sigma\rho\mu}F_{\sigma\nu} 
 + \;\frac{62}{495}\;\mbox{i}\;F_{\kappa\lambda}F_{\lambda\mu}F_{\nu\rho}F_{\sigma\rho\kappa}F_{\mu\sigma\nu} 
 + \;\frac{64}{693}\;F_{\kappa\lambda}F_{\lambda\mu}F_{\kappa\nu}F_{\rho\sigma}F_{\nu\mu}F_{\sigma\rho} 
 \nonumber \\ 
  &&  
 + \;\frac{64}{1155}\;\mbox{i}\;F_{\kappa\lambda}F_{\lambda\mu}F_{\nu\mu\rho}F_{\sigma\nu\kappa}F_{\sigma\rho} 
 - \;\frac{65}{693}\;\mbox{i}\;F_{\kappa\lambda}F_{\lambda\mu\nu}F_{\nu\rho}F_{\kappa\sigma}F_{\mu\sigma\rho} 
 - \;\frac{67}{693}\;\mbox{i}\;F_{\kappa\lambda}F_{\lambda\mu}F_{\nu\mu\rho}F_{\kappa\sigma}F_{\nu\sigma\rho} 
 \nonumber \\ 
  &&  
 - \;\frac{67}{693}\;\mbox{i}\;F_{\kappa\lambda}F_{\lambda\mu}F_{\nu\rho\sigma}F_{\sigma\mu}F_{\nu\rho\kappa} 
 - \;\frac{71}{630}\;\mbox{i}\;F_{\kappa\lambda}F_{\mu\lambda\nu}F_{\mu\kappa}F_{\nu\rho\sigma}F_{\sigma\rho} 
 + \;\frac{71}{1386}\;\mbox{i}\;F_{\kappa\lambda}F_{\lambda\mu}F_{\nu\rho}F_{\kappa\sigma}F_{\sigma\mu\rho\nu} 
 \nonumber \\ 
  &&  
 - \;\frac{73}{990}\;F_{\kappa\lambda}F_{\mu\nu\rho}F_{\lambda\sigma\rho\nu}F_{\sigma\mu\kappa} 
 - \;\frac{79}{630}\;\mbox{i}\;F_{\kappa\lambda}F_{\mu\nu}F_{\rho\lambda\sigma}F_{\rho\kappa}F_{\sigma\nu\mu} 
 - \;\frac{82}{693}\;F_{\kappa\lambda}F_{\lambda\mu\nu\rho}F_{\sigma\rho\kappa}F_{\mu\sigma\nu} 
 \nonumber \\ 
  &&  
 + \;\frac{82}{693}\;\mbox{i}\;F_{\kappa\lambda}F_{\lambda\mu\nu}F_{\nu\rho}F_{\sigma\rho\kappa}F_{\sigma\mu} 
 - \;\frac{83}{3465}\;\mbox{i}\;F_{\kappa\lambda}F_{\lambda\mu}F_{\kappa\nu\rho}F_{\sigma\rho\nu}F_{\sigma\mu} 
 - \;\frac{83}{3465}\;\mbox{i}\;F_{\kappa\lambda}F_{\lambda\mu}F_{\kappa\nu}F_{\nu\rho\sigma}F_{\mu\sigma\rho} 
 \nonumber \\ 
  &&  
 + \;\frac{85}{693}\;F_{\kappa\lambda}F_{\mu\nu\lambda\rho}F_{\nu\kappa\sigma}F_{\mu\sigma\rho} 
 - \;\frac{85}{693}\;F_{\kappa\lambda}F_{\mu\nu\rho}F_{\rho\mu\lambda\sigma}F_{\kappa\sigma\nu} 
 + \;\frac{85}{693}\;F_{\kappa\lambda}F_{\mu\nu\rho}F_{\sigma\rho\lambda}F_{\sigma\mu\nu\kappa} 
 \nonumber \\ 
  &&  
 - \;\frac{85}{1386}\;\mbox{i}\;F_{\kappa\lambda}F_{\lambda\kappa}F_{\mu\nu\rho}F_{\sigma\rho\nu}F_{\sigma\mu} 
 - \;\frac{85}{1386}\;\mbox{i}\;F_{\kappa\lambda}F_{\mu\lambda\kappa}F_{\nu\rho\sigma}F_{\nu\mu}F_{\sigma\rho} 
 - \;\frac{86}{385}\;\mbox{i}\;F_{\kappa\lambda}F_{\lambda\mu\nu\rho}F_{\mu\sigma}F_{\rho\kappa}F_{\sigma\nu} 
 \nonumber \\ 
  &&  
 + \;\frac{86}{693}\;F_{\kappa\lambda}F_{\mu\nu}F_{\lambda\rho}F_{\nu\sigma}F_{\rho\kappa}F_{\sigma\mu} 
 - \;\frac{89}{693}\;F_{\kappa\lambda}F_{\lambda\mu\nu\rho}F_{\sigma\rho\nu}F_{\kappa\sigma\mu} 
 - \;\frac{89}{693}\;F_{\kappa\lambda}F_{\lambda\mu\nu}F_{\nu\rho\sigma}F_{\mu\kappa\sigma\rho} 
 \nonumber \\ 
  &&  
 - \;\frac{89}{693}\;F_{\kappa\lambda}F_{\mu\nu\rho}F_{\sigma\rho\lambda}F_{\mu\kappa\sigma\nu} 
 - \;\frac{89}{1386}\;\mbox{i}\;F_{\kappa\lambda}F_{\lambda\kappa}F_{\mu\nu\rho}F_{\mu\rho\sigma}F_{\sigma\nu} 
 - \;\frac{89}{1386}\;\mbox{i}\;F_{\kappa\lambda}F_{\lambda\kappa}F_{\mu\nu}F_{\rho\nu\sigma}F_{\rho\sigma\mu} 
 \nonumber \\ 
  &&  
 - \;\frac{94}{1155}\;\mbox{i}\;F_{\kappa\lambda}F_{\lambda\mu\nu\rho}F_{\kappa\sigma}F_{\rho\nu}F_{\sigma\mu} 
 + \;\frac{97}{3080}\;F_{\kappa\lambda}F_{\mu\nu}F_{\lambda\kappa}F_{\rho\sigma}F_{\nu\mu}F_{\sigma\rho} 
 + \;\frac{100}{693}\;\mbox{i}\;F_{\kappa\lambda}F_{\mu\nu}F_{\rho\lambda\sigma}F_{\rho\nu}F_{\sigma\mu\kappa} 
 \nonumber \\ 
  &&  
 + \;\frac{101}{1386}\;F_{\kappa\lambda\mu}F_{\nu\mu\lambda}F_{\kappa\rho\sigma}F_{\nu\sigma\rho} 
 - \;\frac{101}{6930}\;F_{\kappa\lambda\mu}F_{\mu\nu\rho}F_{\rho\lambda\sigma}F_{\sigma\nu\kappa} 
 - \;\frac{101}{6930}\;F_{\kappa\lambda\mu}F_{\nu\kappa\rho}F_{\sigma\nu\mu}F_{\lambda\sigma\rho} 
 \nonumber \\ 
  &&  
 + \;\frac{101}{9240}\;F_{\kappa\lambda\mu}F_{\nu\rho\sigma}F_{\kappa\mu\lambda}F_{\nu\sigma\rho} 
 - \;\frac{103}{630}\;\mbox{i}\;F_{\kappa\lambda}F_{\lambda\mu\nu}F_{\nu\kappa}F_{\mu\rho\sigma}F_{\sigma\rho} 
 - \;\frac{103}{693}\;F_{\kappa\lambda}F_{\lambda\mu\nu\rho}F_{\mu\sigma\rho\kappa}F_{\sigma\nu} 
 \nonumber \\ 
  &&  
 + \;\frac{103}{693}\;\mbox{i}\;F_{\kappa\lambda}F_{\lambda\mu}F_{\mu\nu\rho}F_{\rho\sigma}F_{\kappa\sigma\nu} 
 - \;\frac{103}{1386}\;F_{\kappa\lambda}F_{\lambda\kappa}F_{\mu\nu}F_{\nu\rho}F_{\rho\sigma}F_{\sigma\mu} 
 - \;\frac{103}{1386}\;\mbox{i}\;F_{\kappa\lambda}F_{\mu\lambda\nu}F_{\rho\sigma}F_{\mu\nu\kappa}F_{\sigma\rho} 
 \nonumber \\ 
  &&  
 - \;\frac{103}{1386}\;\mbox{i}\;F_{\kappa\lambda}F_{\mu\nu}F_{\rho\lambda\sigma}F_{\nu\mu}F_{\rho\sigma\kappa} 
 - \;\frac{109}{1260}\;\mbox{i}\;F_{\kappa\lambda}F_{\mu\nu}F_{\lambda\rho\sigma}F_{\sigma\rho}F_{\kappa\nu\mu} 
 + \;\frac{109}{3465}\;\mbox{i}\;F_{\kappa\lambda}F_{\lambda\mu}F_{\nu\rho}F_{\sigma\rho\nu}F_{\sigma\mu\kappa} 
 \nonumber \\ 
  &&  
 - \;\frac{115}{693}\;\mbox{i}\;F_{\kappa\lambda}F_{\mu\lambda\nu}F_{\rho\kappa\sigma}F_{\rho\mu}F_{\sigma\nu} 
 - \;\frac{122}{3465}\;\mbox{i}\;F_{\kappa\lambda}F_{\lambda\mu\nu}F_{\rho\nu\sigma}F_{\rho\kappa}F_{\sigma\mu} 
 - \;\frac{122}{3465}\;\mbox{i}\;F_{\kappa\lambda}F_{\lambda\mu}F_{\mu\nu}F_{\rho\nu\sigma}F_{\rho\sigma\kappa} 
 \nonumber \\ 
  &&  
 - \;\frac{124}{3465}\;F_{\kappa\lambda}F_{\mu\nu\rho}F_{\rho\mu\lambda\sigma}F_{\sigma\nu\kappa} 
 - \;\frac{128}{693}\;F_{\kappa\lambda}F_{\lambda\mu}F_{\nu\rho}F_{\mu\kappa}F_{\rho\sigma}F_{\sigma\nu} 
 + \;\frac{128}{1155}\;\mbox{i}\;F_{\kappa\lambda}F_{\mu\lambda\nu}F_{\kappa\rho}F_{\mu\nu\sigma}F_{\sigma\rho} 
 \nonumber \\ 
  &&  
 - \;\frac{128}{3465}\;\mbox{i}\;F_{\kappa\lambda}F_{\lambda\mu\nu}F_{\rho\sigma}F_{\kappa\sigma\nu}F_{\rho\mu} 
 + \;\frac{130}{693}\;\mbox{i}\;F_{\kappa\lambda}F_{\lambda\mu\nu}F_{\kappa\rho}F_{\nu\sigma}F_{\mu\sigma\rho} 
 - \;\frac{134}{3465}\;F_{\kappa\lambda}F_{\mu\lambda\nu}F_{\mu\kappa\rho\sigma}F_{\nu\sigma\rho} 
 \nonumber \\ 
  &&  
 + \;\frac{151}{13860}\;F_{\kappa\lambda}F_{\lambda\kappa}F_{\mu\nu}F_{\rho\sigma}F_{\nu\mu}F_{\sigma\rho} 
 + \;\frac{152}{3465}\;F_{\kappa\lambda}F_{\mu\nu}F_{\rho\sigma\nu\mu}F_{\sigma\rho\lambda\kappa} 
 + \;\frac{157}{3465}\;\mbox{i}\;F_{\kappa\lambda}F_{\lambda\mu\nu}F_{\nu\rho}F_{\mu\sigma}F_{\sigma\rho\kappa} 
 \nonumber \\ 
  &&  
 - \;\frac{163}{1155}\;F_{\kappa\lambda}F_{\mu\lambda\nu}F_{\rho\mu\sigma}F_{\sigma\rho\nu\kappa} 
 - \;\frac{163}{1155}\;F_{\kappa\lambda}F_{\mu\nu\lambda\rho}F_{\nu\mu\sigma}F_{\sigma\rho\kappa} 
 - \;\frac{163}{3465}\;F_{\kappa\lambda}F_{\lambda\mu\nu\rho}F_{\rho\mu\kappa\sigma}F_{\sigma\nu} 
 \nonumber \\ 
  &&  
 + \;\frac{163}{5544}\;F_{\kappa\lambda}F_{\lambda\kappa}F_{\mu\nu}F_{\rho\sigma}F_{\sigma\rho}F_{\nu\mu} 
 + \;\frac{164}{3465}\;F_{\kappa\lambda}F_{\lambda\mu}F_{\nu\rho}F_{\kappa\sigma}F_{\rho\mu}F_{\sigma\nu} 
 - \;\frac{166}{3465}\;F_{\kappa\lambda}F_{\lambda\mu\nu\rho}F_{\mu\rho\sigma}F_{\sigma\nu\kappa} 
 \nonumber \\ 
  &&  
 - \;\frac{166}{3465}\;F_{\kappa\lambda}F_{\lambda\mu\nu}F_{\rho\nu\sigma}F_{\sigma\rho\mu\kappa} 
 - \;\frac{166}{3465}\;F_{\kappa\lambda}F_{\mu\lambda\nu}F_{\rho\mu\sigma}F_{\rho\kappa\sigma\nu} 
 - \;\frac{166}{3465}\;F_{\kappa\lambda}F_{\mu\nu\lambda\rho}F_{\nu\mu\sigma}F_{\kappa\sigma\rho} 
 \nonumber \\ 
  &&  
 + \;\frac{169}{6930}\;\mbox{i}\;F_{\kappa\lambda}F_{\mu\nu}F_{\lambda\rho\sigma}F_{\nu\kappa}F_{\mu\sigma\rho} 
 - \;\frac{178}{3465}\;\mbox{i}\;F_{\kappa\lambda}F_{\lambda\mu}F_{\nu\rho\sigma}F_{\mu\kappa}F_{\nu\sigma\rho} 
 + \;\frac{179}{1386}\;F_{\kappa\lambda}F_{\mu\nu\lambda\kappa}F_{\nu\rho\sigma}F_{\mu\sigma\rho} 
 \nonumber \\ 
  &&  
 + \;\frac{181}{2772}\;F_{\kappa\lambda}F_{\mu\lambda\kappa}F_{\nu\rho\sigma}F_{\nu\mu\sigma\rho} 
 + \;\frac{181}{2772}\;F_{\kappa\lambda}F_{\mu\nu\rho\sigma}F_{\nu\sigma\rho}F_{\mu\lambda\kappa} 
 + \;\frac{212}{3465}\;\mbox{i}\;F_{\kappa\lambda}F_{\lambda\mu\nu}F_{\kappa\rho}F_{\sigma\nu\mu}F_{\sigma\rho} 
 \nonumber \\ 
  &&  
 + \;\frac{218}{3465}\;F_{\kappa\lambda}F_{\lambda\mu}F_{\kappa\nu}F_{\mu\rho}F_{\rho\sigma}F_{\sigma\nu} 
 + \;\frac{218}{3465}\;F_{\kappa\lambda}F_{\lambda\mu}F_{\mu\nu}F_{\kappa\rho}F_{\nu\sigma}F_{\sigma\rho} 
 - \;\frac{218}{3465}\;F_{\kappa\lambda}F_{\lambda\mu}F_{\nu\rho\sigma}F_{\nu\mu\kappa\sigma\rho} 
 \nonumber \\ 
  &&  
 - \;\frac{218}{3465}\;\mbox{i}\;F_{\kappa\lambda}F_{\lambda\mu\nu}F_{\nu\rho\sigma}F_{\sigma\kappa}F_{\rho\mu} 
 + \;\frac{221}{1980}\;F_{\kappa\lambda}F_{\mu\nu\rho}F_{\sigma\rho\nu}F_{\sigma\mu\lambda\kappa} 
 + \;\frac{227}{6930}\;\mbox{i}\;F_{\kappa\lambda}F_{\lambda\mu}F_{\mu\nu\rho\sigma}F_{\nu\kappa}F_{\sigma\rho} 
 \nonumber \\ 
  &&  
 - \;\frac{233}{83160}\;F_{\kappa\lambda}F_{\mu\nu}F_{\rho\sigma}F_{\lambda\kappa}F_{\nu\mu}F_{\sigma\rho} 
 - \;\frac{235}{2772}\;\mbox{i}\;F_{\kappa\lambda}F_{\mu\lambda\kappa}F_{\nu\rho\sigma}F_{\sigma\rho}F_{\nu\mu} 
 - \;\frac{236}{3465}\;\mbox{i}\;F_{\kappa\lambda}F_{\mu\nu}F_{\rho\lambda\sigma}F_{\nu\kappa}F_{\rho\sigma\mu} 
 \nonumber \\ 
  &&  
 + \;\frac{256}{3465}\;F_{\kappa\lambda}F_{\lambda\mu}F_{\kappa\nu}F_{\nu\rho}F_{\mu\sigma}F_{\sigma\rho} 
 - \;\frac{262}{3465}\;\mbox{i}\;F_{\kappa\lambda}F_{\lambda\mu}F_{\nu\mu\rho}F_{\nu\kappa\sigma}F_{\sigma\rho} 
 + \;\frac{263}{3465}\;\mbox{i}\;F_{\kappa\lambda}F_{\lambda\mu}F_{\nu\rho}F_{\sigma\rho\kappa}F_{\sigma\nu\mu} 
 \nonumber \\ 
  &&  
 + \;\frac{269}{3465}\;F_{\kappa\lambda}F_{\mu\lambda\nu}F_{\mu\rho\kappa\sigma}F_{\rho\sigma\nu} 
 + \;\frac{269}{3465}\;F_{\kappa\lambda}F_{\mu\nu\rho}F_{\mu\sigma\rho\lambda}F_{\sigma\nu\kappa} 
 - \;\frac{274}{3465}\;F_{\kappa\lambda}F_{\mu\lambda\nu}F_{\rho\sigma\mu\kappa}F_{\sigma\rho\nu} 
 \nonumber \\ 
  &&  
 - \;\frac{284}{3465}\;\mbox{i}\;F_{\kappa\lambda}F_{\lambda\mu}F_{\mu\kappa}F_{\nu\rho\sigma}F_{\nu\sigma\rho} 
 - \;\frac{284}{3465}\;\mbox{i}\;F_{\kappa\lambda}F_{\mu\lambda\nu}F_{\rho\sigma}F_{\mu\kappa}F_{\nu\sigma\rho} 
 - \;\frac{289}{2310}\;\mbox{i}\;F_{\kappa\lambda}F_{\mu\lambda\nu}F_{\mu\rho\sigma}F_{\nu\kappa}F_{\sigma\rho} 
 \nonumber \\ 
  &&  
 - \;\frac{326}{3465}\;\mbox{i}\;F_{\kappa\lambda}F_{\mu\lambda\nu}F_{\mu\rho}F_{\nu\kappa\sigma}F_{\sigma\rho} 
 + \;\frac{331}{6930}\;F_{\kappa\lambda}F_{\mu\nu}F_{\rho\sigma\lambda\kappa}F_{\sigma\rho\nu\mu} 
 + \;\frac{349}{1155}\;\mbox{i}\;F_{\kappa\lambda}F_{\lambda\mu}F_{\nu\rho}F_{\rho\mu\sigma}F_{\kappa\sigma\nu} 
 \nonumber \\ 
  &&  
 - \;\frac{349}{3465}\;\mbox{i}\;F_{\kappa\lambda}F_{\lambda\mu}F_{\nu\mu\kappa}F_{\nu\rho\sigma}F_{\sigma\rho} 
 - \;\frac{356}{3465}\;\mbox{i}\;F_{\kappa\lambda}F_{\mu\nu}F_{\lambda\rho}F_{\sigma\nu\kappa}F_{\sigma\rho\mu} 
 + \;\frac{358}{3465}\;\mbox{i}\;F_{\kappa\lambda}F_{\lambda\mu}F_{\nu\rho}F_{\rho\mu\sigma}F_{\sigma\nu\kappa} 
 \nonumber \\ 
  &&  
 - \;\frac{368}{3465}\;F_{\kappa\lambda}F_{\mu\nu\rho}F_{\lambda\sigma\rho\nu}F_{\kappa\sigma\mu} 
 - \;\frac{376}{3465}\;\mbox{i}\;F_{\kappa\lambda}F_{\lambda\mu}F_{\kappa\nu\rho}F_{\rho\sigma}F_{\mu\sigma\nu} 
 - \;\frac{397}{3465}\;\mbox{i}\;F_{\kappa\lambda}F_{\mu\nu}F_{\lambda\rho\sigma}F_{\sigma\nu}F_{\kappa\rho\mu} 
 \nonumber \\ 
  &&  
 + \;\frac{398}{3465}\;F_{\kappa\lambda}F_{\lambda\mu}F_{\kappa\nu}F_{\mu\rho}F_{\nu\sigma}F_{\sigma\rho} 
 + \;\frac{404}{3465}\;\mbox{i}\;F_{\kappa\lambda}F_{\mu\lambda\nu}F_{\kappa\rho}F_{\mu\sigma}F_{\nu\sigma\rho} 
 + \;\frac{404}{3465}\;\mbox{i}\;F_{\kappa\lambda}F_{\mu\nu}F_{\lambda\rho\sigma}F_{\sigma\nu}F_{\rho\mu\kappa} 
 \nonumber \\ 
  &&  
 - \;\frac{421}{3465}\;\mbox{i}\;F_{\kappa\lambda}F_{\mu\nu}F_{\lambda\rho\sigma}F_{\sigma\kappa}F_{\rho\nu\mu} 
 - \;\frac{428}{1155}\;F_{\kappa\lambda\mu}F_{\kappa\nu\rho}F_{\rho\mu\sigma}F_{\lambda\sigma\nu} 
 - \;\frac{452}{3465}\;\mbox{i}\;F_{\kappa\lambda}F_{\mu\nu}F_{\lambda\rho}F_{\kappa\sigma\rho\nu}F_{\sigma\mu} 
 \nonumber \\ 
  &&  
 - \;\frac{458}{3465}\;F_{\kappa\lambda}F_{\lambda\mu\nu\rho}F_{\sigma\rho\nu}F_{\sigma\mu\kappa} 
 - \;\frac{458}{3465}\;F_{\kappa\lambda}F_{\mu\lambda\nu}F_{\mu\rho\sigma}F_{\nu\kappa\sigma\rho} 
 - \;\frac{479}{6930}\;\mbox{i}\;F_{\kappa\lambda}F_{\mu\nu}F_{\lambda\rho}F_{\kappa\sigma\nu\mu}F_{\sigma\rho} 
 \nonumber \\ 
  &&  
 - \;\frac{499}{6930}\;\mbox{i}\;F_{\kappa\lambda}F_{\lambda\mu\nu}F_{\rho\sigma}F_{\nu\mu}F_{\kappa\sigma\rho} 
 - \;\frac{512}{3465}\;\mbox{i}\;F_{\kappa\lambda}F_{\lambda\mu\nu}F_{\kappa\rho}F_{\sigma\rho\nu}F_{\sigma\mu} 
 + \;\frac{541}{3465}\;\mbox{i}\;F_{\kappa\lambda}F_{\lambda\mu}F_{\nu\rho}F_{\sigma\rho\nu}F_{\kappa\sigma\mu} 
 \nonumber \\ 
  &&  
 - \;\frac{541}{3465}\;\mbox{i}\;F_{\kappa\lambda}F_{\mu\lambda\kappa}F_{\nu\rho}F_{\sigma\rho\mu}F_{\sigma\nu} 
 - \;\frac{541}{6930}\;F_{\kappa\lambda\mu}F_{\kappa\mu\nu}F_{\nu\rho\sigma}F_{\lambda\sigma\rho} 
 - \;\frac{541}{6930}\;F_{\kappa\lambda}F_{\lambda\mu\nu}F_{\rho\sigma\nu\mu}F_{\sigma\rho\kappa} 
 \nonumber \\ 
  &&  
 - \;\frac{541}{6930}\;F_{\kappa\lambda}F_{\mu\lambda\nu}F_{\nu\mu\rho\sigma}F_{\kappa\sigma\rho} 
 + \;\frac{554}{3465}\;\mbox{i}\;F_{\kappa\lambda}F_{\lambda\mu}F_{\mu\nu}F_{\nu\rho\sigma}F_{\kappa\sigma\rho} 
 - \;\frac{569}{3465}\;\mbox{i}\;F_{\kappa\lambda}F_{\lambda\mu\nu}F_{\nu\rho}F_{\mu\kappa\sigma}F_{\sigma\rho} 
 \nonumber \\ 
  &&  
 + \;\frac{569}{13860}\;F_{\kappa\lambda}F_{\mu\lambda\kappa}F_{\mu\nu\rho\sigma}F_{\nu\sigma\rho} 
 + \;\frac{569}{13860}\;F_{\kappa\lambda}F_{\mu\nu\rho}F_{\mu\sigma\rho\nu}F_{\sigma\lambda\kappa} 
 + \;\frac{593}{6930}\;\mbox{i}\;F_{\kappa\lambda}F_{\lambda\mu}F_{\mu\nu}F_{\nu\kappa\rho\sigma}F_{\sigma\rho} 
 \nonumber \\ 
  &&  
 + \;\frac{652}{3465}\;\mbox{i}\;F_{\kappa\lambda}F_{\lambda\mu}F_{\nu\rho}F_{\kappa\sigma\rho\mu}F_{\sigma\nu} 
 + \;\frac{697}{3465}\;\mbox{i}\;F_{\kappa\lambda}F_{\lambda\mu}F_{\nu\rho}F_{\rho\sigma}F_{\mu\kappa\sigma\nu} 
 - \;\frac{743}{3465}\;F_{\kappa\lambda\mu}F_{\kappa\nu\rho}F_{\rho\mu\sigma}F_{\sigma\nu\lambda} 
 \nonumber \\ 
  &&  
 - \;\frac{743}{3465}\;F_{\kappa\lambda\mu}F_{\kappa\nu\rho}F_{\sigma\rho\mu}F_{\lambda\sigma\nu} 
 - \;\frac{761}{13860}\;\mbox{i}\;F_{\kappa\lambda}F_{\lambda\mu\nu}F_{\nu\mu}F_{\kappa\rho\sigma}F_{\sigma\rho} 
 - \;\frac{793}{6930}\;\mbox{i}\;F_{\kappa\lambda}F_{\mu\nu}F_{\lambda\rho}F_{\sigma\nu\mu}F_{\sigma\rho\kappa} 
 \nonumber \\ 
  &&  
 - \;\frac{794}{3465}\;F_{\kappa\lambda}F_{\lambda\mu}F_{\nu\rho}F_{\mu\sigma}F_{\rho\kappa}F_{\sigma\nu} 
 - \;\frac{838}{3465}\;\mbox{i}\;F_{\kappa\lambda}F_{\mu\lambda\nu}F_{\kappa\rho}F_{\sigma\rho\mu}F_{\sigma\nu} 
 - \;\frac{851}{6930}\;\mbox{i}\;F_{\kappa\lambda}F_{\lambda\mu}F_{\nu\rho\sigma}F_{\sigma\rho}F_{\nu\mu\kappa} 
 \nonumber \\ 
  &&  
 + \;\frac{877}{13860}\;F_{\kappa\lambda\mu}F_{\kappa\nu\rho}F_{\sigma\rho\nu}F_{\sigma\mu\lambda} 
 + \;\frac{949}{3465}\;\mbox{i}\;F_{\kappa\lambda}F_{\lambda\mu}F_{\nu\rho}F_{\rho\sigma\mu\kappa}F_{\sigma\nu} 
 + \;\frac{953}{13860}\;F_{\kappa\lambda}F_{\mu\nu\rho}F_{\mu\sigma\lambda\kappa}F_{\sigma\rho\nu} 
 \nonumber \\ 
  &&  
 + \;\frac{1718}{10395}\;F_{\kappa\lambda}F_{\lambda\mu}F_{\nu\rho}F_{\rho\kappa}F_{\mu\sigma}F_{\sigma\nu} 
\nonumber
\ear

\newpage

\renewcommand{\large}{\normalsize}


\begin{thebibliography}{99}
\bibitem{feynman}R. P. Feynman, {\sl Phys. Rev.} {\bf 80}
(1950) 440.
\bibitem{fradkin}E. S. Fradkin, {\sl Nucl. Phys.} {\bf 76}
(1966) 588.
\bibitem{brink} L. Brink, P. Di Vecchia, P. Howe, {\sl Nucl. Phys.} 
{\bf B118} (1977) 76.
\bibitem{bordi} F. Bordi, R, Casalbuoni, {\sl Phys. Lett.} {\bf B93} (1980) 
308.
\bibitem{barducci} A. Barducci, F. Bordi, R, Casalbuoni, {\sl Nuovo Cim.}
{\bf 64B} (1981) 287.
\bibitem{gitman} E. S. Fradkin, D. M. Gitman, {\sl Phys. Rev.} {\bf D44} 
(1991) 3230.
\bibitem{polybook}A. M. Polyakov, Gauge Fields and Strings, 
Harwood 1987, and references therein.
\bibitem{cks}F. Cooper, A. Khare, U. Sukhatme,
{\sl Phys. Rep.} {\bf 251} (1995) 267.
\bibitem{hjs}M. B. Halpern, A. Jevicki, P. Senjanovic,
{\sl Phys. Rev.} {\bf D16} (1977) 2476.
\bibitem{alv-g}
L. Alvarez-Gaum\'e, {\sl Commun. Math. Phys.} {\bf 90} (1983) 161. 
\bibitem{alvwit}L. Alvarez-Gaum\'e, E. Witten, {\sl Nucl. Phys.}
{\bf B234} (1984) 269.
\bibitem{fw}
D. Friedan, P. Windey, {\sl Nucl. Phys.} {\bf B235} (1984) 395.
\bibitem{bast}F. Bastianelli, {\sl Nucl. Phys.} {\bf B376} (1992) 113;
F. Bastianelli, P. van Nieuwenhuizen, {\sl Nucl. Phys.} {\bf B389} 
(1993) 53.
\bibitem{boer} J. de Boer, B. Peeters, K. Skenderis, P. van Nieuwenhuizen,
{\sl Nucl. Phys.} {\bf B446} (1995) 211; Nucl. Phys. {\bf B459} (1996) 631.
\bibitem{abp}
M. F. Atiyah, R. Bott, V. K. Patodi, {\sl Inv. Math.} {\bf 19} (1973) 279.
\bibitem{bk}
Z. Bern, D. A. Kosower, {\sl Phys. Rev. Lett.} {\bf 66} (1991) 1669;
{\sl Nucl. Phys.} {\bf B379} (1992) 451.
\bibitem{bd}Z. Bern, D. C. Dunbar, {\sl Nucl. Phys.} {\bf B379} (1992) 562.
\bibitem{5glu}Z. Bern, L. Dixon, D. A. Kosower, 
{\sl Phys. Rev. Lett.} {\bf 70} (1993) 2677. 
\bibitem{dannor}D. C. Dunbar, P. S. Norridge, {\sl Nucl. Phys.} 
{\bf B433} (1995) 181. 
\bibitem{str1}M. J. Strassler, {\sl Nucl. Phys.} {\bf B385} (1992) 145.
\bibitem{str2}M. J. Strassler, SLAC-PUB-5978 (1992) (unpublished).
\bibitem{ss}M. G. Schmidt, C. Schubert, 
{\sl Phys. Lett.} {\bf B318} (1993) 438. 
\bibitem{chd}D. Cangemi, E. D'Hoker, G. Dunne,
{\sl Phys. Rev.} {\bf D51} (1995) 2513.
\bibitem{rss}M. Reuter, M. G. Schmidt, C. Schubert,
IASSNS-HEP-96-90, to appear in {\sl Ann. Phys.} (hep-th/9610191).
\bibitem{adlsch}S.L. Adler, C. Schubert, 
{\sl Phys. Rev. Lett.} {\bf 77} (1996) 1695. 
\bibitem{sch}C. Schubert, {\sl Act. Phys. Pol.} {\bf B27} (1996) 2965.
\bibitem{mckeon}D.G.C. McKeon, {\sl Ann. Phys.} {\bf 224} (1993) 139; 
D.G.C. McKeon, A. Rebhan, {\sl Phys.Rev.} {\bf D48} (1993) 2891;
V.P. Gusynin, I.A. Shovkovy, {\sl Can. J. Phys.} {\bf 74} (1996) 282. 
\bibitem{fss}D. Fliegner, M. G. Schmidt, C. Schubert, 
{\sl Z. Phys.} {\bf C64} (1994) 111.
\bibitem{fhss1}
D. Fliegner, P. Haberl, M. G. Schmidt, C. Schubert,
Discourses in Mathematics and its Applications, No. 4, p. 87,
Texas A\&M 1995 (hep-th/9411177).
\bibitem{fhss2}
D. Fliegner, P. Haberl, M. G. Schmidt, C. Schubert,
New Computing Techniques in Physics Research, p. 199, 
World Scientific 1996 (hep-th/9505077).
\bibitem{fhss3}
D. Fliegner, P. Haberl, M.G. Schmidt, C. Schubert,
HD-THEP-96/54, to appear in {\sl Nucl. Instrum. Meth.}
(hep-th/9702092). 
\bibitem{flinter} 
http://www.thphys.uni-heidelberg.de/$\tilde{\phantom{x}}$fliegner \,.
\bibitem{ball}R. D. Ball, {\sl Phys. Rep.} {\bf 182} (1989) 1.
\bibitem{klsch} J. Kripfganz, A. Laser, M.G. Schmidt, {\sl Nucl. Phys.} 
{\bf B433} (1995) 467; {\sl Z. Phys.} {\bf C73} (1997) 353.
\bibitem{khsch} J. Kripfganz, Hellmund, M.G. Schmidt, {\sl Phys. Rev.} 
{\bf D50} (1994) 7650.
\bibitem{dodgrad}
S. Dodelsohn, B. Gradwohl, {\sl Nucl. Phys.} {\bf B400} (1993) 435.
\bibitem{zhitnitsky}A. R. Zhitnitsky, {\sl Phys. Rev.} {\bf D54} (1996) 5148. 
\bibitem{dunne}G. Dunne, {\sl Int. J. Mod. Phys.} {\bf A12} (1997) 1143.
\bibitem{carson}L. Carson, {\sl Phys. Rev.} {\bf D42} (1990) 2853.
\bibitem{vdv}A. van de Ven, {\sl Nucl. Phys.} {\bf B250} (1985) 593.
\bibitem{abc}P. Amsterdamski, A. L. Berkin, D. J. Connor, 
{\sl Class. Quant. Grav.} {\bf 6} (1989) 1981. 
\bibitem{avra}
I. G. Avramidi, {\sl Nucl. Phys.} {\bf B355} (1991) 712.
\bibitem{bles}A. A. Belkov, D. Ebert, A. V. Lanyov, A. Schaale,  
{\sl Int. Journ. Mod. Phys.} {\bf C4} (1993) 775;
{\sl Int. Journ. Mod. Phys.} {\bf A8} (1993) 1313.
\bibitem{carsal}J. Caro, L. L. Salcedo, 
{\sl Phys. Lett.} {\bf B309} (1993) 359.
\bibitem{bargus}A. O. Barvinsky, Yu. V. Gusev, MANITOBA-11-94
(gr-qc/9507026).
\bibitem{form}J. A. M. Vermaseren, Symbolic Manipulation with
FORM (Version 2), CAN 1991.
\bibitem{PERL} L. Wall, R.L. Schwartz, Programming PERL,
O'Reilly 1991.
\bibitem{M} P. Overmann, M --- Reference Manual 1997;\\
http://www.thphys.uni-heidelberg.de/$\tilde{\phantom{x}}$overmann \,.
\bibitem{lanyov}A. A. Belkov, A. V. Lanyov, A. Schaale, 
{\sl Comp. Phys. Commun.} {\bf 95} (1996) 123.
\bibitem{kikkawa}K. Kikkawa, {\sl Prog. Theor. Phys.} {\bf 56} (1976) 947.  
\bibitem{mwz} R. MacKenzie, F. Wilczek, A. Zee, 
{\sl Phys. Rev. Lett.} {\bf 53} (1984) 2203.
\bibitem{fraser}
C. M. Fraser, {\sl Z. Phys.} {\bf C28} (1985) 101.   
\bibitem{tH}G. 't Hooft, {\sl Nucl. Phys.} {\bf B62} (1973) 444.
\bibitem{onofri}E. Onofri, {\sl Am. J. Phys.} {\bf 46} (1978) 379.
\bibitem{fow}Y. Fujiwara, T. A. Osborn, S. F. J. Wilk, 
{\sl Phys. Rev.} {\bf A25} (1982) 14.
\bibitem{zuk1}J. A. Zuk, {\sl Journ. Phys.} {\bf A18} (1985) 1795;
{\sl Phys. Rev.} {\bf D34} (1986) 1791.
\bibitem{zuk2}J. A. Zuk, {\sl Nucl. Phys.} {\bf 280} (1987) 125.
\bibitem{nepo}R. I. Nepomechie, {\sl Phys. Rev.} {\bf D31} (1985) 3291.
\bibitem{chan}L. Chan, {\sl Phys. Rev.} {\bf D38} (1988) 3739.
\bibitem{bog}A. O. Barvinsky, T. A. Osborn, Yu. V. Gusev,
{\sl J. Math. Phys.} {\bf 36} (1995) 30. 
\bibitem{bv}A. O. Barvinsky, G. A. Vilkovisky, {\sl Nucl. Phys.}
{\bf B282} (1987) 163; {\sl Nucl. Phys.} {\bf B333} (1990) 471; 
{\sl Nucl. Phys.} {\bf B333} (1990) 512;
A. O. Barvinsky, Yu. V. Gusev, V. V. Zhytnikov, G. A. Vilkovisky, 
{\sl J. Math. Phys.} {\bf 35} (1994) 3543.
\bibitem{carmcl}L. Carson, L. McLerran, {\sl Phys. Rev.}
{\bf D41} (1990) 647.
\bibitem{shif}M. A. Shifman, {\sl Nucl. Phys.} {\bf B173} (1980) 12.
\bibitem{muellerbasis}U. M\"uller, New Computing Techniques in
Physics Research IV, p. 193, World Scientific 1996; DESY-96-154
(hep-th/9701124).
\bibitem{vandevennew} A. van de Ven, to be published.
\bibitem{mnss} M. Mondrag\'on, L. Nellen, M. G. Schmidt, C. Schubert,
{\sl Phys. Lett.} {\bf B351} (1995) 200.
\bibitem{fnms} D. Fliegner, M. Mondrag\'on, L. Nellen, M. G. Schmidt,
work in progress.
\bibitem{ss2}M. G. Schmidt, C. Schubert, {\sl Phys. Lett.} {\bf B331} 
(1994) 69. 
\end{thebibliography}
\end{document}